\documentclass[prd, preprint, tightenlines, superscriptaddress, showpacs, byrevtex]{revtex4-1}

\usepackage[normalem]{ulem}

\usepackage{xcolor}
\usepackage{amsmath}
\usepackage{graphicx}
\usepackage[colorlinks=true, allcolors=blue]{hyperref} 
\usepackage{multirow}
\usepackage{comment}
\usepackage{enumitem}
\usepackage{lineno}
\usepackage{textcomp}
\usepackage{tikz}
\usepackage{xcolor}
\usepackage{geometry} 
\usepackage{siunitx}
\usepackage {cancel}
\usepackage{bm}
\usepackage{ulem}

\numberwithin{equation}{section}
\newcommand{\EE}{e^+e^-}

\newcommand{\pp}{\pi^+\pi^-}
\newcommand{\gev}{\rm GeV}

\newcommand{\gevcs}{{\rm GeV}/c^2}
\newcommand{\mev}{\rm MeV}

\newcommand{\kev}{\rm keV}

\newcommand{\pb}{\si{\pico\barn}}
\newcommand{\fb}{\si{\femto\barn}}
\newcommand{\ab}{\si{\atto\barn}}

\newcommand{\infb}{\fb^{-1}}
\newcommand{\inpb}{\pb^{-1}}
\newcommand{\inab}{\ab^{-1}}

\newcommand{\s}{\si{\second }}
\newcommand{\cm}{\si{\centi\metre}}

\newcommand{\dd}{\textit{D}\bar{D}}
\newcommand{\ndd}{\mbox{\text{non}-}\dd}
\newcommand{\psinn}{\psi(3770)\to\mbox{\text{non}-}\dd}

\newcommand{\bb}{B\Bar{B}}
\newcommand{\upsilonnn}{\Upsilon(4S)\to\mbox{\text{non}-}\bb}
\newcommand{\nbb}{\mbox{\text{non}-}\bb}
\newcommand{\psip}{\psi(3770)}
\newcommand{\ufs}{\Upsilon(4S)}
\newcommand{\BR}{\mathcal{B}}
\newcommand{\sexp}{\sigma_{\rm{exp}}}

\newcommand{\ecm}{E_{\rm{c.m.}}}
\newcommand{\beq}{\begin{equation}}
\newcommand{\eeq}{\end{equation}}
\newcommand{\npt}{N_{\rm{pt}}}
\newcommand{\nbr}{N_{\rm{br}}}

\newcommand{\nsamp}{N_{\rm{samp}}} 
 
\newcommand{\llow}{L_{\rm{low}}} 
\newcommand{\lhigh}{L_{\rm{high}}} 
\newcommand{\xexp}{x_{i}^{\rm{exp}}}

\def\Journal#1#2#3#4{{#1} {\bf #2}, #3 (#4)}

\def\PLB{Phys. Lett. B}
\def\PRL{Phys. Rev. Lett.}
\def\PRD{Phys. Rev. D}

\def\CPC{Chin. Phys. C}

\def\PTEP{Prog. Theor. Exp. Phys.}

\def\PTPS{Prog. Theor. Phys. Suppl.}

\begin{document}

\preprint{}

\title{\bf\boldmath Data taking strategy for $\psi(3770)$ and $\Upsilon(4S)$ branching fraction measurements at
$e^+e^-$ colliders}

\noaffiliation
\affiliation{Institute of Modern Physics, Fudan University, Shanghai 200443, China}
\affiliation{Key Laboratory of Nuclear Physics and Ion-beam Application, Ministry of Education,
China}
\affiliation{Institute of High Energy Physics, Chinese Academy of Sciences, 19B Yuquan Road, Beijing, 100049, China }
\affiliation{University of Chinese Academy of Sciences, 19A Yuquan Road, Beijing, 100049, China}


\author{Jiaxin Li}
\affiliation{Institute of Modern Physics, Fudan University, Shanghai 200443, China}
\affiliation{Key Laboratory of Nuclear Physics and Ion-beam Application, Ministry of Education,
China}

\author{Xiantao Hou}
\affiliation{Institute of High Energy Physics, Chinese Academy of Sciences, 19B Yuquan Road, Beijing, 100049, China }
\affiliation{University of Chinese Academy of Sciences, 19A Yuquan Road, Beijing, 100049, China}

\author{Junli Ma}
\affiliation{Institute of High Energy Physics, Chinese Academy of Sciences, 19B Yuquan Road, Beijing, 100049, China }
\affiliation{University of Chinese Academy of Sciences, 19A Yuquan Road, Beijing, 100049, China}

\author{Changzheng Yuan}
\affiliation{Institute of High Energy Physics, Chinese Academy of Sciences, 19B Yuquan Road, Beijing, 100049, China }
\affiliation{University of Chinese Academy of Sciences, 19A Yuquan Road, Beijing, 100049, China}

\author{Xiaolong Wang} \email{Contact author: xiaolong@fudan.edu.cn} 
\affiliation{Institute of Modern Physics, Fudan University, Shanghai 200443, China}
\affiliation{Key Laboratory of Nuclear Physics and Ion-beam Application, Ministry of Education,
China}


\begin{abstract}

The $\psi(3770)$ and $\Upsilon(4S)$ states predominantly decay into open-flavor meson pairs, whereas the decays of
$\psi(3770) \to \mbox{\text{non}-}D\bar{D}$ and $\Upsilon(4S) \to \mbox{\text{non}-}B\bar{B}$ are rare but crucial
for elucidating the inner structure and decay dynamics of heavy quarkonium states. To achieve precise branching
fraction measurements for $\psi(3770) \to \mbox{\text{non}-}D\bar{D}$ and $\Upsilon(4S) \to
\mbox{\text{non}-}B\bar{B}$ decays at the high luminosity $e^+e^-$ annihilation experiments, we employed Monte Carlo
simulations and Fisher information to evaluate various data taking scenarios, ultimately determining the optimal
scheme. The consistent results of both methodologies indicate that the optimal energy points for studies of $\psi(3770)
\to \mbox{\text{non}-}D\bar{D}$ decays are $3.769~\gev$ and $3.781~\gev$, whereas those for $\Upsilon(4S) \to
\mbox{\text{non}-}B\bar{B}$ decays are $10.574~\gev$ and $10.585~\gev$. In addition, we studied the dependence of the precision in branching fraction measurements on the integrated luminosity, with the branching fractions spanning several orders of magnitude.
\end{abstract}
 
 
\maketitle

\section{Introduction}
\label{sec:1}

The $\psip$ state above the $\dd$ threshold and the $\ufs$ state above the $\bb$ threshold are two typical broad resonances.
Their dominant decay modes are open-flavor meson pair final states and hadronic or radiative transitions to low lying
quarkonia. Their decays into light hadronic ($\ndd$, $\nbb$) final states are suppressed according to the OZI
rule~\cite{ref::ozi}.  
Most of these studies were performed using $\EE$ collision experiments, and the total cross
sections of the resonance production and continuum process were of the same order of magnitude. In this case, the
interference between the continuum and resonance amplitudes might be large and can-not be ignored. The mishandling of the
interference effect will lead to a systematic bias with a much larger size compared with the statistical uncertainty
and comparable to or even larger than all the other sources of systematic uncertainties~\cite{ref::guo}. The
optimal solution to this problem is to take data with a strategy of accumulating data at the resonance peak
and energy points off the peak (called continuum data).

The BESIII and Belle II are $\EE$ collision experiments that operate in the charmonium and bottomonium energy regions.
At the BESIII experiment, $20~\infb$ data were accumulated by 2024 at the center-of-mass energy of $\sqrt{s} =
3.773~\gev$, which is the peak position of the $\psip$ state~\cite{ref::bes3}. This data sample is at least six times
larger than the one collected by BESIII up to 2019 and dozens of times larger than those in the CLEOc and BESII
measurements~\cite{pdg}. The Belle II experiment plans to capture about $50~\inab$ data at the peak position of the $\ufs$ ($\sqrt{s}
= 10.58~\gev$) state~\cite{ref::belle2}, which are $10^2$ ($10^3$) of times larger than the data samples of the Belle and
BaBar (CLEO) experiments~\cite{pdg}. With the currently available BESIII and  expected Belle II data
samples, the precision of the branching fraction of $\psinn$ and $\upsilonnn$ decays can be improved significantly.
Therefore, determining an optimal data taking strategy is essential.

This work focuses on the relative uncertainties of the branching fractions of $\psinn$ and $\upsilonnn$ decays. We
take the optimization of $\psip/\ufs \to \eta^{\prime}\phi$ measurement as an example. Monte Carlo (MC) simulation
and Fisher information methods were employed to simulate various data taking cases with two parameters: the branching fraction
of $\ndd$ or $\nbb$ decays denoted as $\BR$, and the relative angle between resonance and continuum amplitudes denoted
as $\phi$. The optimal data taking scheme obtained from both methods is consistent. Herein, to achieve the measurements
of branching fractions of $\psinn$ and $\upsilonnn$ decays, we aim to determine the following:
\begin{itemize}
\item What are the optimal $\sqrt{s}$ values for data taking around the resonances?
\item How many energy points around an optimal $\sqrt{s}$ are necessary for the scan?
\item What are the required data sizes to achieve a specific precision in the branching fraction measurements?
\end{itemize}

\section{MC simulation Methodology}
\label{sec:2}

\subsection{Analysis Framework}
\label{sec2a}

Within a specified period of data taking time or equivalently for a given integrated luminosity, we attempt to determine
the scheme that can provide the best precision for branching fraction measurements of $\psinn$ or $\upsilonnn$ decays.
The sampling technique is utilized to simulate various data taking strategies, among which the optimal one will be selected.
For specific simulations, the likelihood function is constructed as
\beq\label{likelihood}
  LF = \prod_{i=1}^{N_{pt}} \frac{1}{\sqrt{2\pi\xexp } }e^{-\frac{(x_{i}-\xexp)^2}{2\xexp}},
\eeq
where $x_i$ and $\xexp$ are the numbers of observed and expected events, respectively, of $\psinn$ or $\upsilonnn$
decays, the subscript $i$ labels the $i$-th scan energy point, and $N_{pt}$ is the number of energy points. For a
certain exclusive final state $f$, the $\xexp$ in a data sample taken at the $i$-th energy point with an integrated
luminosity of $\mathcal{L}_i$ is given by
\beq\label{expn}
  \xexp(\BR, \phi) = \mathcal{L}_{i} \cdot \sexp (\BR, \phi, \sqrt{s}) \cdot \epsilon,
\eeq 
where $\epsilon$ is the selection efficiency of signal reconstruction, and $\sexp$, with $\BR$ and $\phi$ as two
parameters, is the expected cross section at $\sqrt{s}$. The maximum value of $\BR$, derived from a conservative
estimate of the summed branching ratios for the resonance decaying to light hadronic final states as reported by the
Particle Data Group (PDG)~\cite{pdg}, is $1 \times 10^{-3}$ for $\psinn$ decays and $1 \times 10^{-6}$ for
$\upsilonnn$ decays. The cross section $\sexp$ can be expressed as
\beq\label{sexpeq}
  \sexp(\sqrt{s})=|a_c^{f}(\sqrt{s}) + e^{i\phi} \cdot a_{R}^{f}(\sqrt{s})|^2,
\eeq
where $a_c^{f}(\sqrt{s})$ is the continuum amplitude of $\EE \to f$, and $a_{R}^{f}(\sqrt{s})$ is the resonance
amplitude of $\EE \to~R \to f$. They are parametrized as
\begin{align}
a_c^{f}(\sqrt{s}) &= \frac{a}{(\sqrt{s})^{n}}\sqrt{PS(\sqrt{s})},\\
a_R^{f}(\sqrt{s}) &= \frac{\sqrt{12\pi\Gamma_{\EE}\Gamma\BR}}{s-M^2+iM\Gamma}\sqrt {\frac{PS(\sqrt s)}{PS(M)}},
\end{align}
where $PS(\sqrt{s})$ is the phase space factor at $\sqrt{s}$. In the continuum amplitude, the constants $a$ and $n$
describe the magnitude and slope of the continuum process, respectively. In the Breit-Wigner function for the resonance amplitude,
$M$, $\Gamma$, and $\Gamma_{\EE}$ are the mass, total width, and partial width to $\EE$ final state of the
resonance $R$, respectively. For the $\eta^\prime \phi$ final state, the phase space factor is $PS(\sqrt{s}) =
~q^3(\sqrt{s})/s$, where $q(\sqrt{s})$ is the momentum of $\eta^\prime$ or $\phi$. The following study focused on
the optimization of statistical uncertainty. The values of the key parameters for the numerical calculations presented
in the following sections are summarized in Table~\ref{para}. In the fitting procedure, $\BR$ and $\phi$ are defined
as two free parameters to fit $\sexp(\sqrt{s})$. The maximization of the likelihood function $LF$ presented in Eq.~(\ref{likelihood}) yields
the optimal estimate for $\BR$.

\begin{table}[htbp]
\caption{Values of key parameters for the numerical calculation. The quantities with $\star$ are set as free
parameters, and the others are fixed for the corresponding study.}
\begin{center}
\begin{tabular}{c|c|c}
\hline 
Parameter       & $\psinn$	      & $\upsilonnn$   \\ \hline
$M$	($\gevcs$)          & 3.7736~\cite{pdg}	   & 10.58~\cite{pdg}    \\
$\Gamma_{\EE}$ ($\kev$) & 0.26~\cite{pdg}	   & 0.28~\cite{pdg}     \\
$\Gamma$  ($\mev$)      & 27.2~\cite{pdg} 	   & 20.5~\cite{pdg}     \\
$\star\BR$              & free                 & free    \\
$\star\phi$             & free                 & free   \\
$\epsilon$     & 10\%~\cite{4sepsilon0, 4sepsilon1, 4sepsilon2, 4sepsilon3, 4sepsilon4}
               & 15\%~\cite{4sepsilon0, 4sepsilon1, 4sepsilon2, 4sepsilon3, 4sepsilon4}           \\
$a$            & 1.82~\cite{4sepsilon0, 4sepsilon1, 4sepsilon2, 4sepsilon3, 4sepsilon4}
     & $3.2\times10^{-2.5}$~\cite{4sepsilon0, 4sepsilon1, 4sepsilon2, 4sepsilon3, 4sepsilon4}       \\
$n$            & 5.82~\cite{4sepsilon0, 4sepsilon1, 4sepsilon2, 4sepsilon3, 4sepsilon4}
               & 3.00~\cite{4sepsilon0, 4sepsilon1, 4sepsilon2, 4sepsilon3, 4sepsilon4}     \\
               \hline
\end{tabular}
\label{para}
\end{center}
\end{table}

In the following analysis, we assume that the value of $\BR$ is known. Under this assumption, we aim to address the
questions posed at the end of Sec.~\ref{sec:1}. However, upon further reflection on the first two questions, we
notice that they are interconnected. Specifically, the optimal number of energy points depends on the distribution of points,
and {\it vice versa}. To resolve this dilemma, we begin with a simple distribution to determine the optimal number of
energy points, enabling us to finalize the total number of energy points required. 

\subsection{Two-parameter fit}

We take $N_{pt} = 100$ energy points and $N_{br} = 300$ branching fractions as a tentative beginning. The values of
energies and branching fractions are calculated via
\begin{align}
   \ecm^i &=   \ecm^0 + (i-1) \times \delta \ecm ~(i = 1,2,...,\npt), \\
 \BR_j &= \BR_0 + (j-1) \times \delta \BR ~(j = 1,2,...,\nbr),
\end{align}
where the first energy point is $\ecm^0 = 3.74~\gev$ with a branching fraction of $\BR_0 = 1\times 10^{-6}$, and the last 
energy point is $\ecm^f = 3.80~\gev$ with a branching fraction of $\BR_f = 1\times 10^{-3}$ for $\psip$ decays. Correspondingly, for
$\ufs$ decays, $\ecm^0 = 10.56~\gev$, $\BR_0 = 1\times 10^{-9}$, $\ecm^f = 10.60~\gev$, and $\BR_f = 1\times 10^{-6}$.
The energy interval and branching fraction interval are calculated with $\delta \ecm = (\ecm^f -
\ecm^0)/\npt$ and $\delta \BR = (\BR_f - \BR_0)/\nbr$.  

\begin{figure}[htp]
\begin{center}
\includegraphics[width=10.0cm]{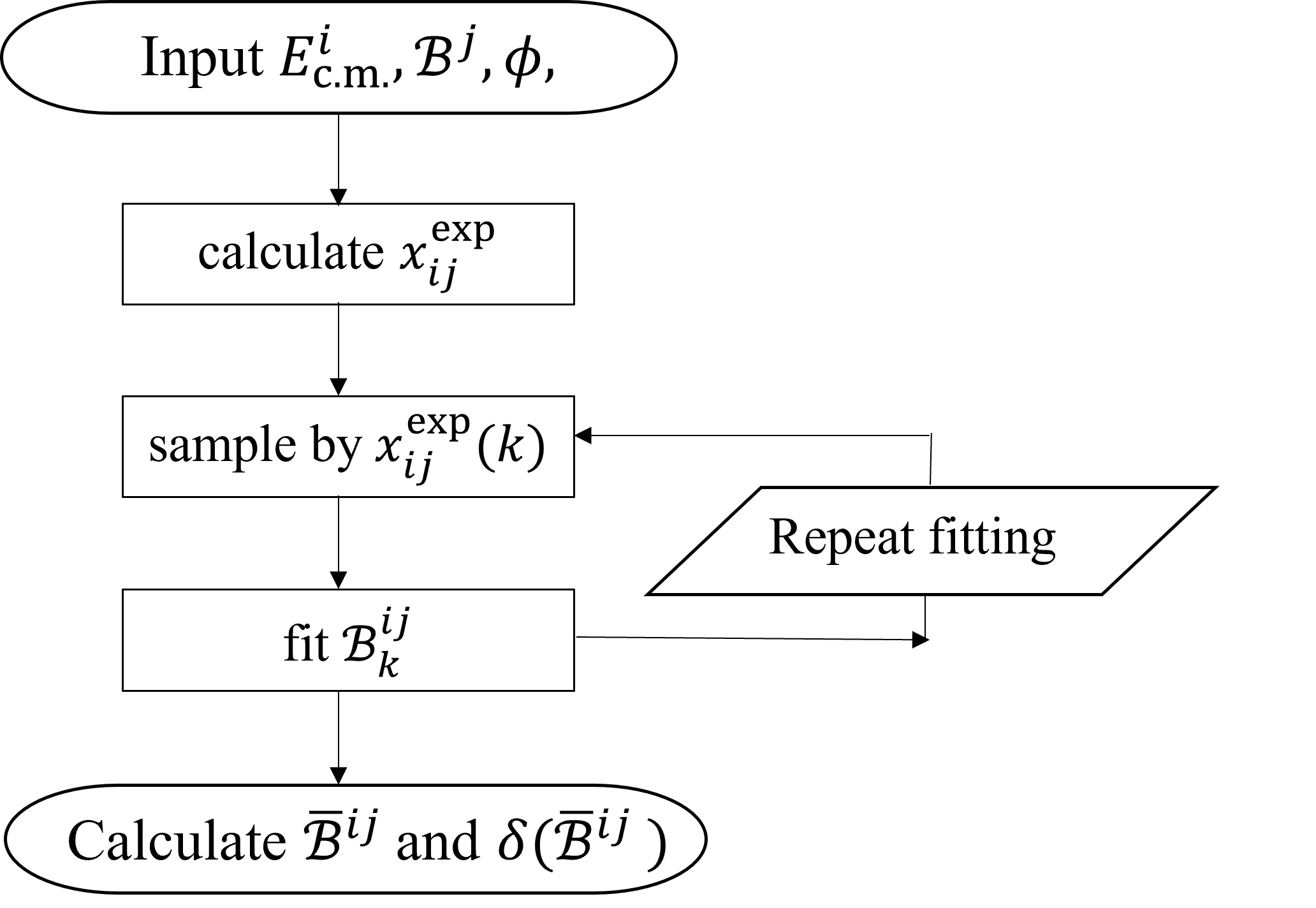}
\caption{Flow-chart of the sampling, where $ij$ ($i = 1,2,...,\npt$; $j = 1,2,...,\nbr$) denotes a specific scheme of
an $\ecm^i$ and $\BR$ combination, and $k$ ($k = 1,2,...,\nsamp$) represents the $k$-th sampling time.}
\label{liucheng}
\end{center}
\end{figure}

To reduce the statistical fluctuation, we repeat sampling $\nsamp = 200$ times for each scheme of an $\ecm^i$ and $\BR$ combination, 
which satisfy a Gaussian distribution with a mean of $\xexp$ and deviation of $\sqrt{\xexp}$. 
The general-flow chart of sampling and fitting is presented in Fig.~\ref{liucheng}. We obtain the
mean value of $\BR^{ij}_k$ and its error $\delta(\BR^{ij}_k)$ from fitting to the $k$-th sample. The average value of
$\BR$ and the corresponding uncertainty from fitting to the 200 samples are calculated via~\cite{tot}
\beq\label{aveb}
 \Bar{B}^{ij} = \frac{1}{\nsamp} \sum_{k=1}^{\nsamp} \BR_k^{ij}, \;\;\;\;
 \delta(\Bar{B}^{ij}) = \frac{1}{\nsamp} \sum_{k=1}^{\nsamp} \delta(\BR_k^{ij}).
\eeq
The average relative uncertainty is calculated via
\beq\label{relb}
E(\BR) = \delta(\Bar{B}^{ij})/{\Bar{B}^{ij}}.
\eeq
Here, note that $ij$ denotes a specific scheme of an $\ecm^i$ and $\BR$ combination. Additionally, $k$
represents the $k$-th sampling time. $\Bar{B}^{ij}$ and $\delta(\Bar{B}^{ij})$ represent the average value and
error of the branching fraction from the fit, respectively. Without the special declaration, the meaning of the
average defined by Eqs.~(\ref{aveb}) and (\ref{relb}) is maintained in the following analysis. 

\begin{figure}[htp]
\begin{center}
\includegraphics[width=4.0cm]{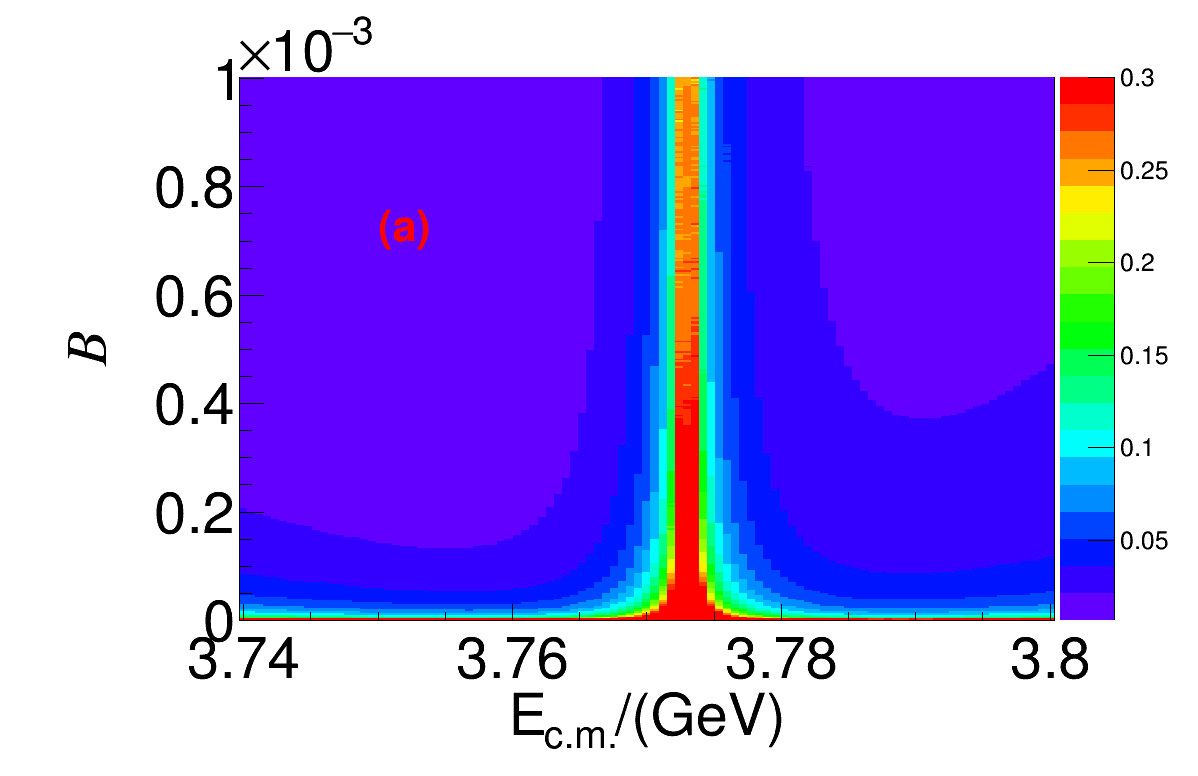}
\includegraphics[width=4.0cm]{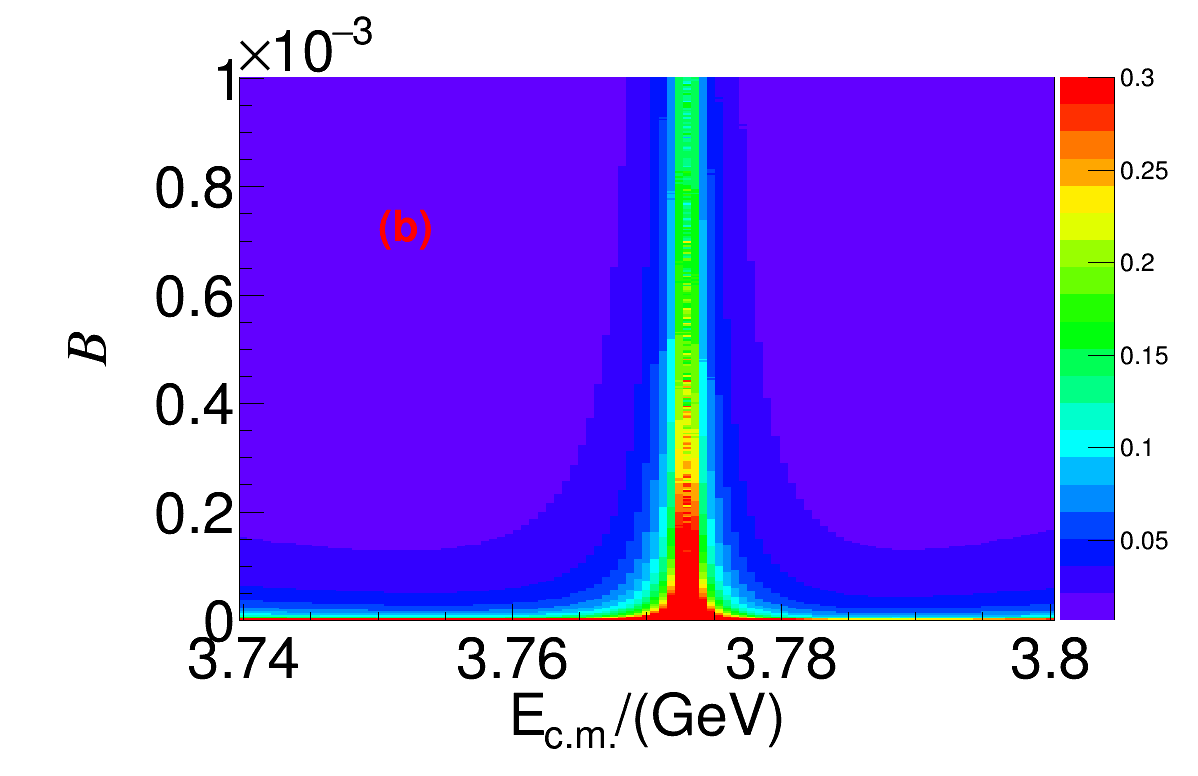} 
\includegraphics[width=4.0cm]{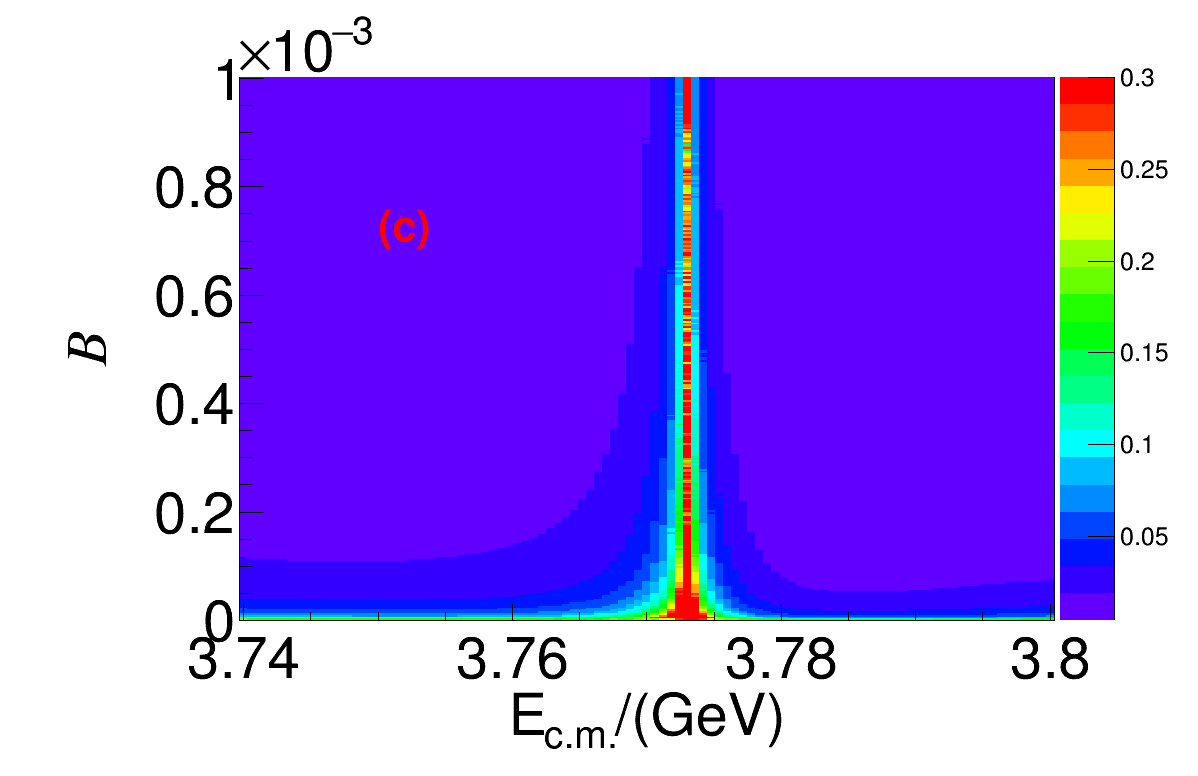} 
\includegraphics[width=4.0cm]{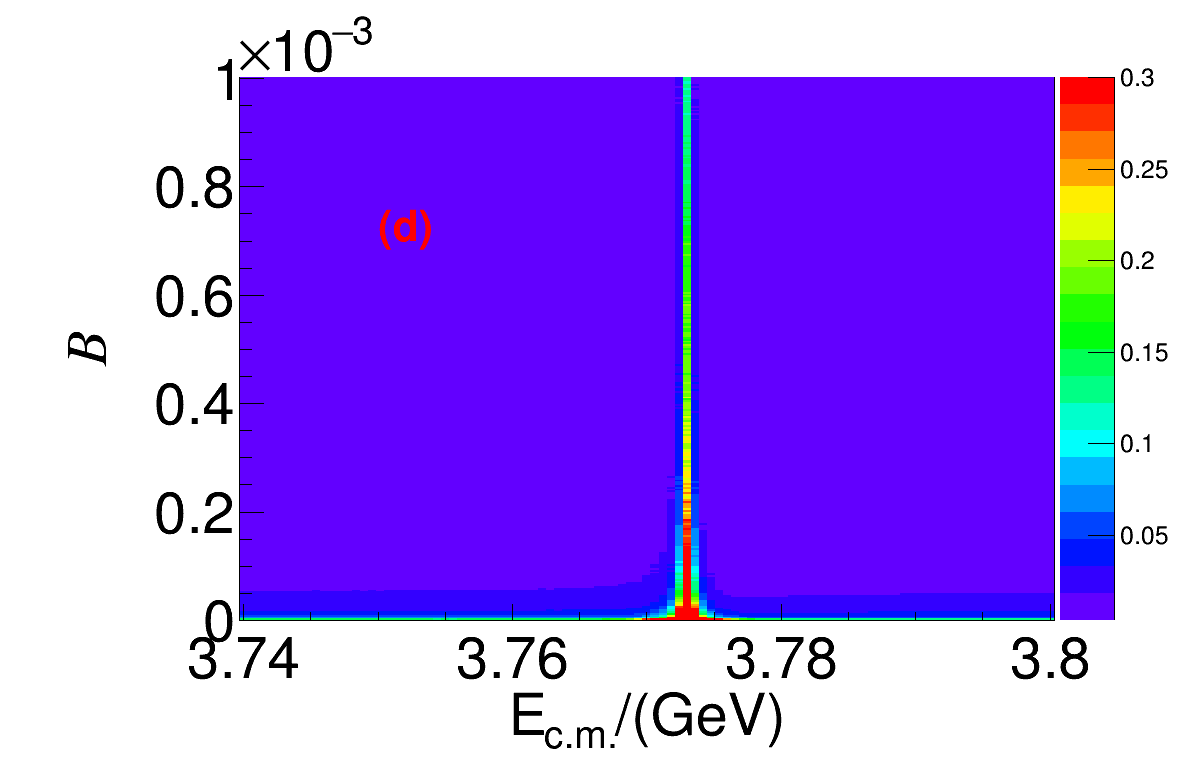} \\
\includegraphics[width=4.0cm]{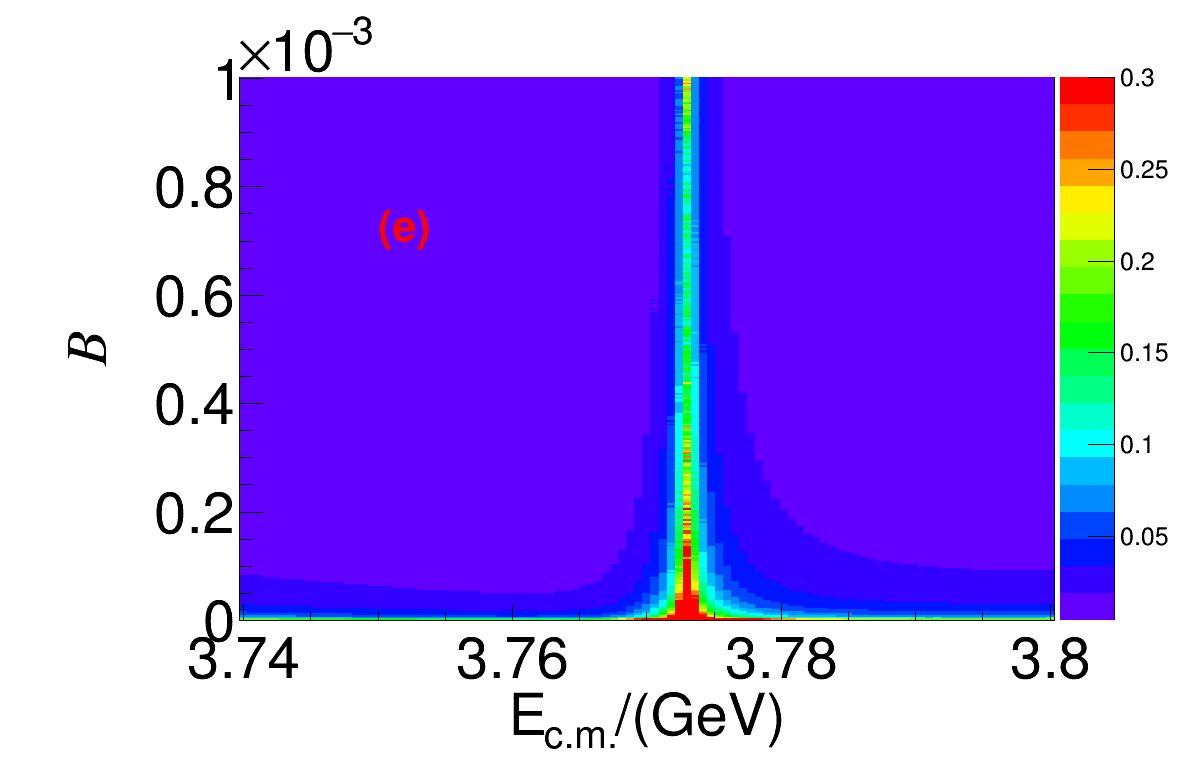} 
\includegraphics[width=4.0cm]{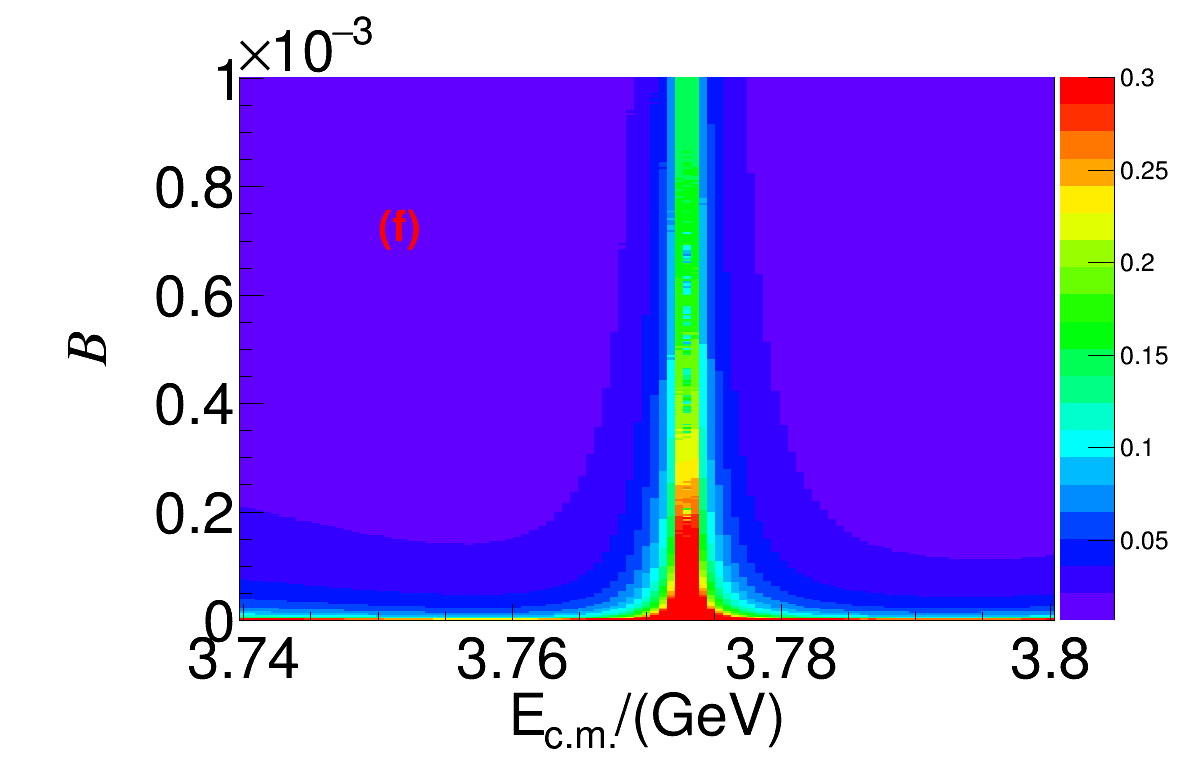} 
\includegraphics[width=4.0cm]{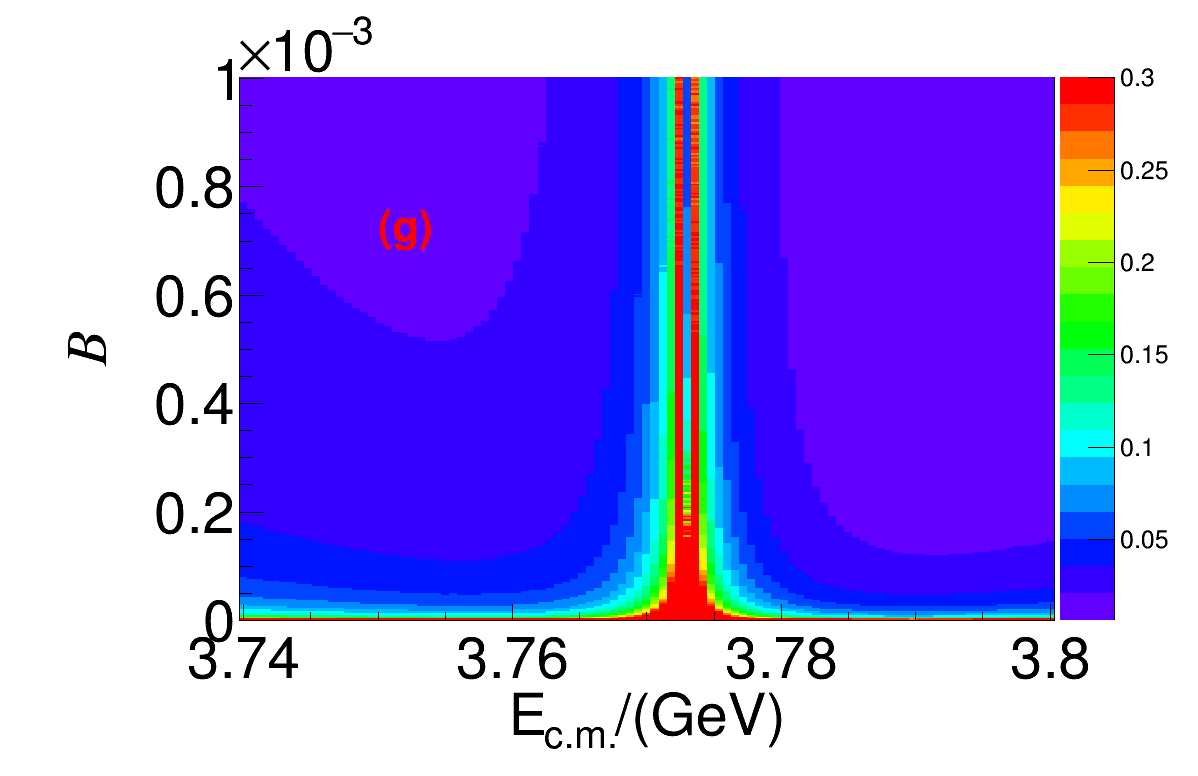}  
\includegraphics[width=4.0cm]{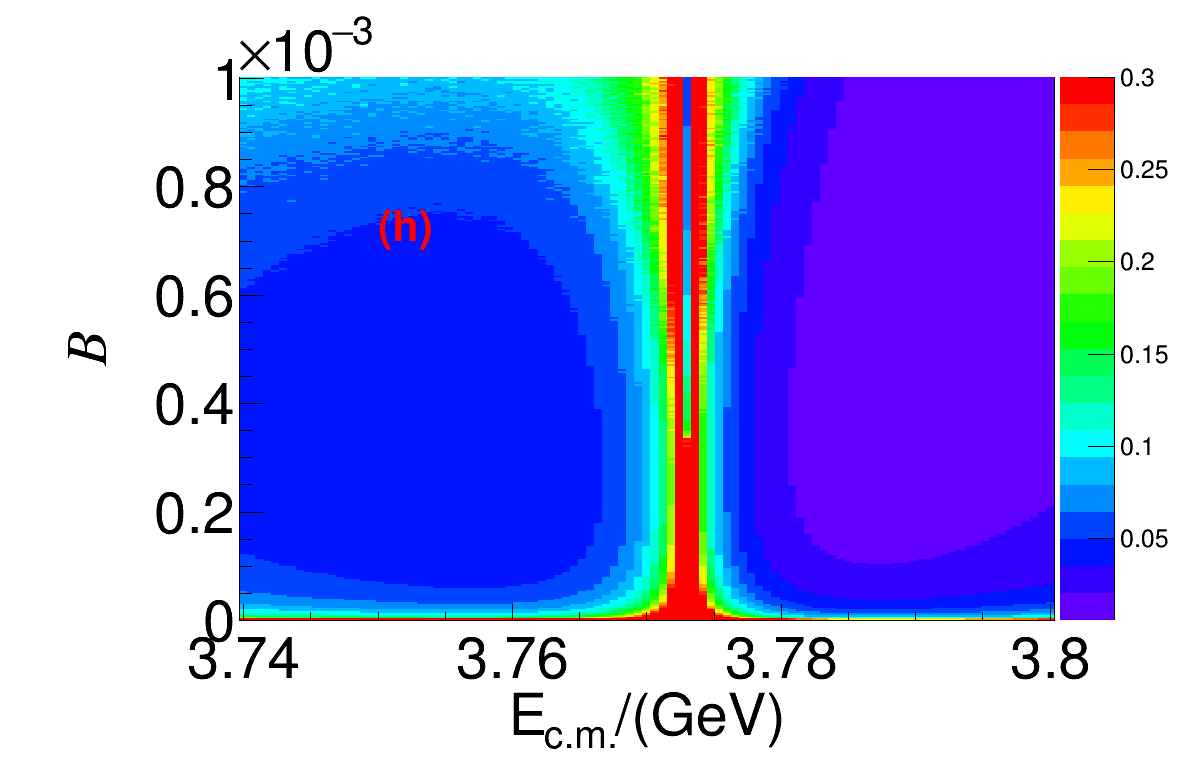}  \\
\includegraphics[width=4.0cm]{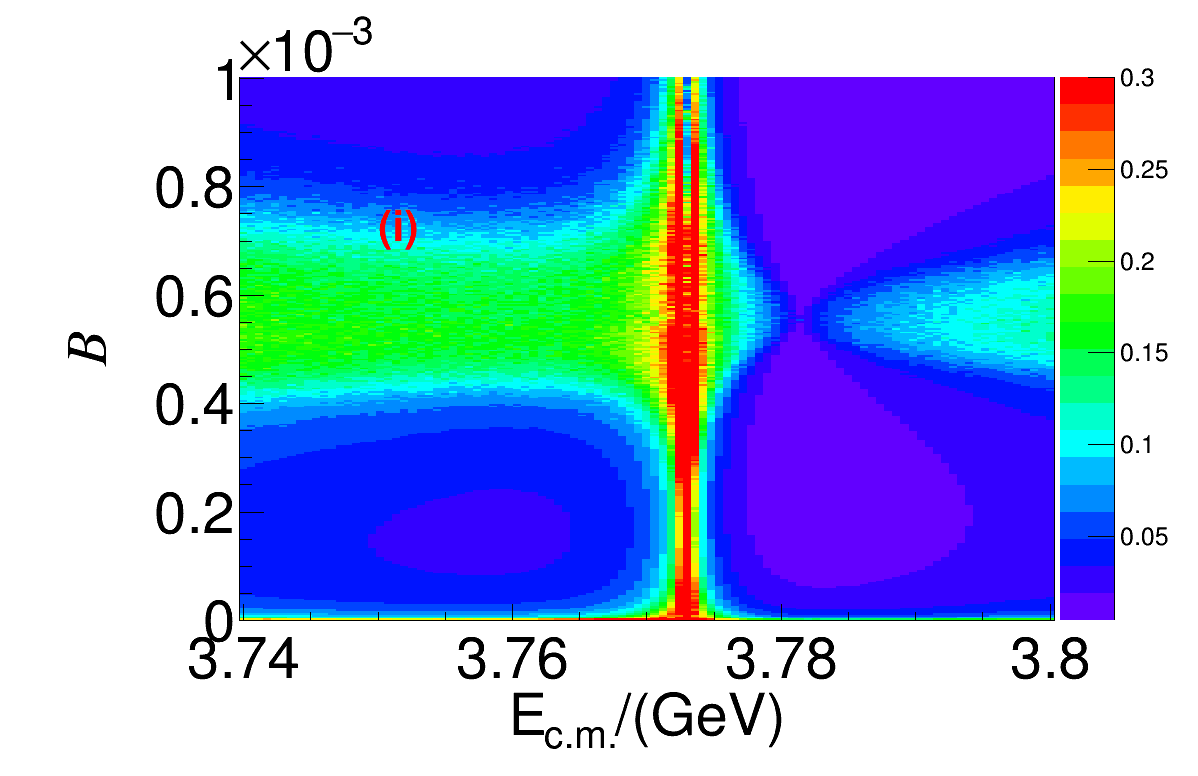}  
\includegraphics[width=4.0cm]{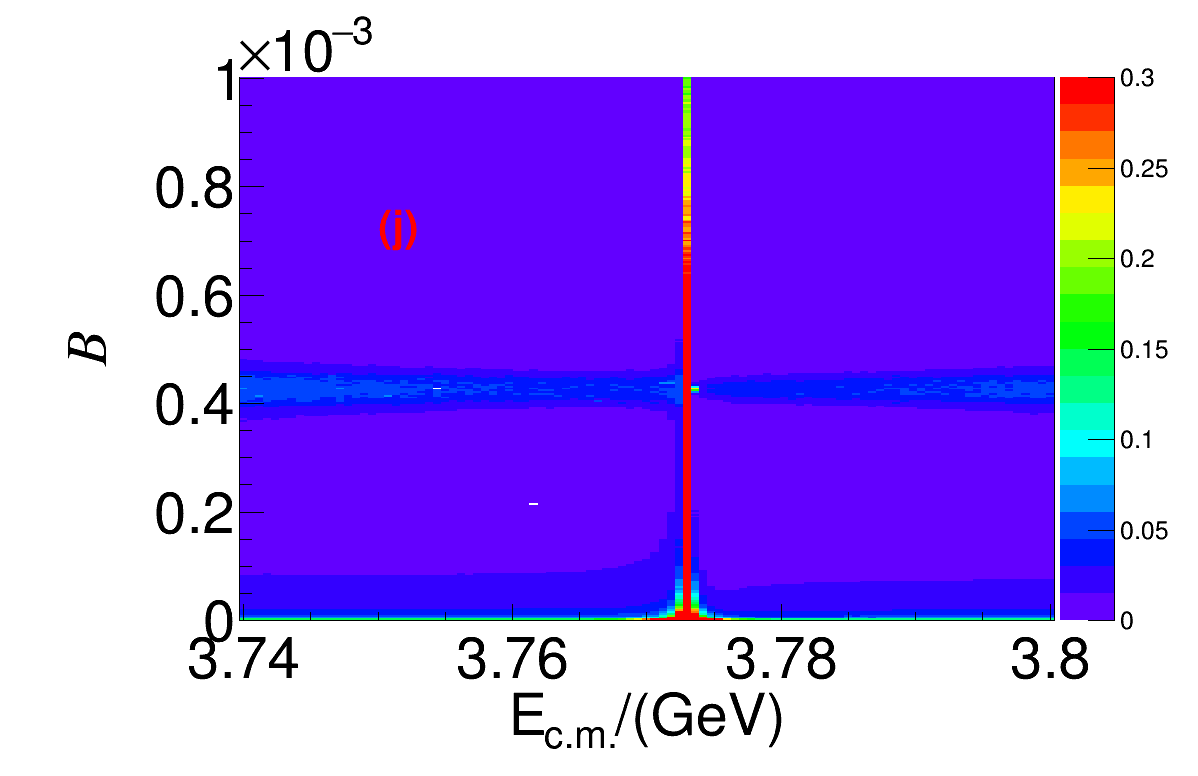}
\includegraphics[width=4.0cm]{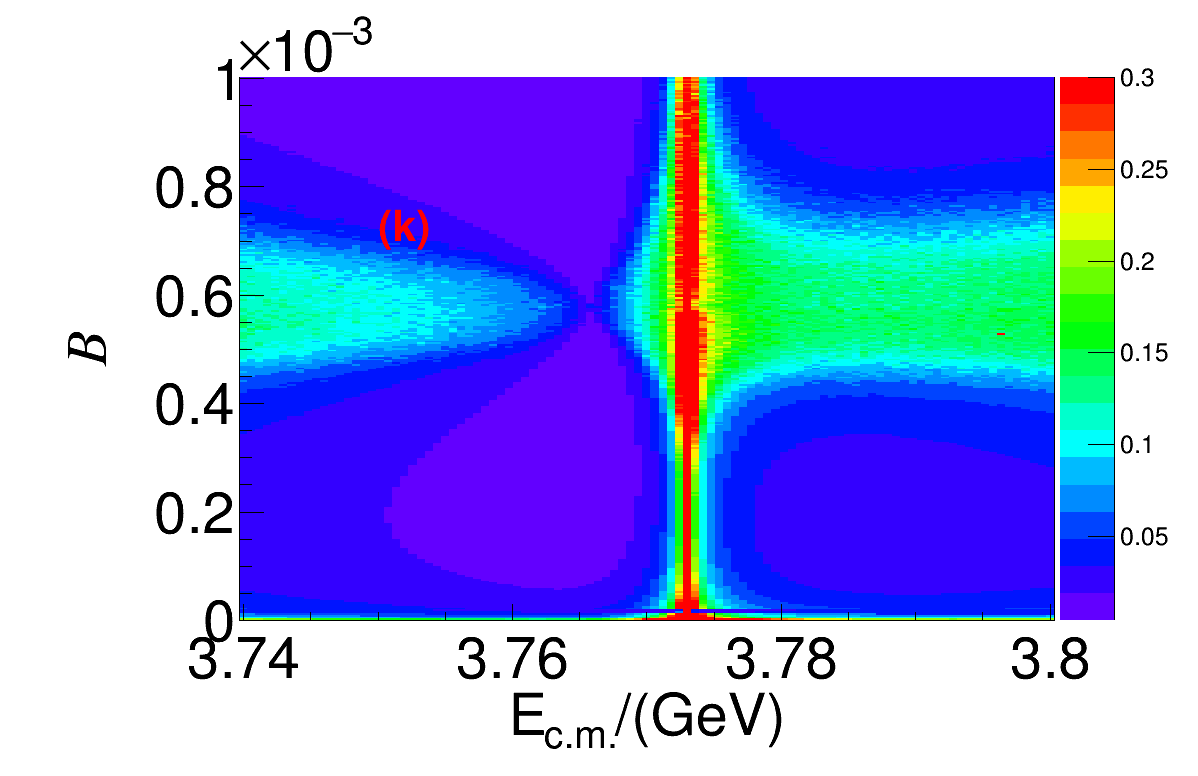}
\includegraphics[width=4.0cm]{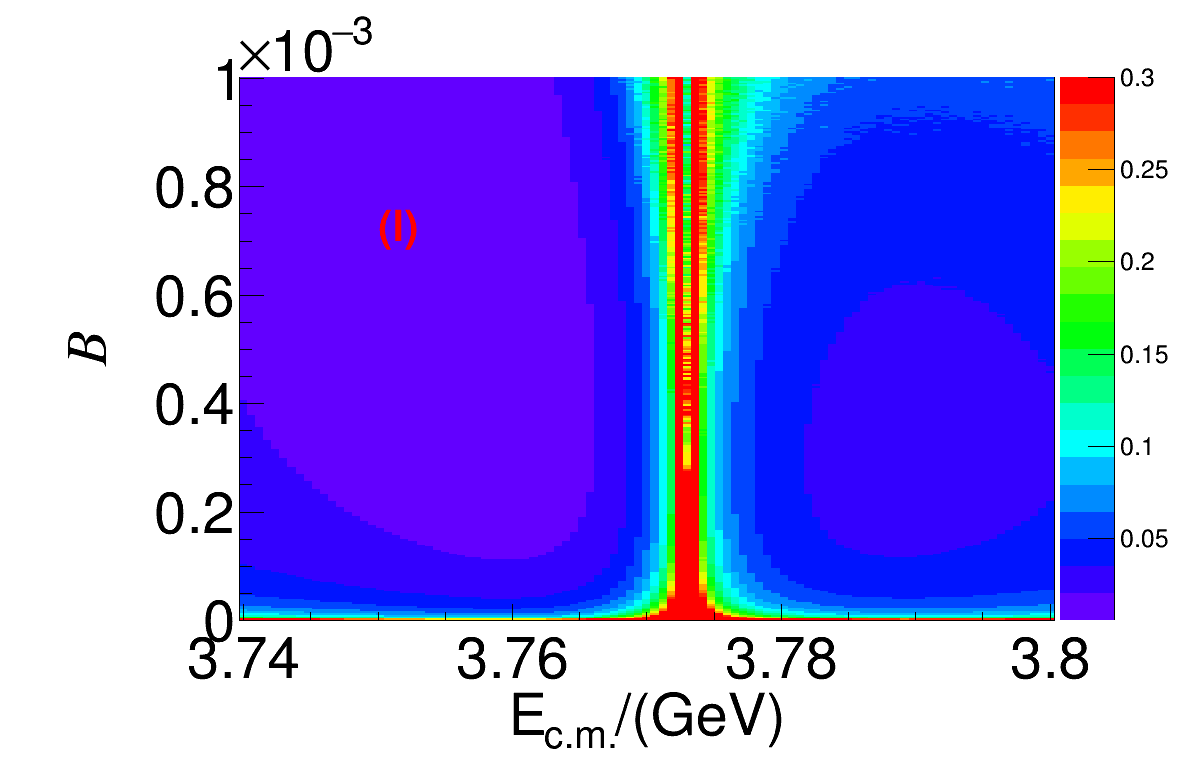}
\caption{Distributions of $E(\BR)$ with $\ecm^i$ and the branching fraction for $\psinn$ decays with
different phase angles $\phi$: (a) $0^\circ$, (b) $30^\circ$, (c) $60^\circ$, (d) $90^\circ$, (e) $120^\circ$, (f)
$150^\circ$, (g) $180^\circ$, (h) $210^\circ$, (i) $240^\circ$, (j) $270^\circ$, (k) $300^\circ$, and (l) $330^\circ$.}
\label{poi3773}
\end{center}
\end{figure}

\begin{figure}[htp]
\begin{center}
\includegraphics[width=4.0cm]{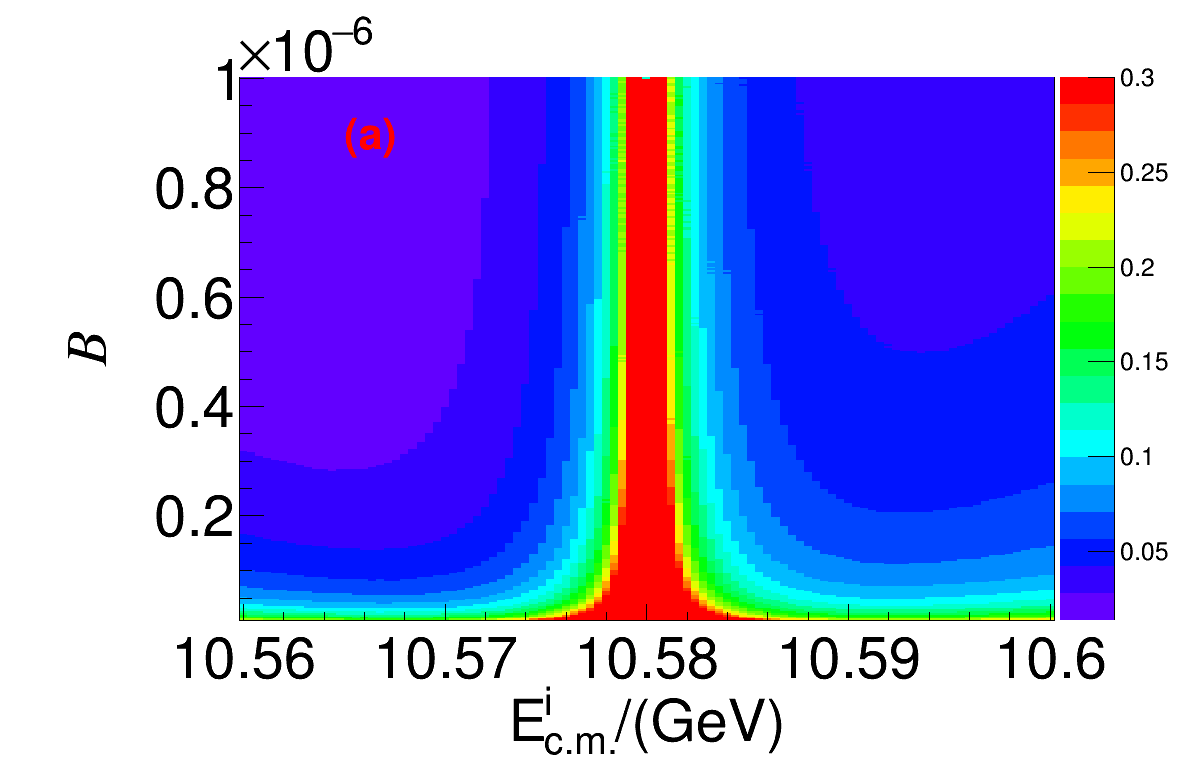} 
\includegraphics[width=4.0cm]{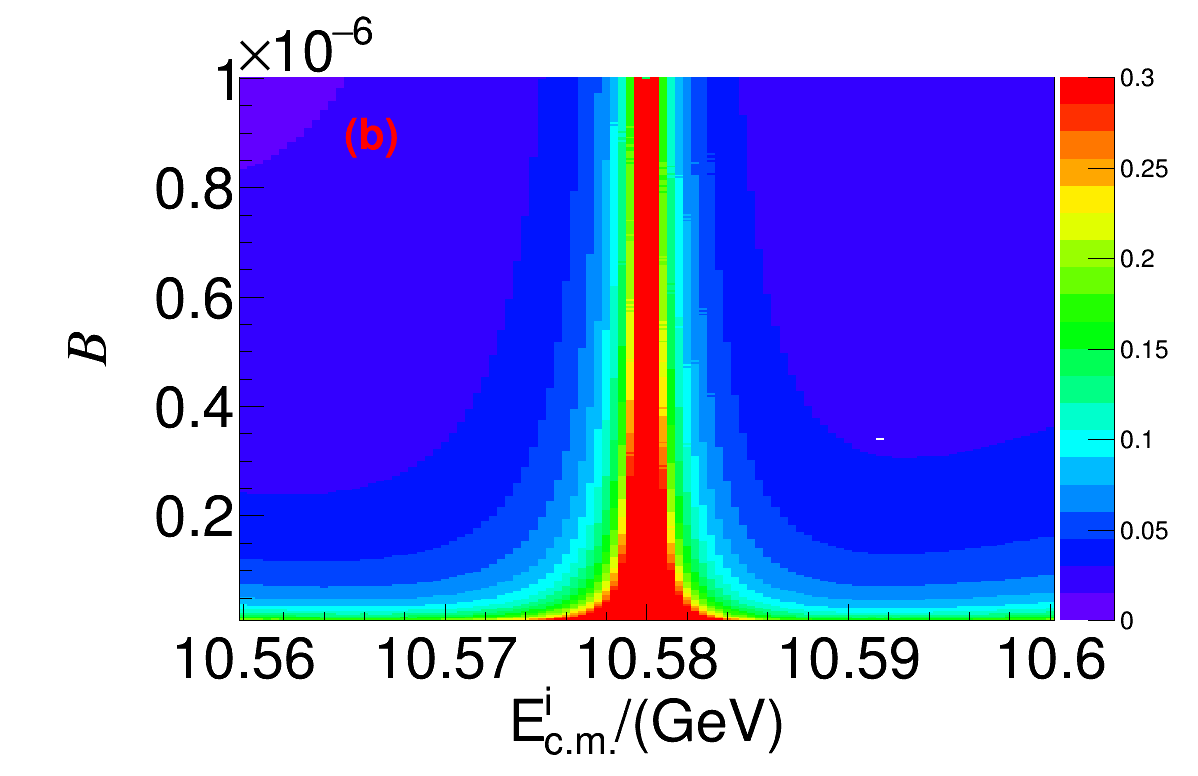} 
\includegraphics[width=4.0cm]{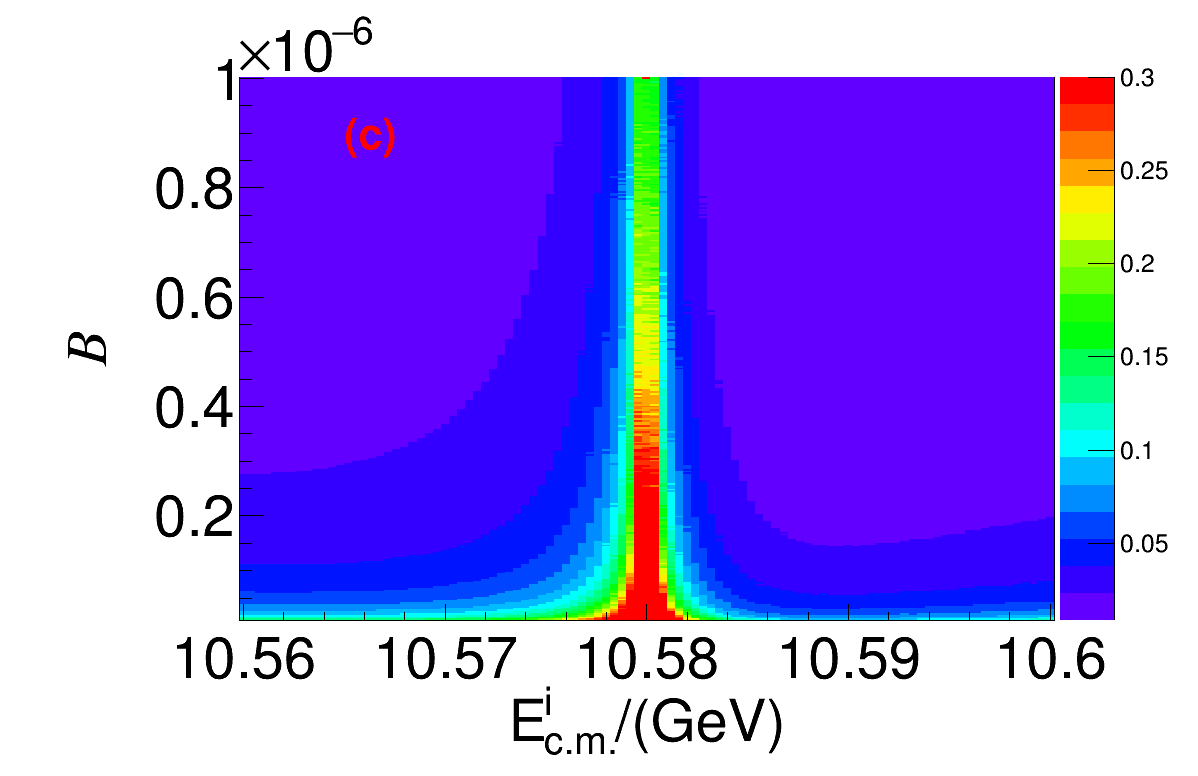} 
\includegraphics[width=4.0cm]{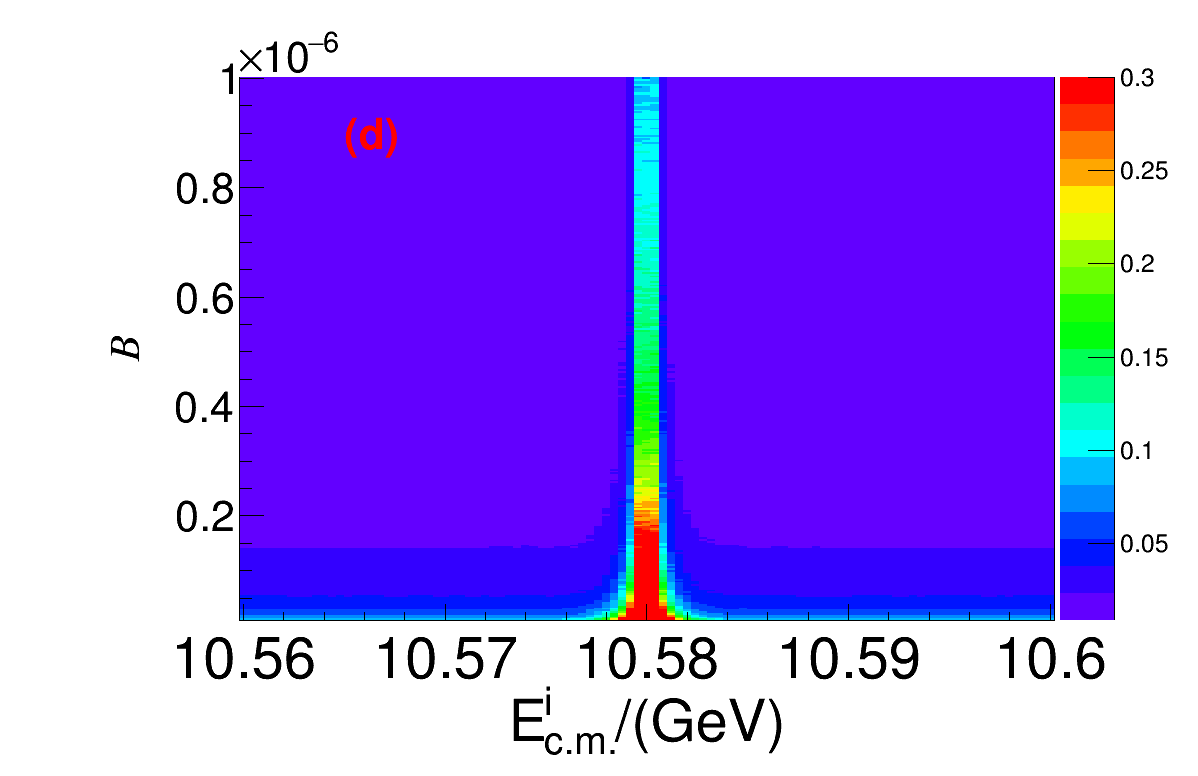} \\
\includegraphics[width=4.0cm]{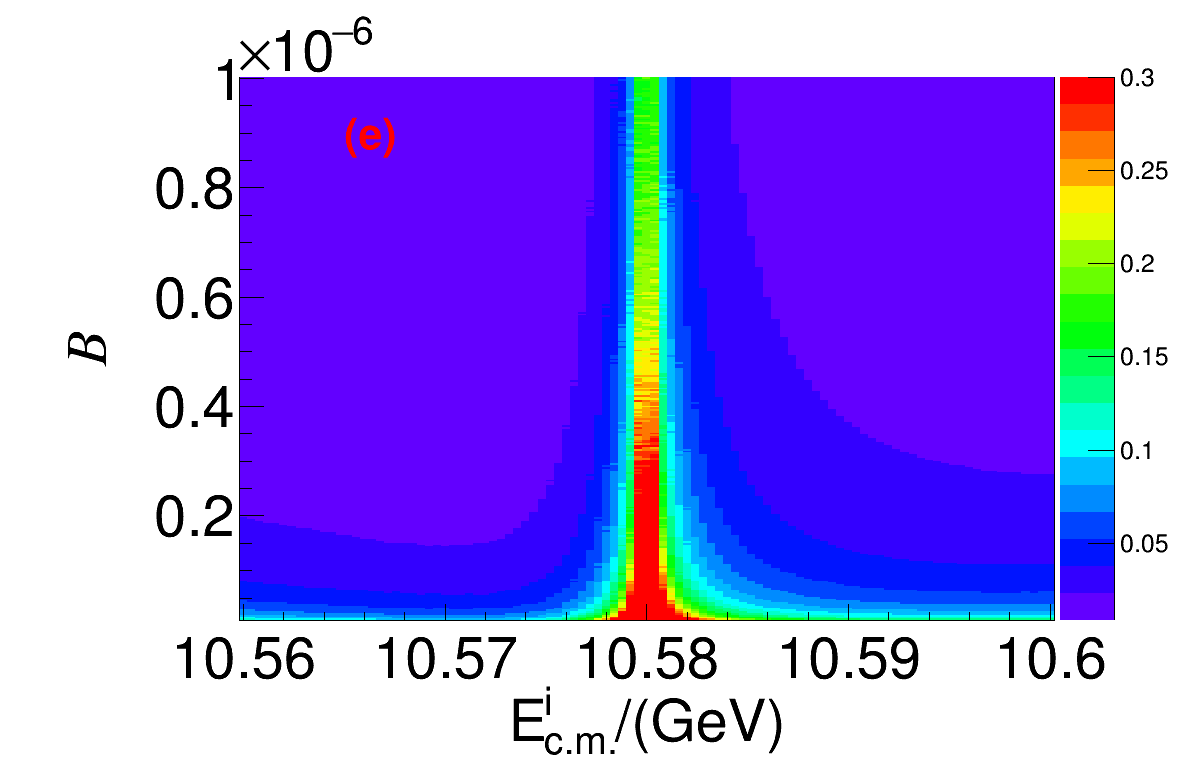} 
\includegraphics[width=4.0cm]{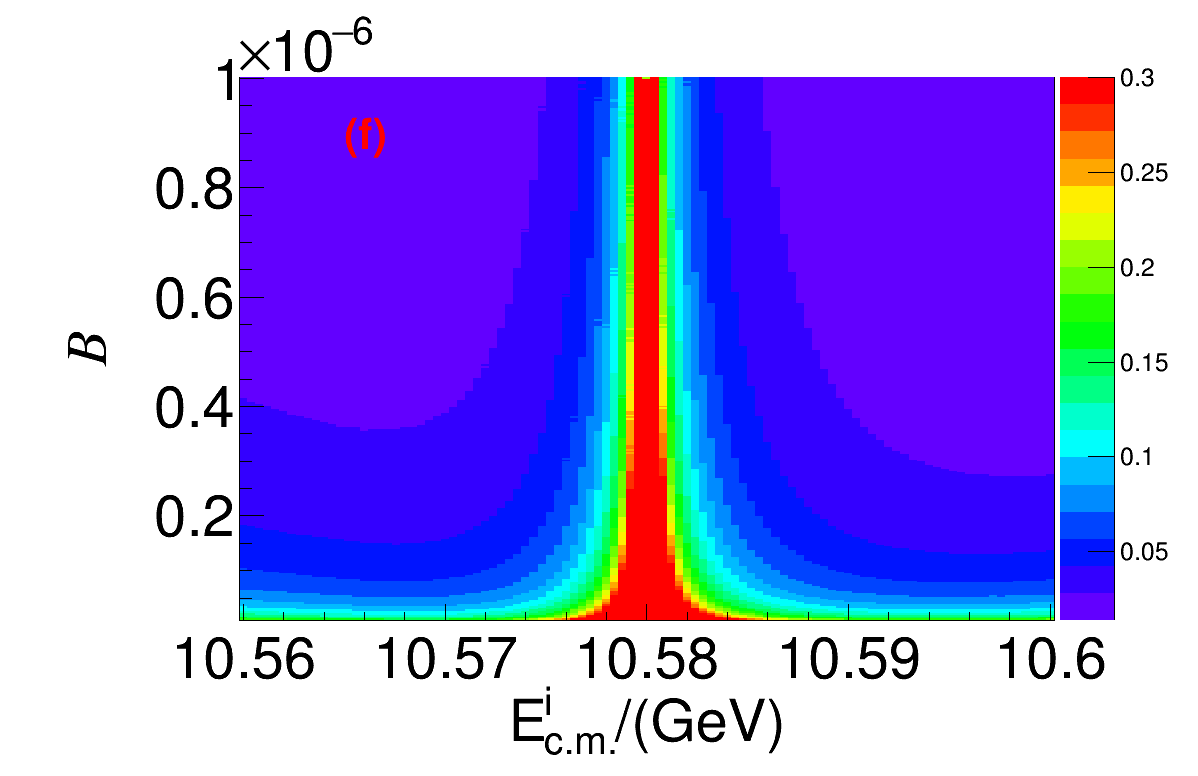} 
\includegraphics[width=4.0cm]{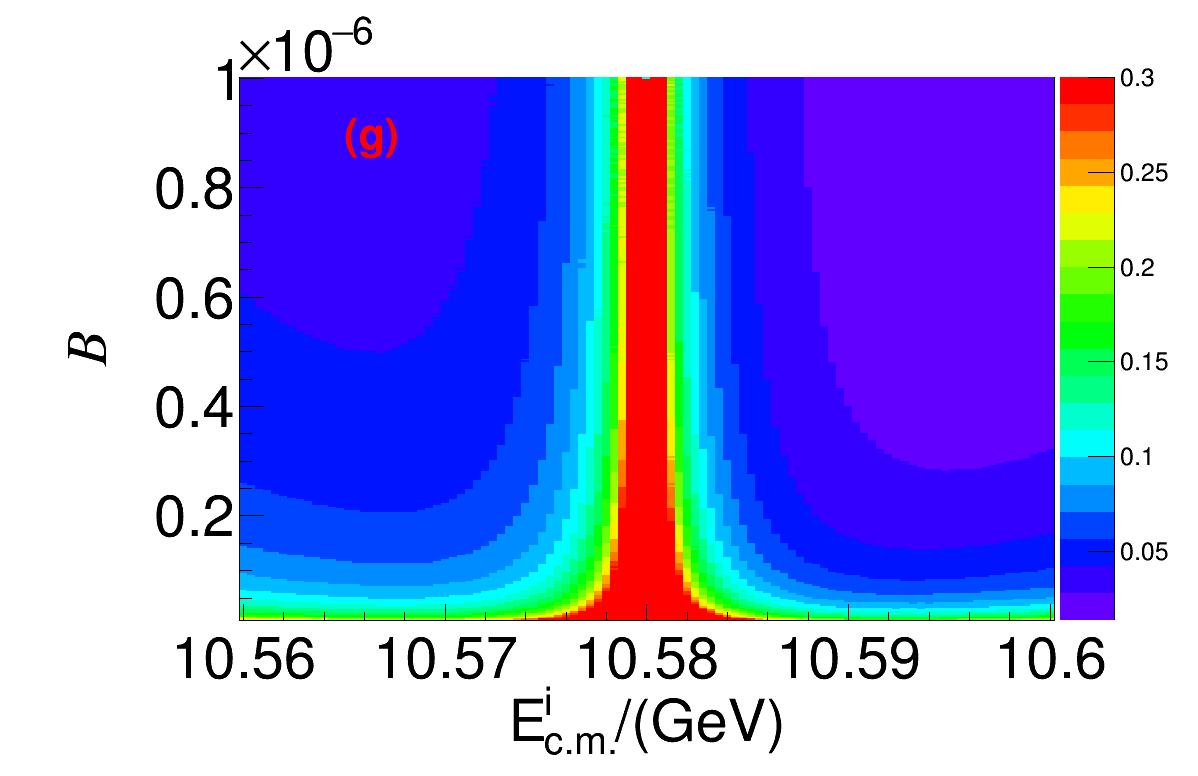}  
\includegraphics[width=4.0cm]{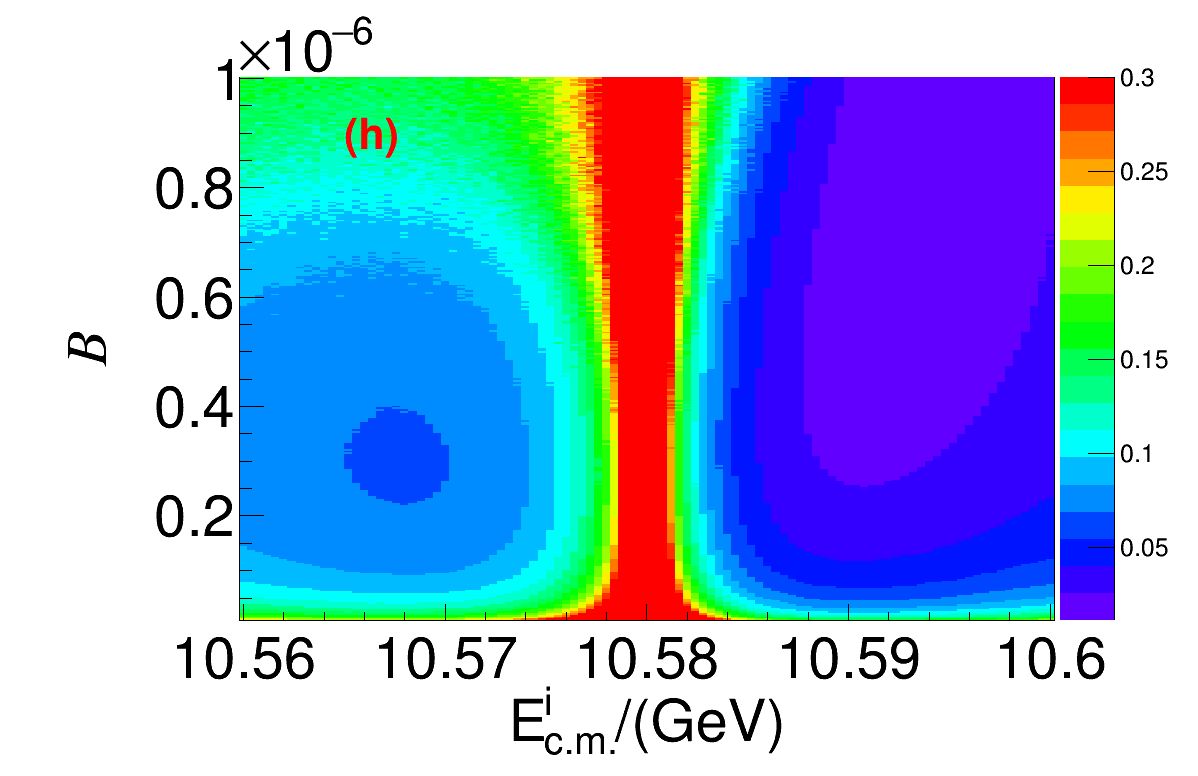}  \\
\includegraphics[width=4.0cm]{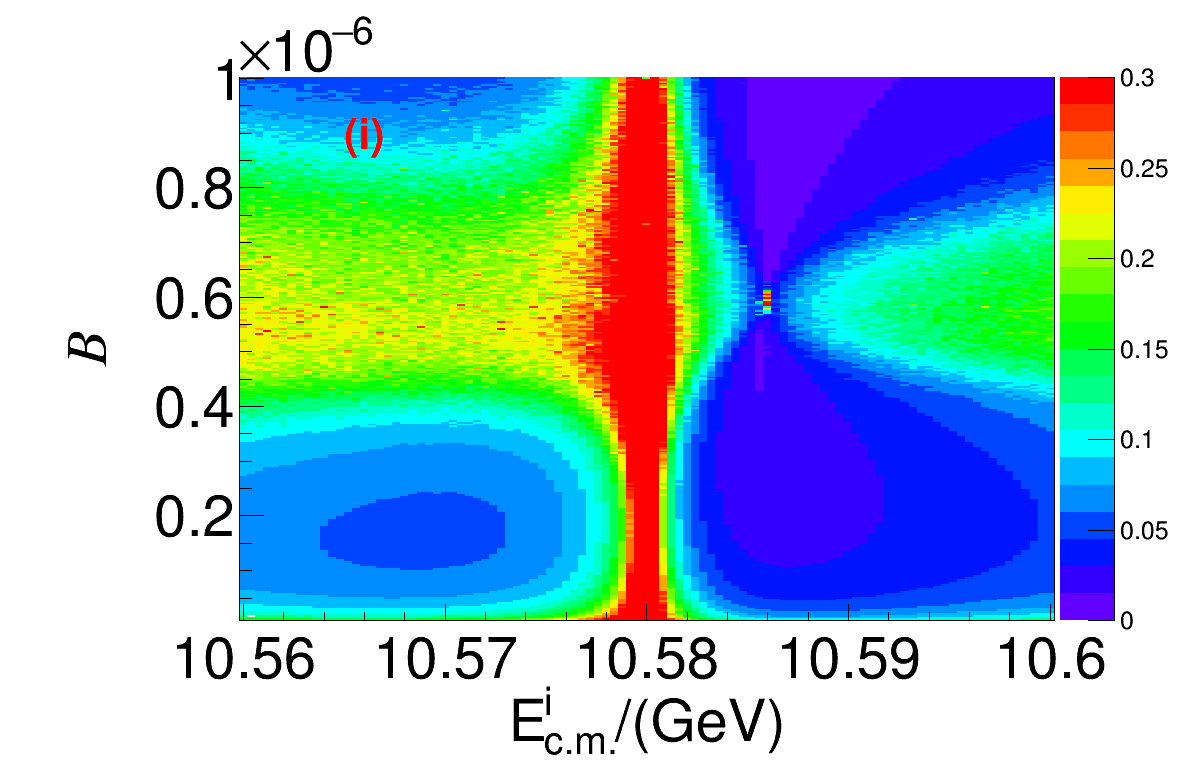}  
\includegraphics[width=4.0cm]{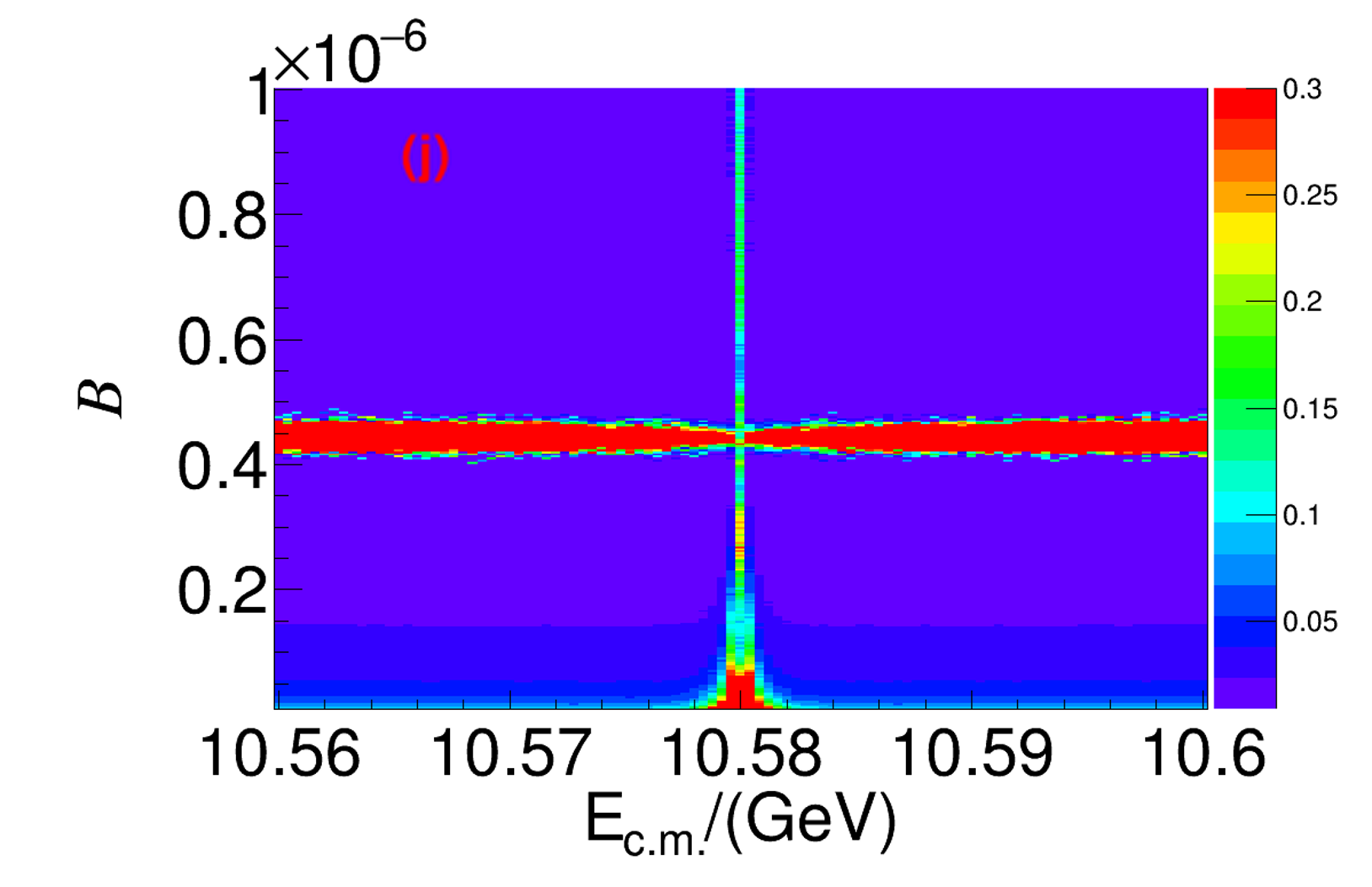}
\includegraphics[width=4.0cm]{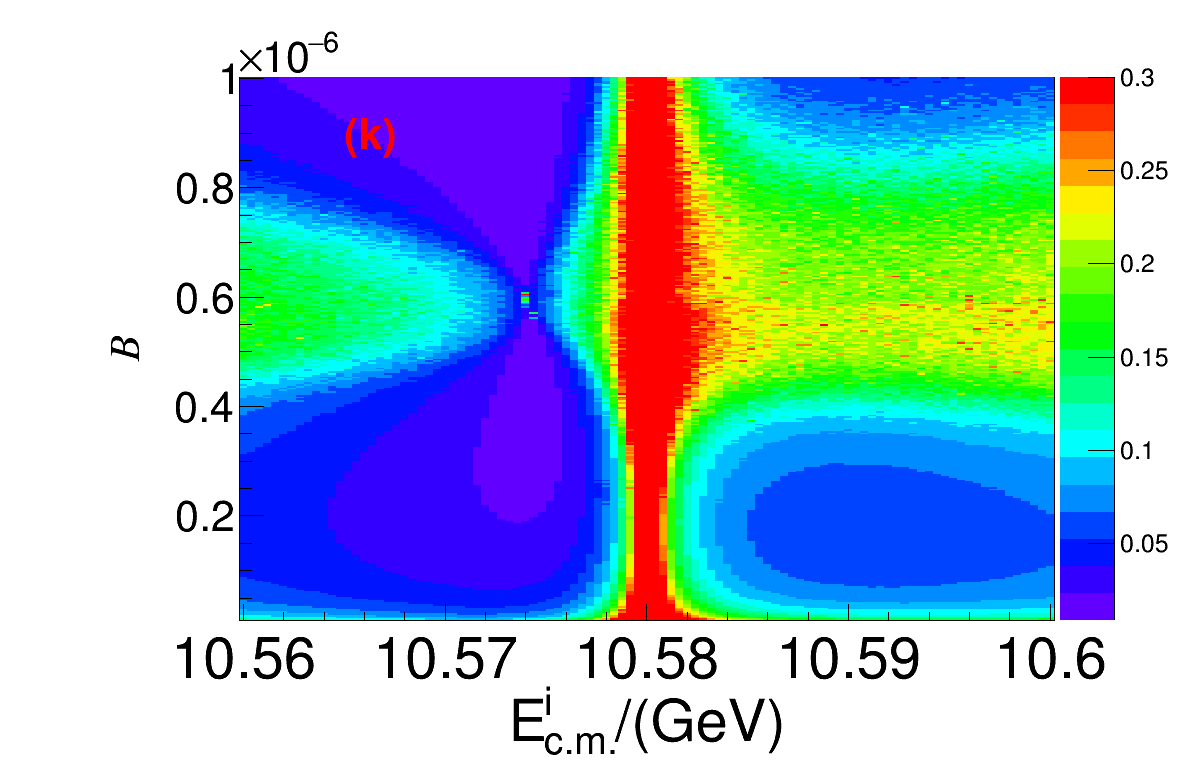}
\includegraphics[width=4.0cm]{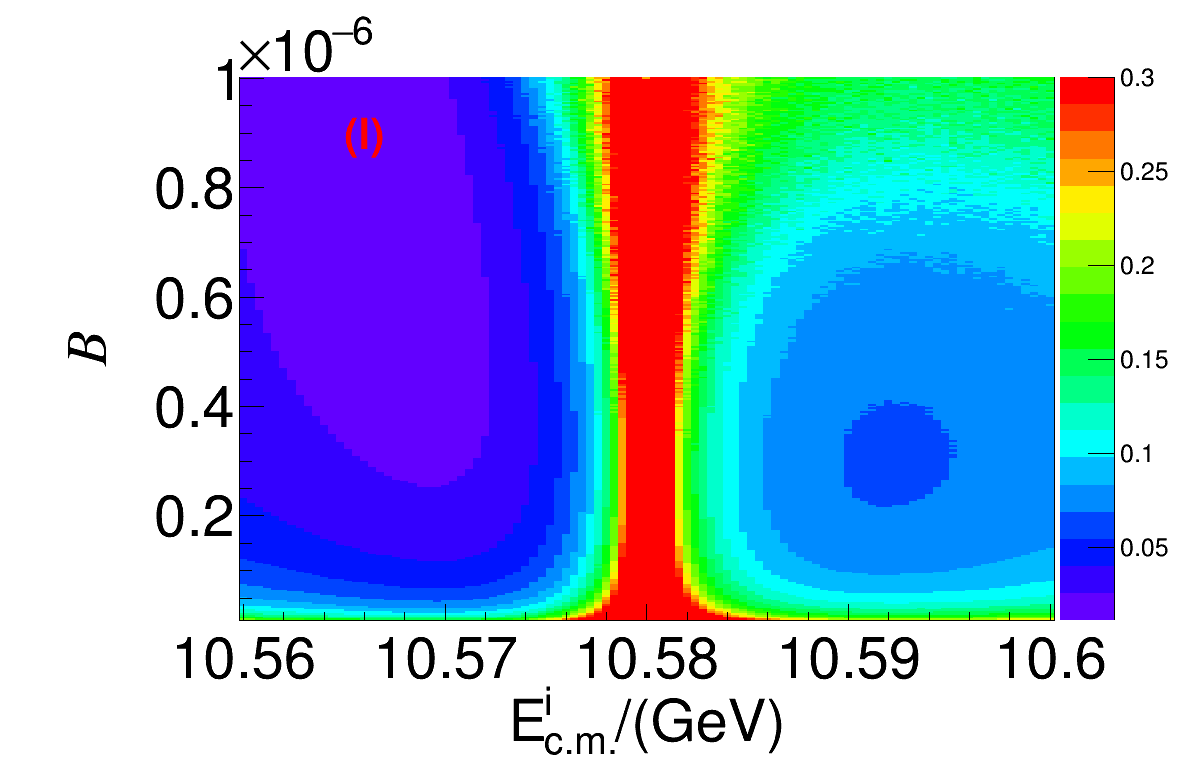}
\caption{Distributions of $E(\BR)$ with $\ecm^i$ and the branching fraction for $\upsilonnn$ decays with
different phase angles $\phi$: (a) $0^\circ$, (b) $30^\circ$, (c) $60^\circ$, (d) $90^\circ$, (e) $120^\circ$,
(f) $150^\circ$, (g) $180^\circ$, (h) $210^\circ$, (i) $240^\circ$, (j) $270^\circ$, (k) $300^\circ$, and (l)
$330^\circ$.}
\label{4s}
\end{center}
\end{figure}

According to the available data at the BESIII experiment and future data acquisition strategies of the Belle II
experiment, the integrated luminosities of the data we use are $20~\infb$ at $\sqrt{s} = 3.773~\gev$ and $50~\inab$
at $\sqrt{s} = 10.58~\gev$, respectively. $\BR$ and $\phi$ as free parameters in the fits, and we obtain the results
for each $30^{\circ}$ variations in the phase angle, as presented in Fig.~\ref{poi3773} for $\psip$ decays and
Fig.~\ref{4s} for $\ufs$ decays. The colors represent the $E(\BR)$ for each energy point and branching fraction
combination at the corresponding phase angle $\phi$.

\section{Fisher information }

In mathematical statistics, Fisher information quantifies the amount of information that a random variable $X$
carries about a known parameter $\theta$~\cite{fisher}. In the context of our study, Fisher information is employed
to estimate the parameter $\theta = (\BR, \phi)$ based on the observed event numbers $\boldsymbol{x} = (x_1, x_2)$ as
described in Sec.~\ref{sec2a}.

\begin{figure}[htp]
\begin{center}
\includegraphics[width=4.0cm]{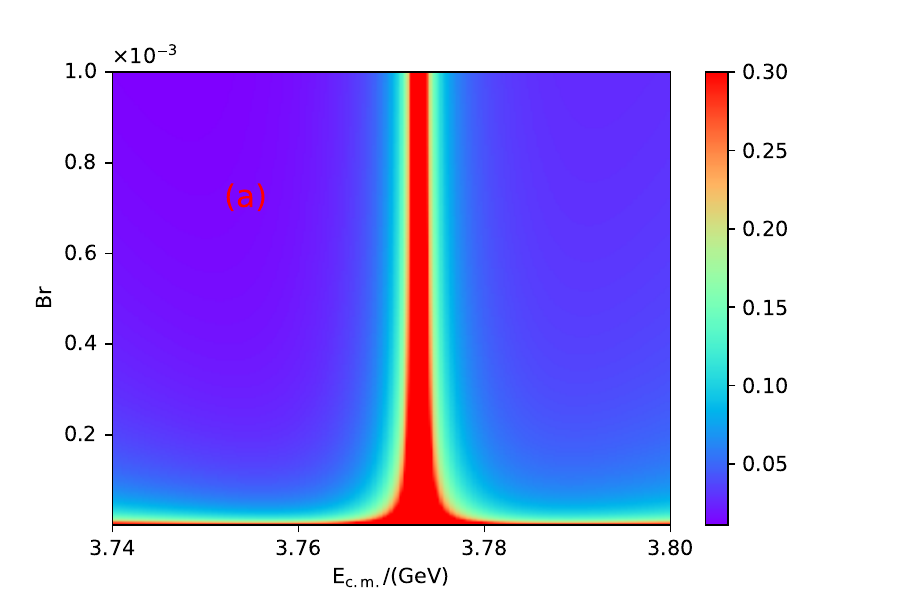} 
\includegraphics[width=4.0cm]{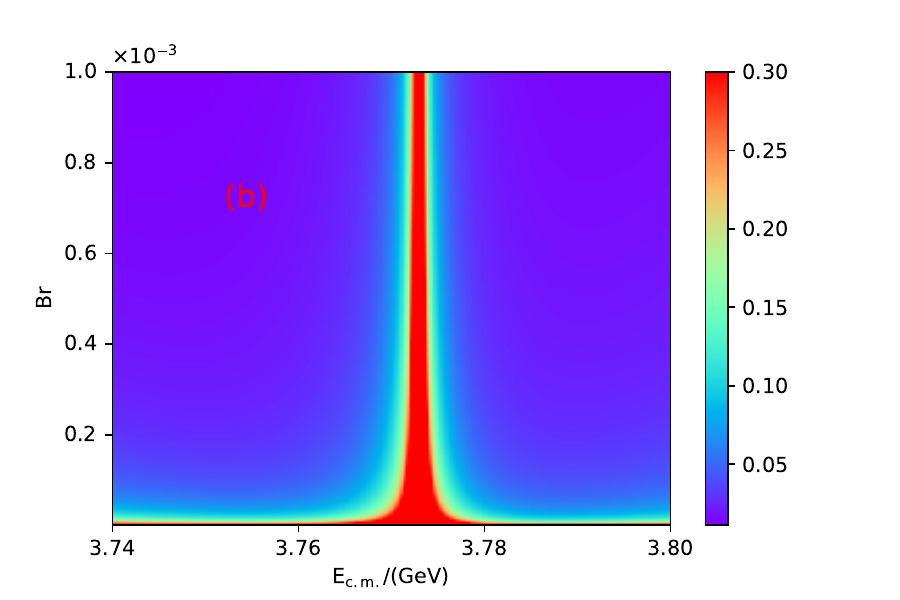} 
\includegraphics[width=4.0cm]{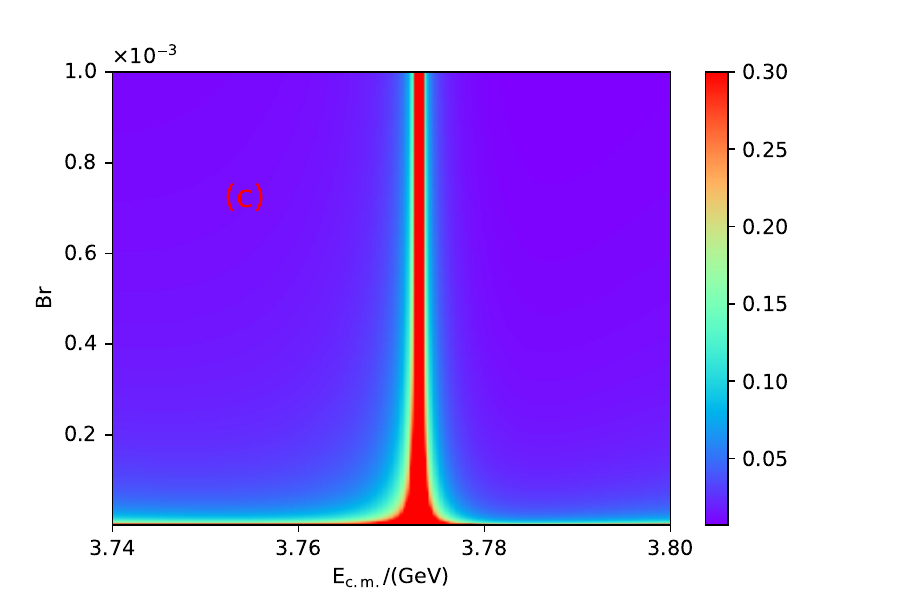} 
\includegraphics[width=4.0cm]{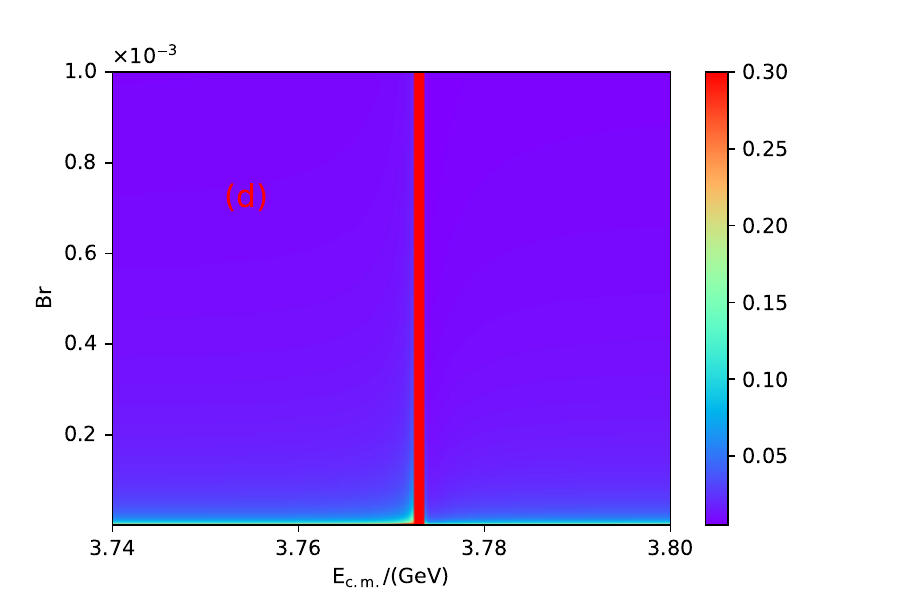} \\
\includegraphics[width=4.0cm]{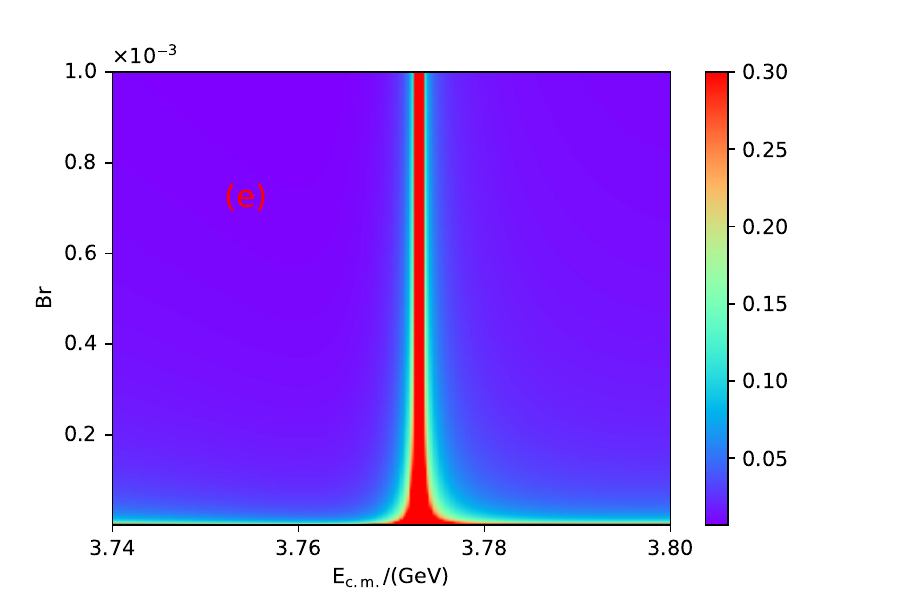} 
\includegraphics[width=4.0cm]{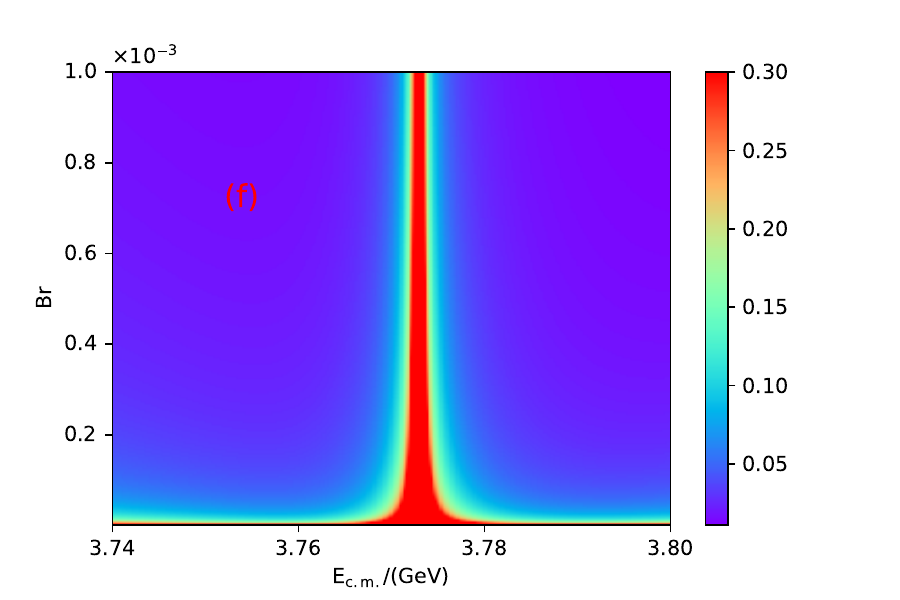} 
\includegraphics[width=4.0cm]{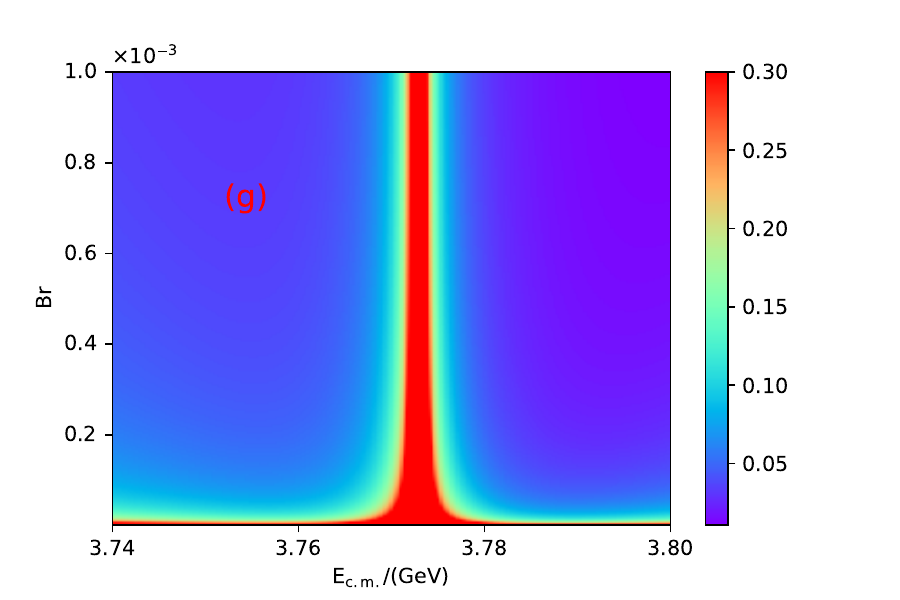}  
\includegraphics[width=4.0cm]{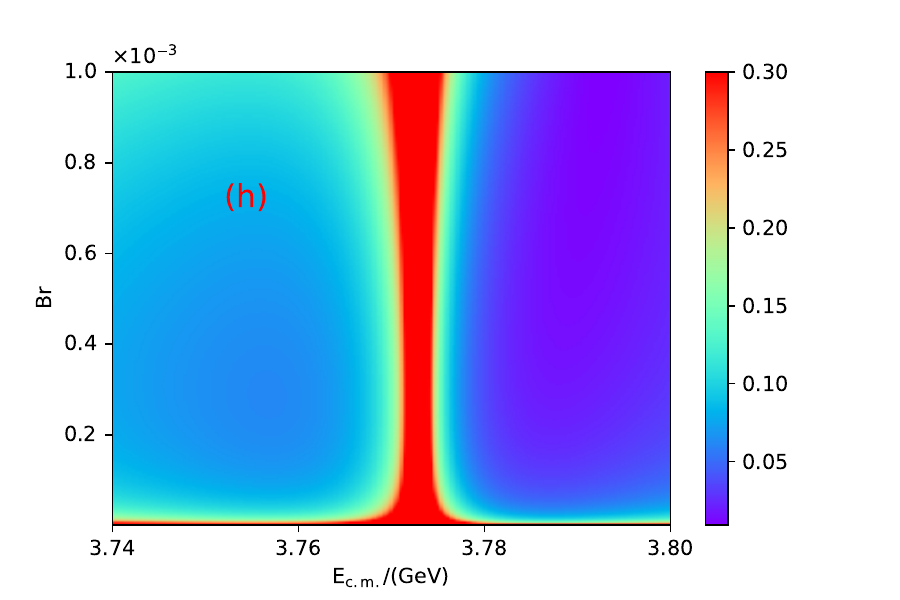}  \\
\includegraphics[width=4.0cm]{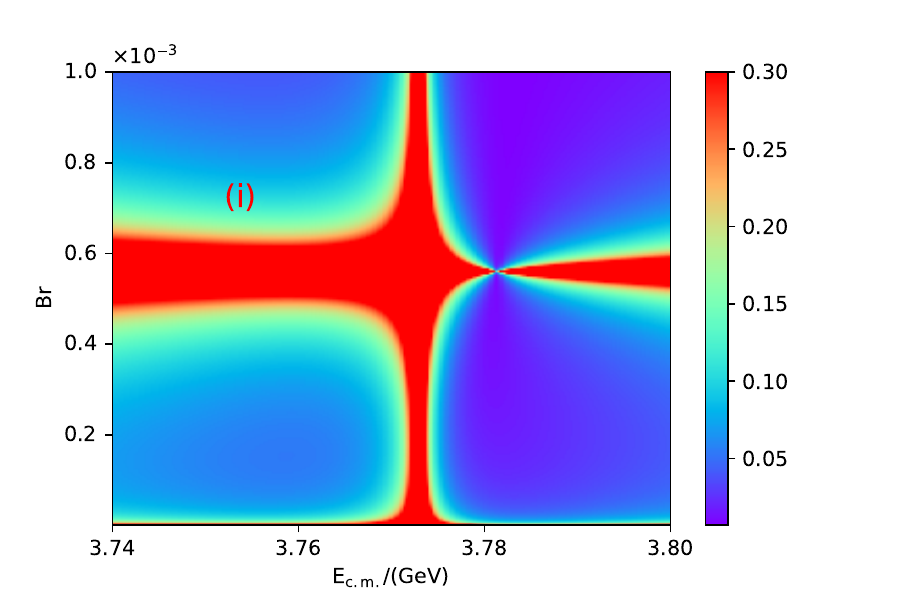}  
\includegraphics[width=4.0cm]{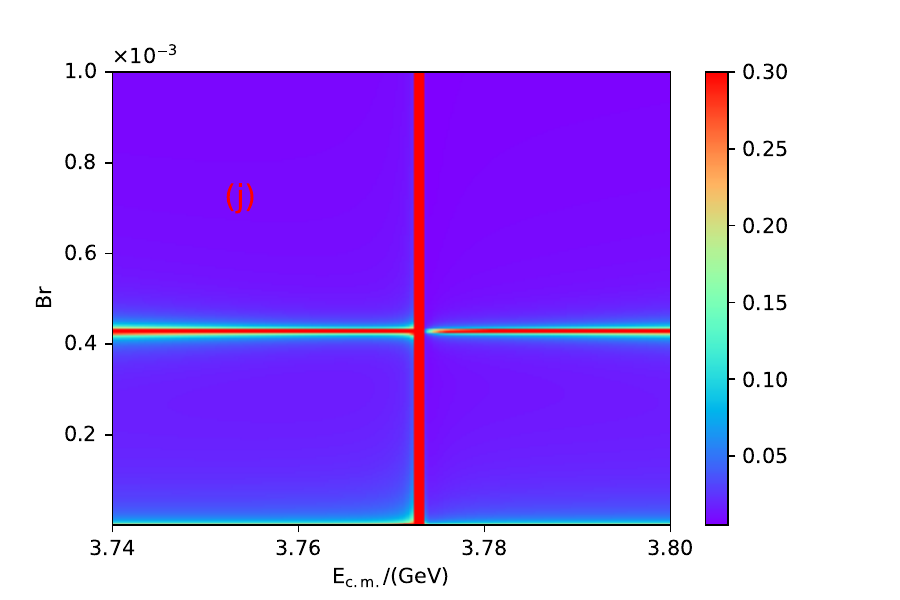}
\includegraphics[width=4.0cm]{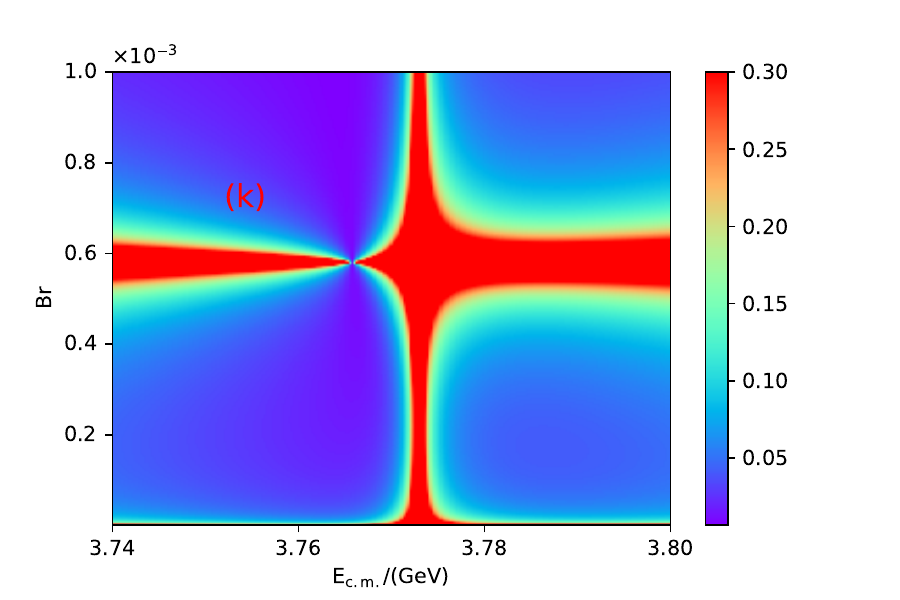}
\includegraphics[width=4.0cm]{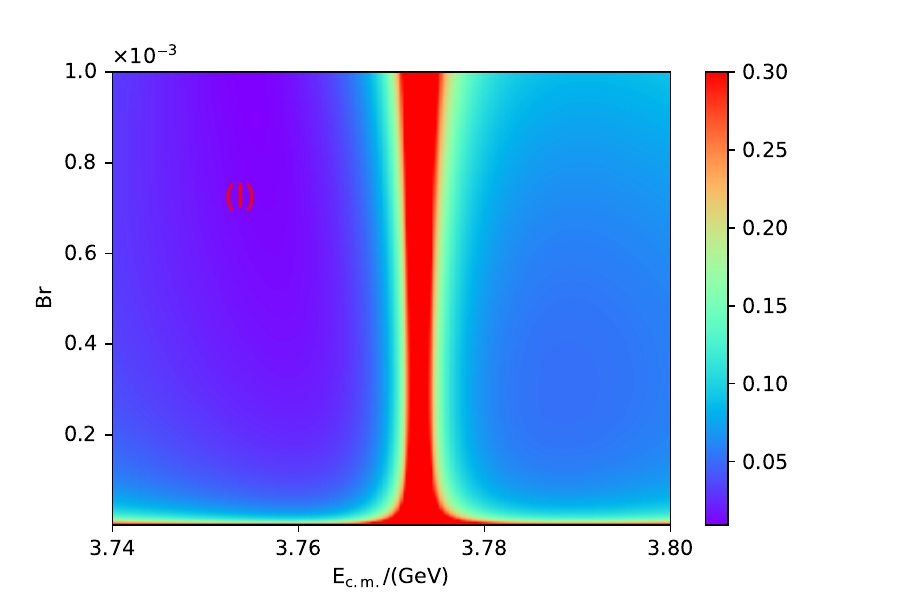}
\caption{Distributions of $E(\BR)$ with $\ecm^i$ and the branching fraction obtained by Fisher information
for $\psinn$ decays with different phase angles $\phi$: (a) $0^\circ$, (b) $30^\circ$, (c) $60^\circ$, (d) $90^\circ$,
(e) $120^\circ$, (f) $150^\circ$, (g) $180^\circ$, (h) $210^\circ$, (i) $240^\circ$, (j) $270^\circ$, (k) $300^\circ$,
and (l) $330^\circ$.}
\label{3773_fisher}
\end{center}
\end{figure}

\begin{figure}[htp]
\begin{center}
\includegraphics[width=4.0cm]{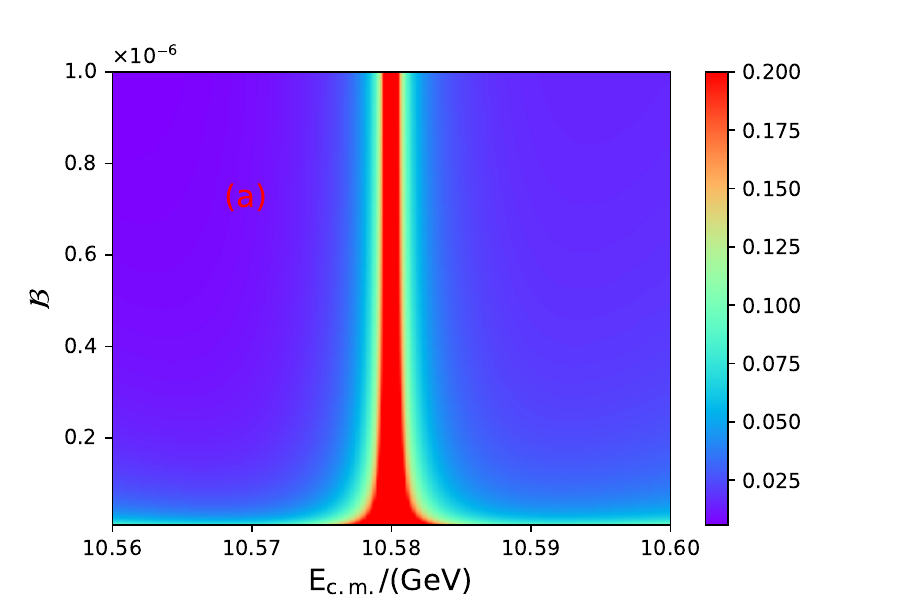} 
\includegraphics[width=4.0cm]{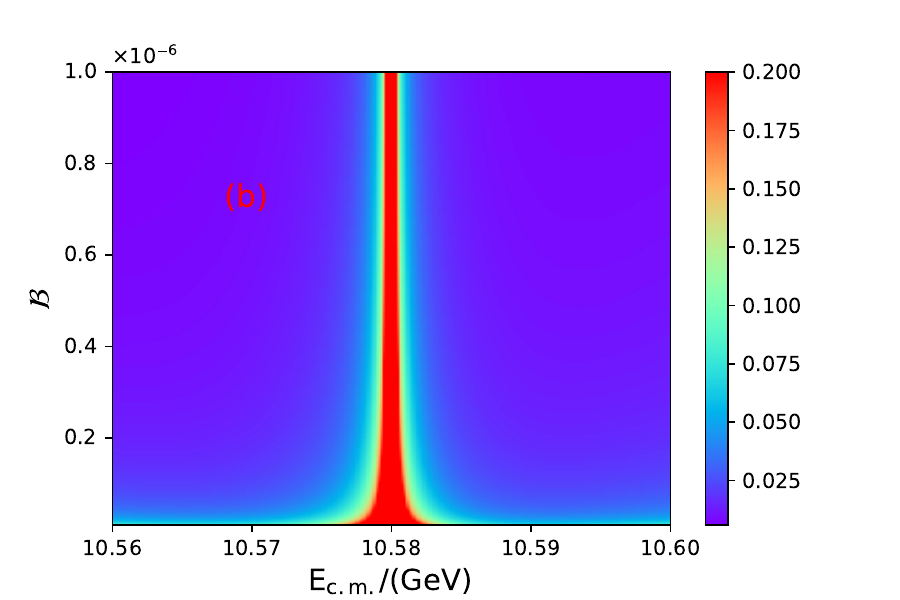} 
\includegraphics[width=4.0cm]{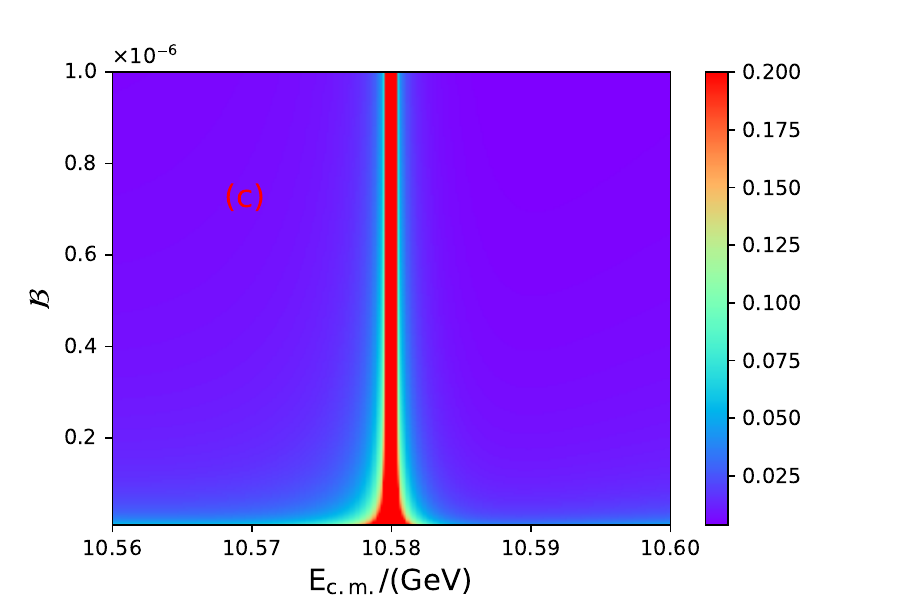} 
\includegraphics[width=4.0cm]{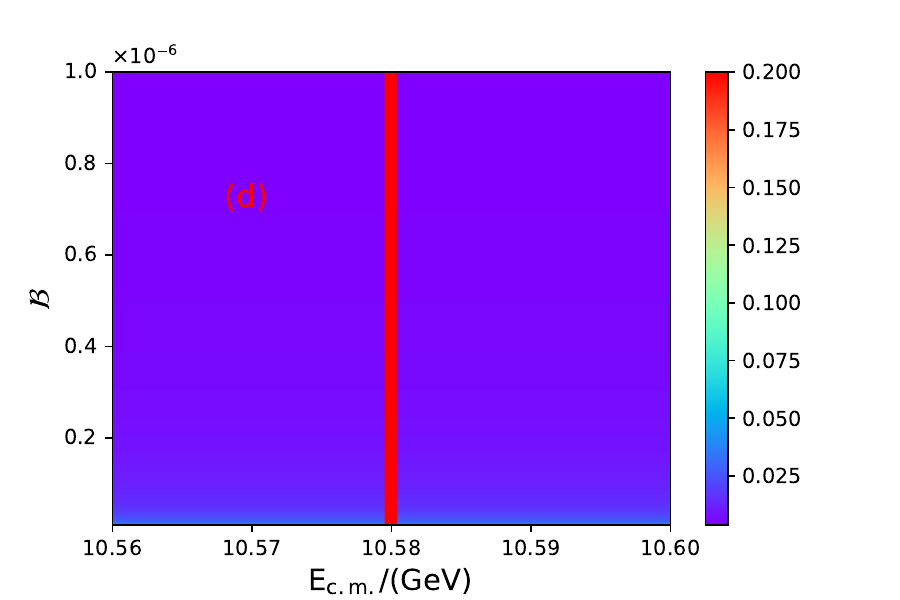} \\
\includegraphics[width=4.0cm]{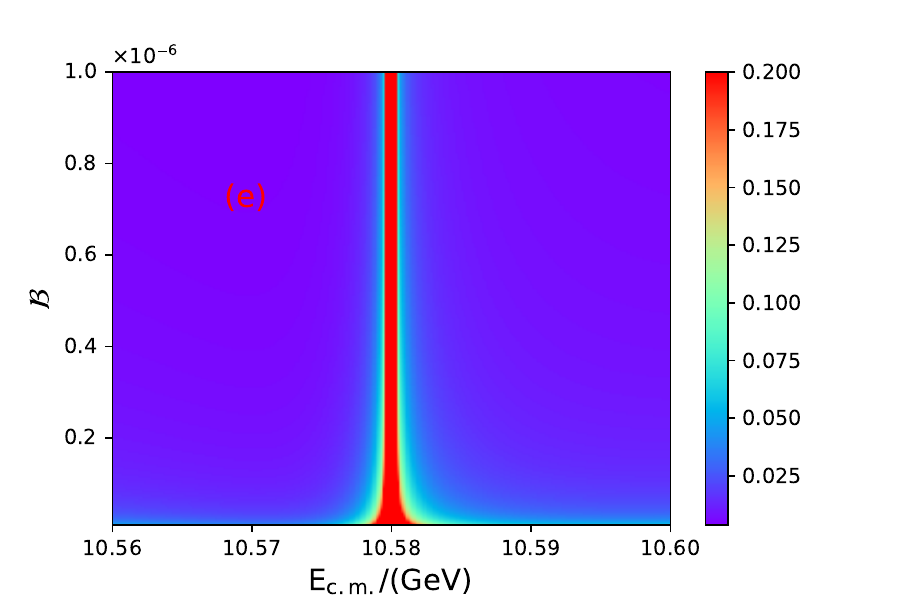} 
\includegraphics[width=4.0cm]{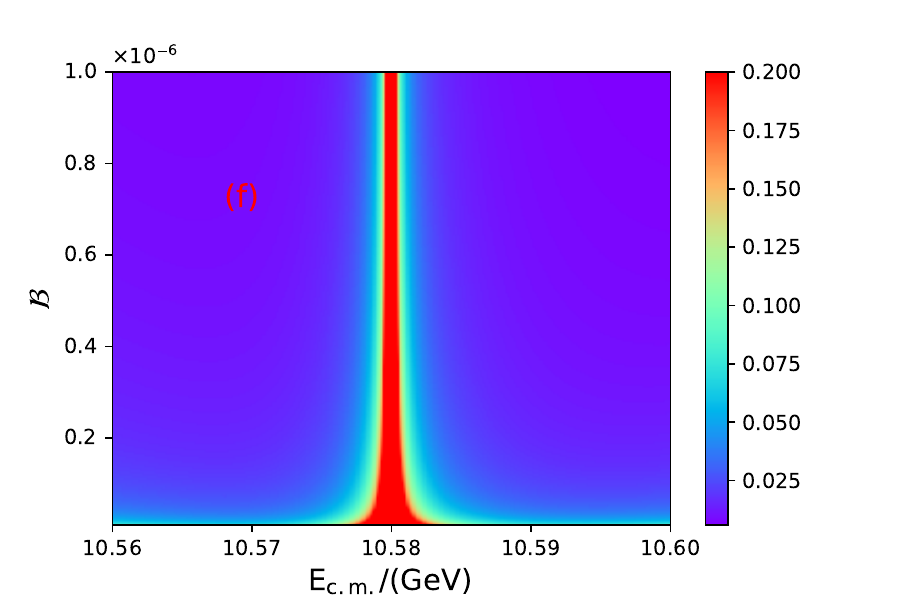} 
\includegraphics[width=4.0cm]{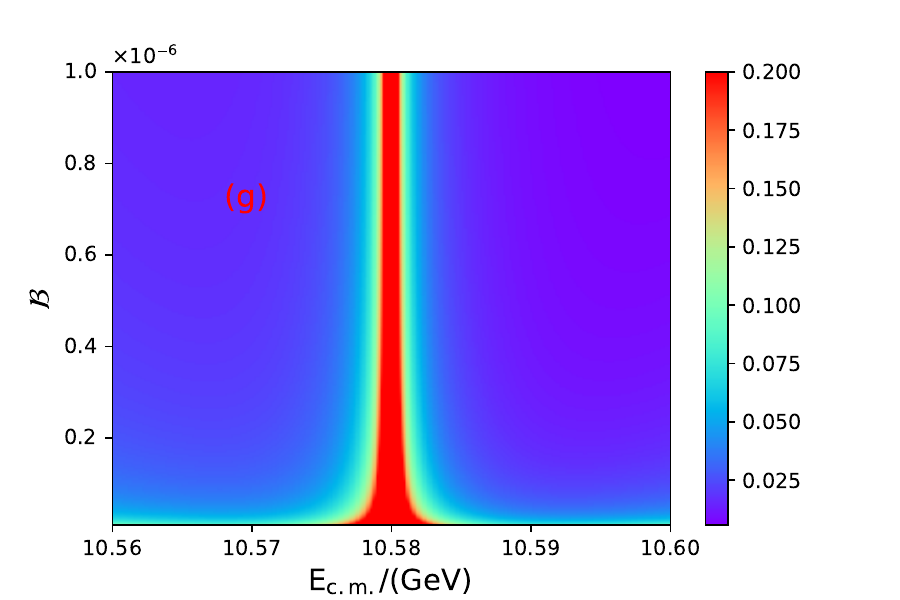}  
\includegraphics[width=4.0cm]{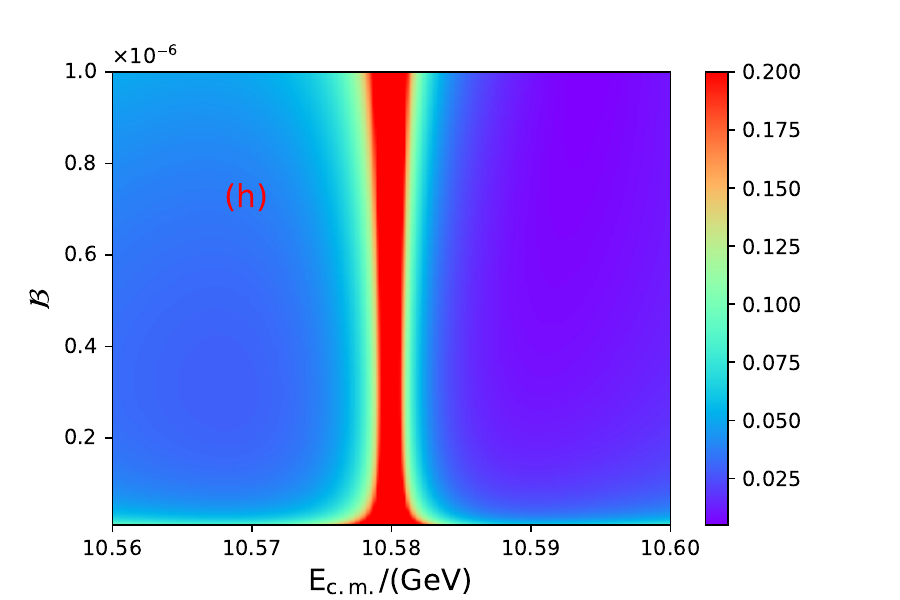}  \\
\includegraphics[width=4.0cm]{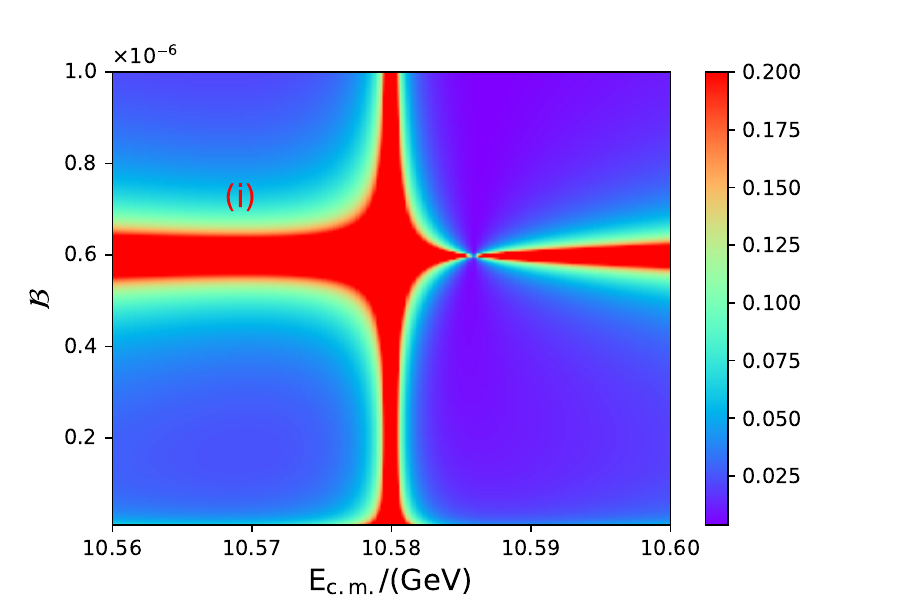}  
\includegraphics[width=4.0cm]{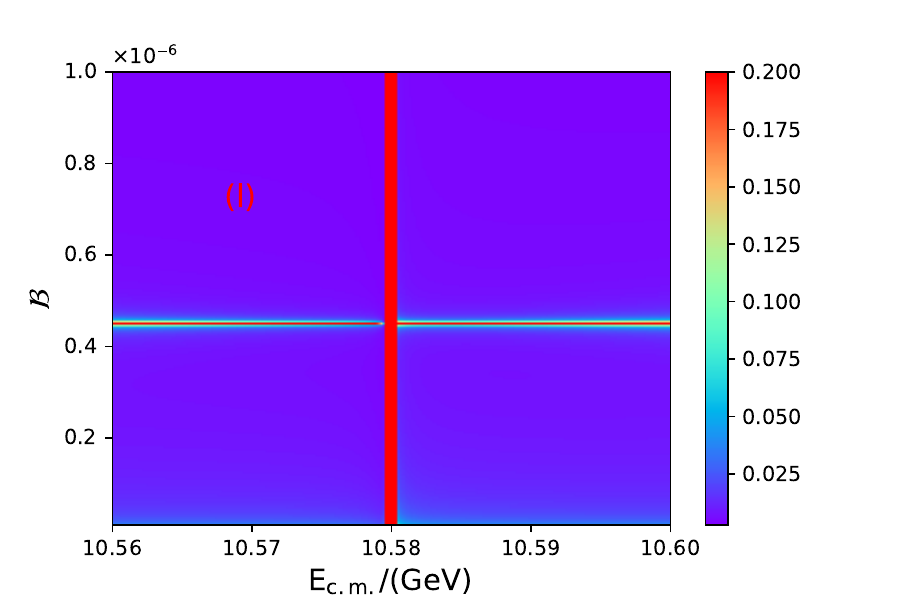}
\includegraphics[width=4.0cm]{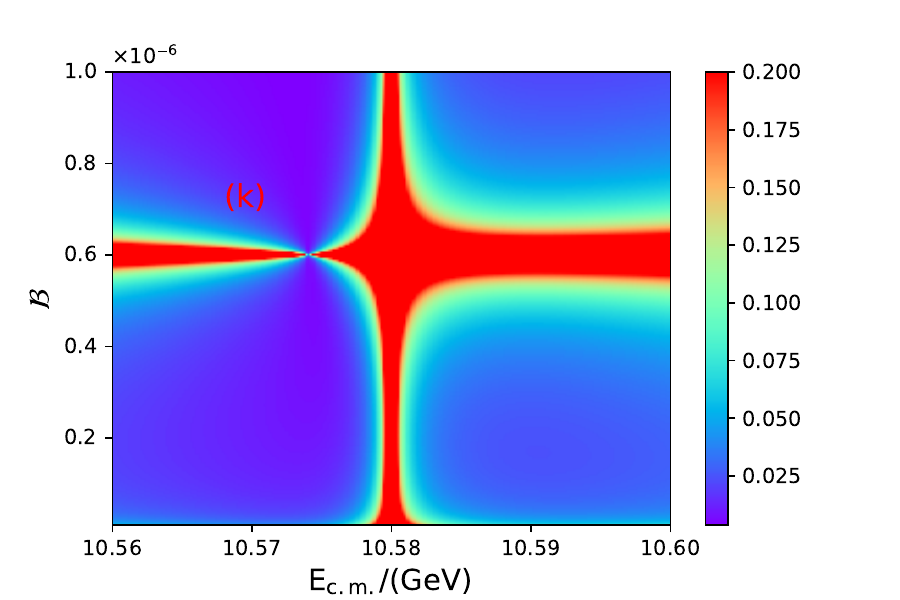}
\includegraphics[width=4.0cm]{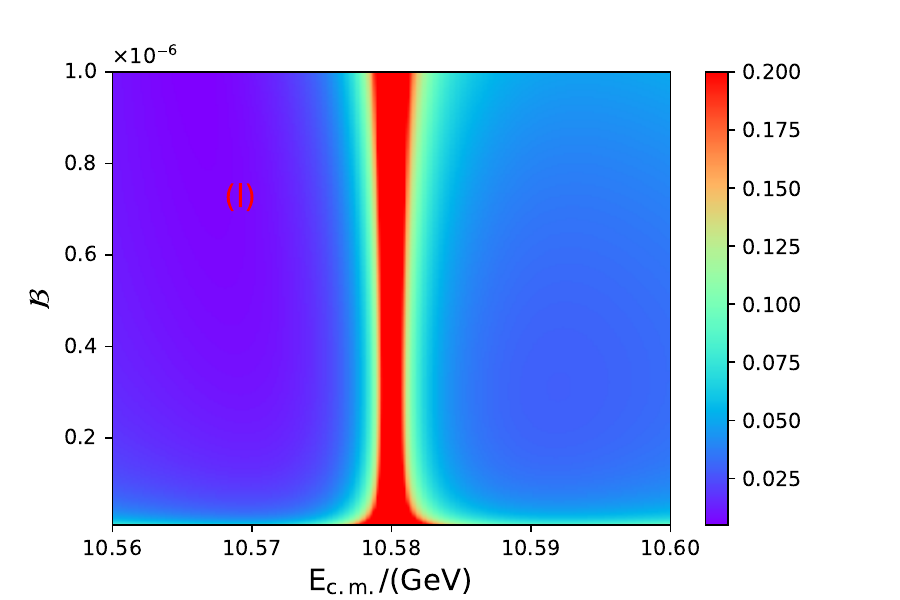}
\caption{Distributions of $E(\BR)$ with $\ecm^i$ and the branching fraction obtained by Fisher information
for $\upsilonnn$ decays with different phase angles $\phi$: (a) $0^\circ$, (b) $30^\circ$, (c) $60^\circ$, (d)
$90^\circ$, (e) $120^\circ$, (f) $150^\circ$, (g) $180^\circ$, (h) $210^\circ$, (i) $240^\circ$, (j) $270^\circ$, (k)
$300^\circ$, and (l) $330^\circ$.}
\label{4s_fisher}
\end{center}
\end{figure}

According to Eq.~\ref{likelihood}, the score function is constructed as 
\beq
  s(\theta, \boldsymbol{x}) = \frac{\partial [\log LF(\boldsymbol{x};\BR,\phi)]}{\partial \theta},
\eeq
and the Fisher information matrix is defined as the covariance matrix of the score function. Concretely,
\beq
  i(\theta) = E_{\theta}(s(\theta,\boldsymbol{x})s(\theta,\boldsymbol{x})^T).
\eeq
where $E_\theta$ represents the expected value of the outer product of the score function $s(\theta,\boldsymbol{x})
s(\theta,\boldsymbol{x})^T$ given the parameter $\theta$. When considering $\BR$ and $\phi$ as two free parameters,
the Fisher information matrix of $\theta = (\BR, \phi)$ is a $2 \times 2$ matrix, and its elements are
\begin{align*}
i_{11} &= \int \int \left(\frac{\partial [\log LF(\boldsymbol{x}; \BR,\phi)]}{\partial\BR}\right)^2 \cdot
LF(\boldsymbol{x}; \BR,\phi) d^2\boldsymbol{x}, \\
i_{12}=i_{21} &= \int \int \left(\frac{\partial [\log LF(\boldsymbol{x}; \BR,\phi)]}{\partial\BR}
\frac{\partial [\log LF(\boldsymbol{x}; \BR,\phi)]}{\partial\phi}\right) \cdot LF(\boldsymbol{x}; \BR,\phi)
d^2\boldsymbol{x}, \\
i_{22} &= \int \int \left(\frac{\partial [\log LF(\boldsymbol{x}; \BR,\phi)]}{\partial\phi}\right)^2 \cdot
LF(\boldsymbol{x}; \BR,\phi) d^2\boldsymbol{x},
\end{align*}
where indices $1$ and $2$ indicate $\BR$ and $\phi$, respectively. Inverting the $2 \times 2$ matrix yields the
asymptotic covariance matrix:
\beq
\begin{bmatrix}
(i_{11}-\frac{i_{12}^2}{i_{22}})^{-1} & (\frac{i_{11}i_{22}}{i_{12}}-i_{12})^{-1} \\
(\frac{i_{11}i_{22}}{i_{12}}-i_{12})^{-1} & (i_{22}-\frac{i_{12}^2}{i_{11}})^{-1}
\end{bmatrix}.
\eeq
Hence, the variance of the branching fraction can be determined as
\beq
 \left(i_{11}-\frac{i_{12}^2}{i_{22}}\right)^{-1} = \frac{2\left( D\phi_1^2 \cdot (1+2x_1^{\rm{exp}})
 (x_2^{\rm{exp}})^2+D\phi_2^2 \cdot (1+2x_2^{\rm{exp}})(x_1^{\rm{exp}})^2\right)}{(DB_2 \cdot D\phi_1 - DB_1 \cdot
 D\phi_2)^2(1+2x_1^{\rm{exp}})(1+2x_2^{\rm{exp}})},
\eeq
Where $x_{1}^{\rm{exp}}$ and $x_2^{\rm{exp}}$ are the expected events at energy points $\ecm^{i}$ and the resonance,
respectively. $DB_i$ and $D\phi_i$ denote $\partial x_i^{\rm{exp}}/\partial \BR$ and $\partial x_i^{\rm{exp}}/\partial
\phi$, respectively. As in Sec.~\ref{sec:2}, we present $E(\BR)$ using Fisher information. The same phase $\phi$
distributions are shown in Figs.~\ref{3773_fisher} and \ref{4s_fisher}. The results are consistent with the
methodology employed in the MC simulation.

\section{Data taking energy points} 

\begin{figure}[htp]
\begin{center}
\includegraphics[width=4.0cm]{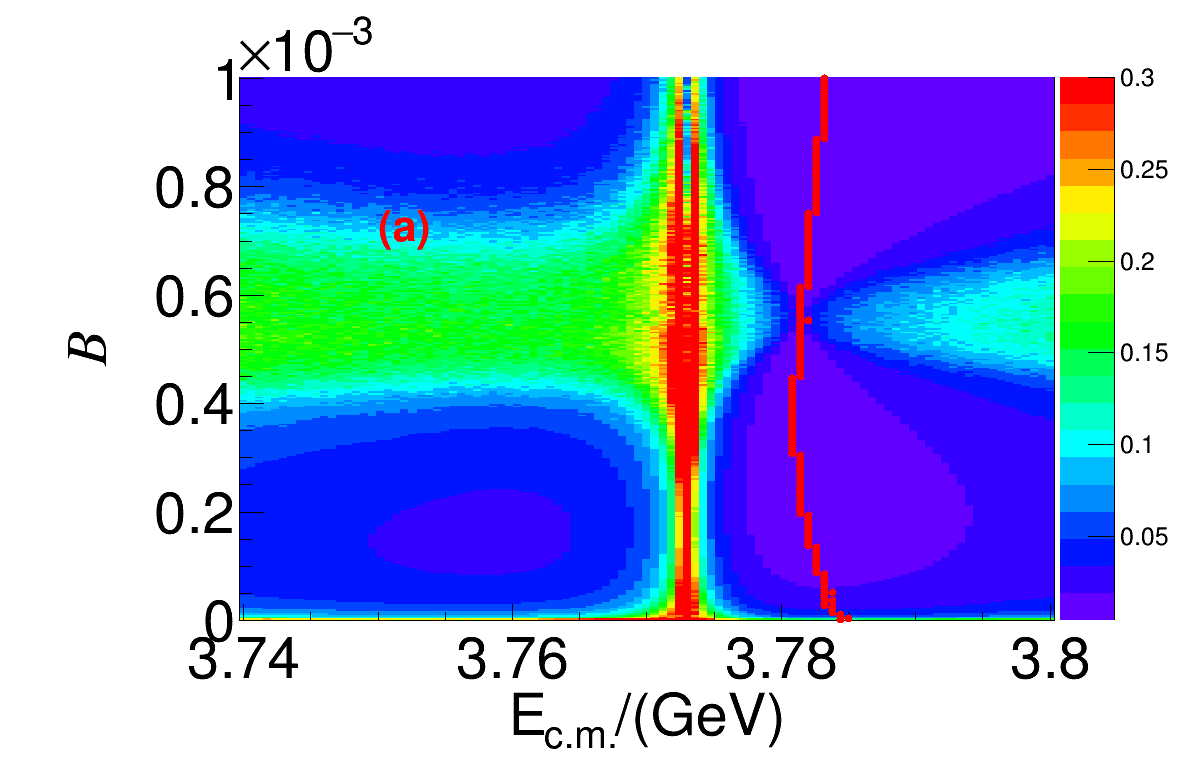} 
\includegraphics[width=4.0cm]{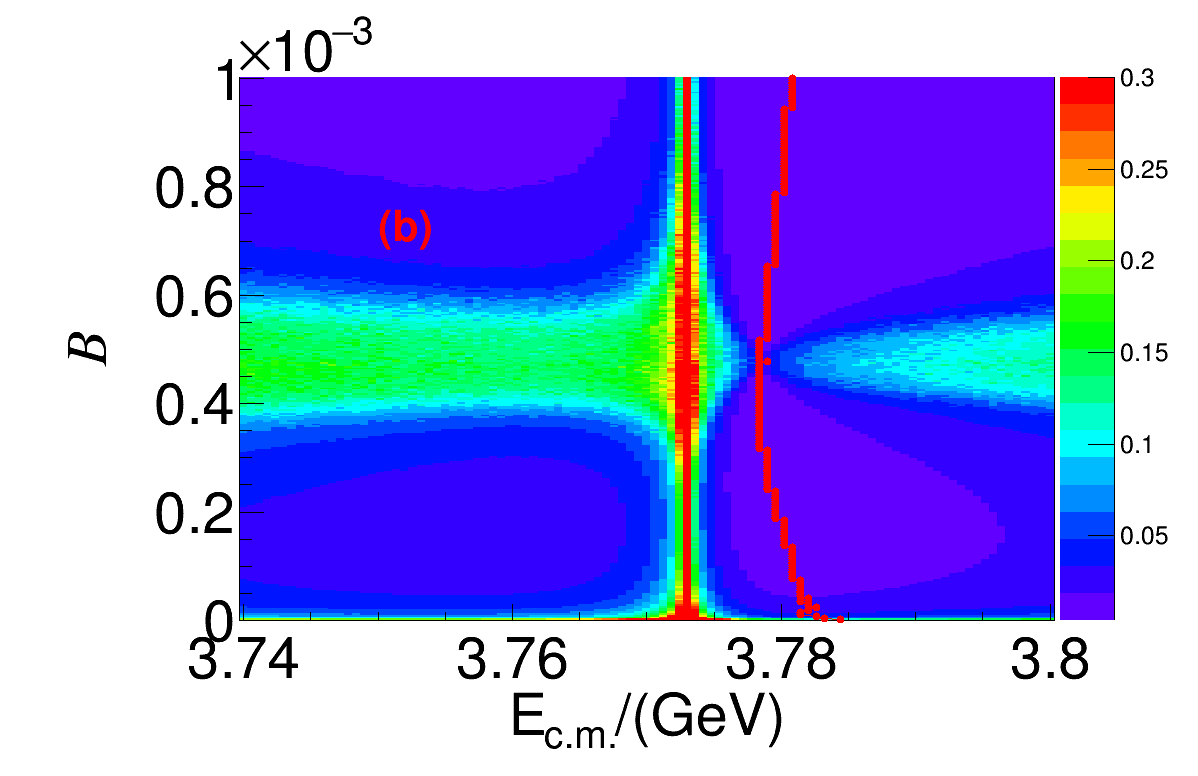} 
\includegraphics[width=4.0cm]{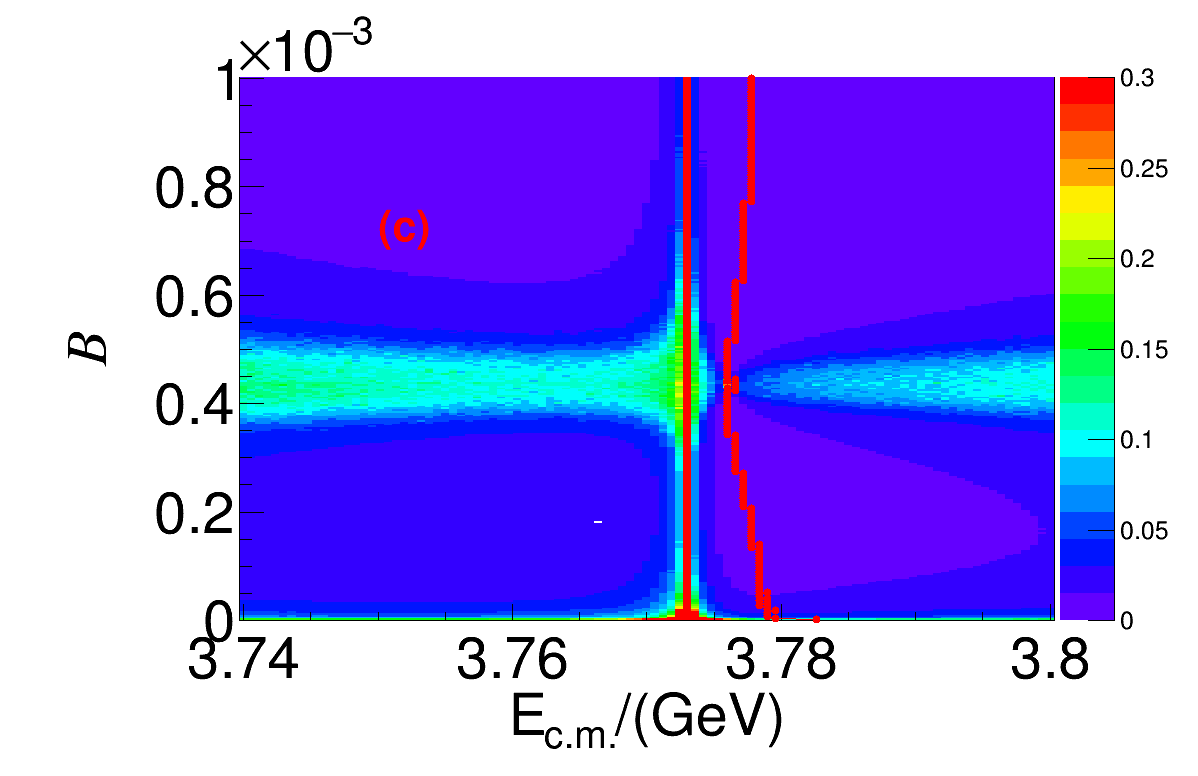} 
\includegraphics[width=4.0cm]{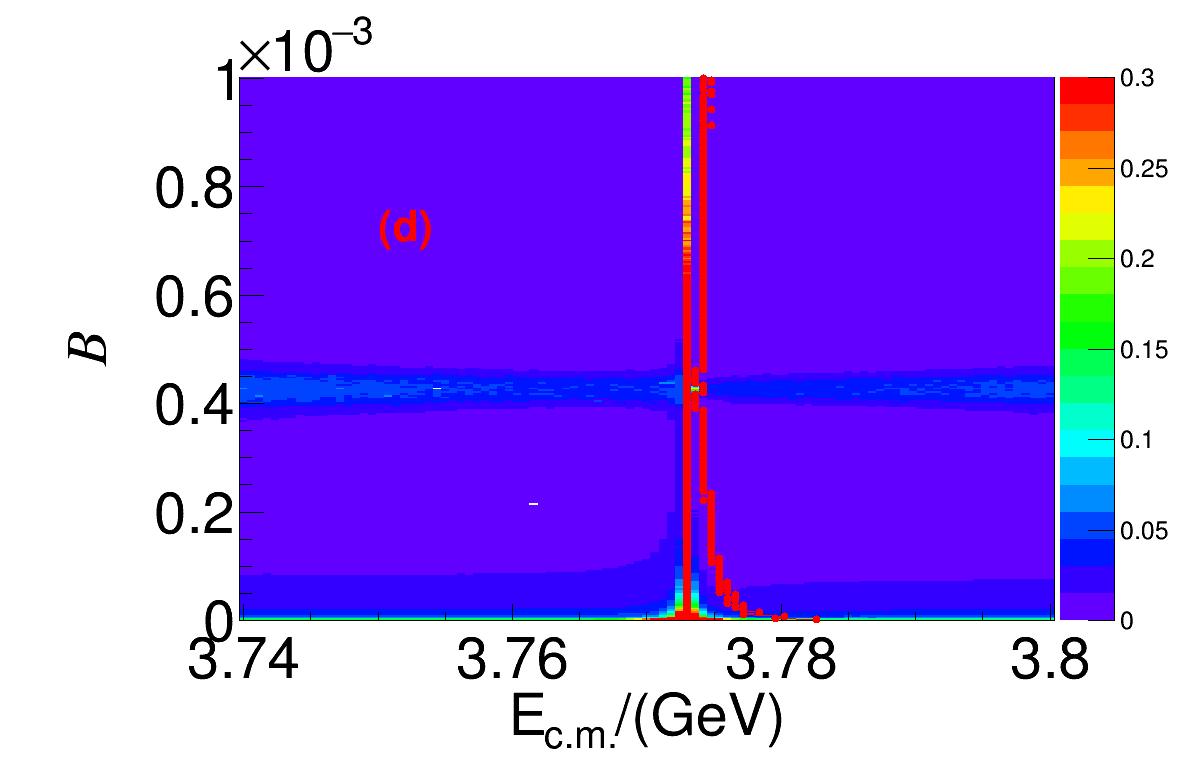} \\
\includegraphics[width=4.0cm]{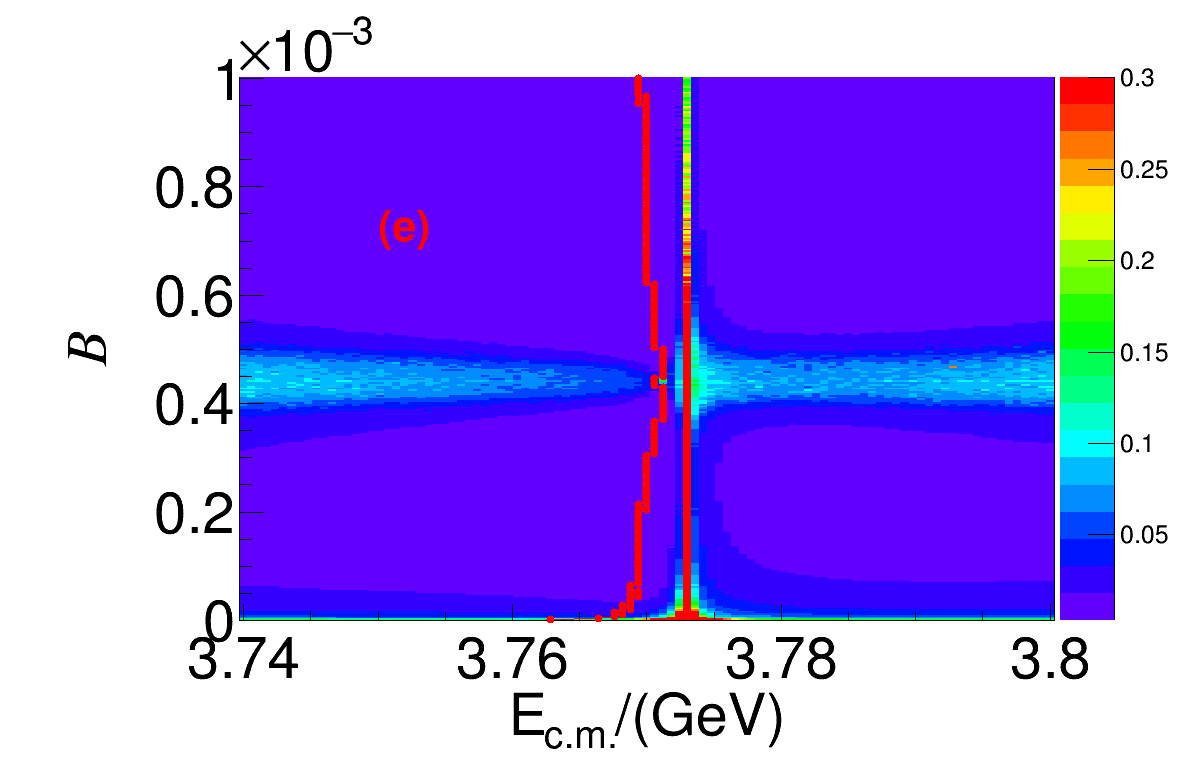} 
\includegraphics[width=4.0cm]{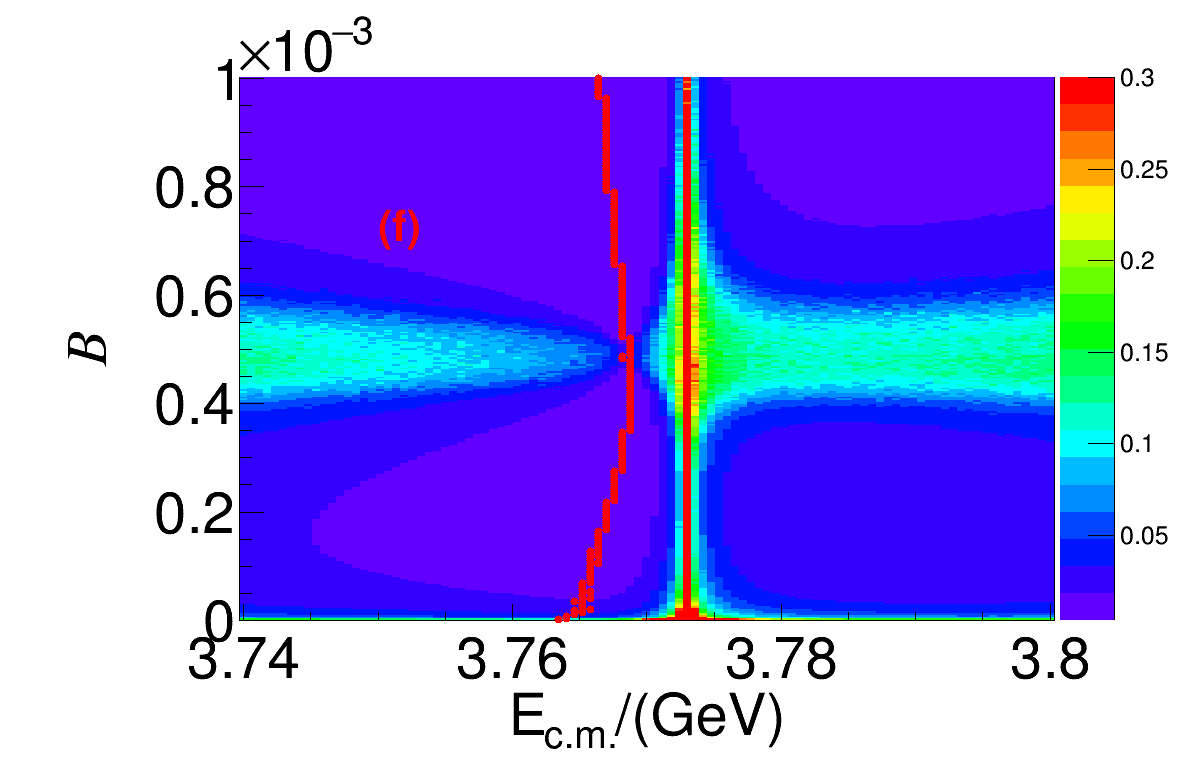} 
\includegraphics[width=4.0cm]{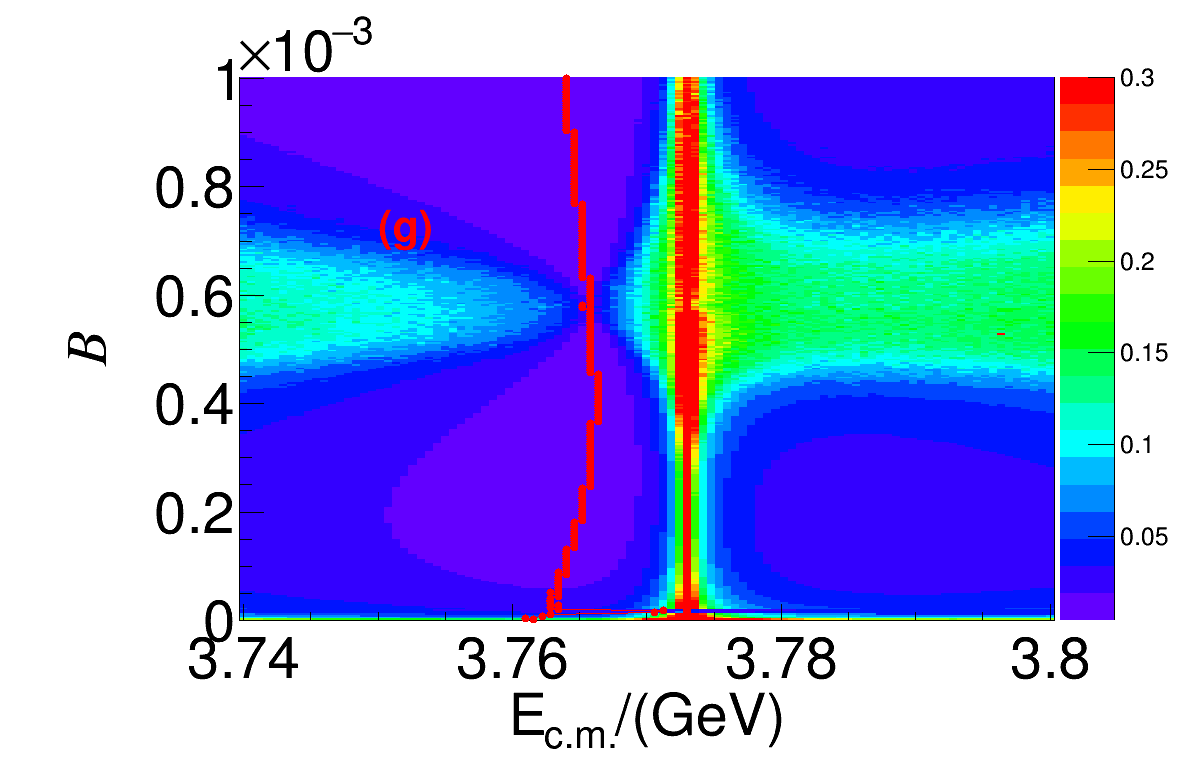}  
\caption{Distributions of $E(\BR)$  with $\ecm^i$ and the branching fraction for $\psinn$ decays from
specific phase angles $\phi$: (a) $240^\circ$, (b) $250^\circ$, (c) $260^\circ$, (d) $270^\circ$, (e) $280^\circ$, (f)
$290^\circ$, and (g) $300^\circ$. The red lines show the best precision of $E(\BR)$ versus the branching ratio $\BR$.}
\label{poimin}
\end{center}
\end{figure}

\begin{figure}[htp]
\begin{center}
\includegraphics[width=4.0cm]{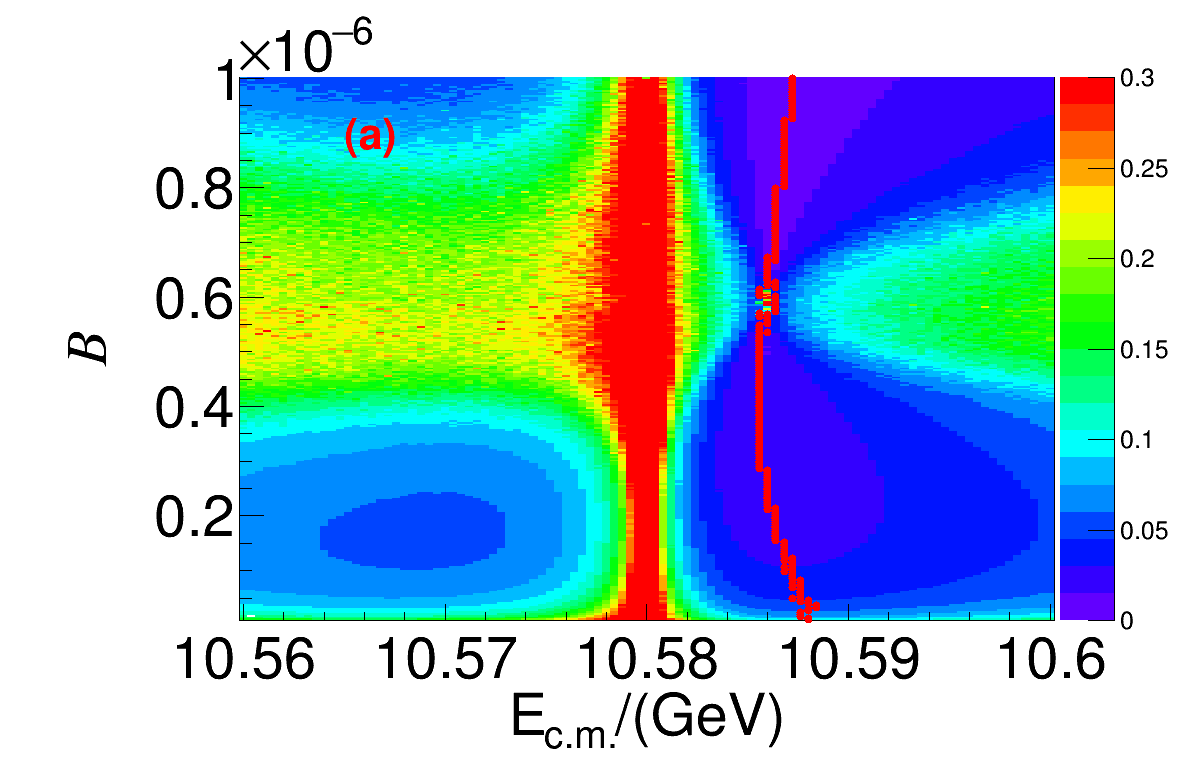} 
\includegraphics[width=4.0cm]{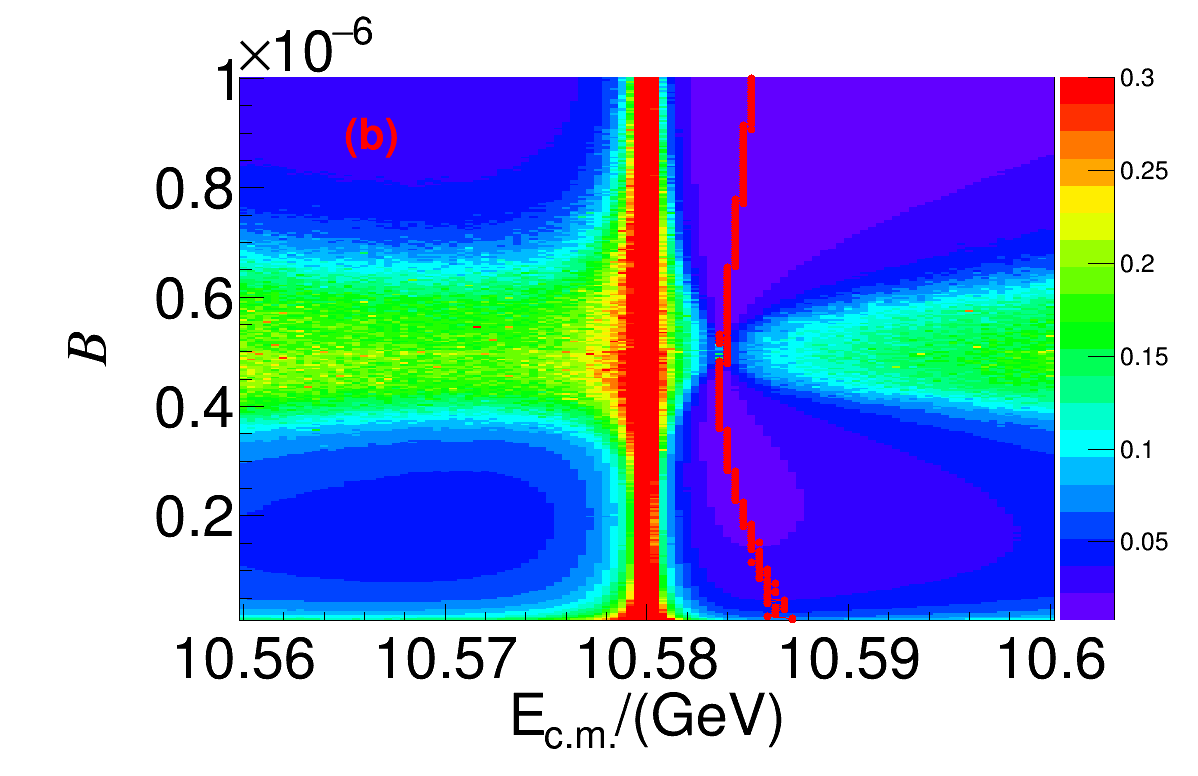} 
\includegraphics[width=4.0cm]{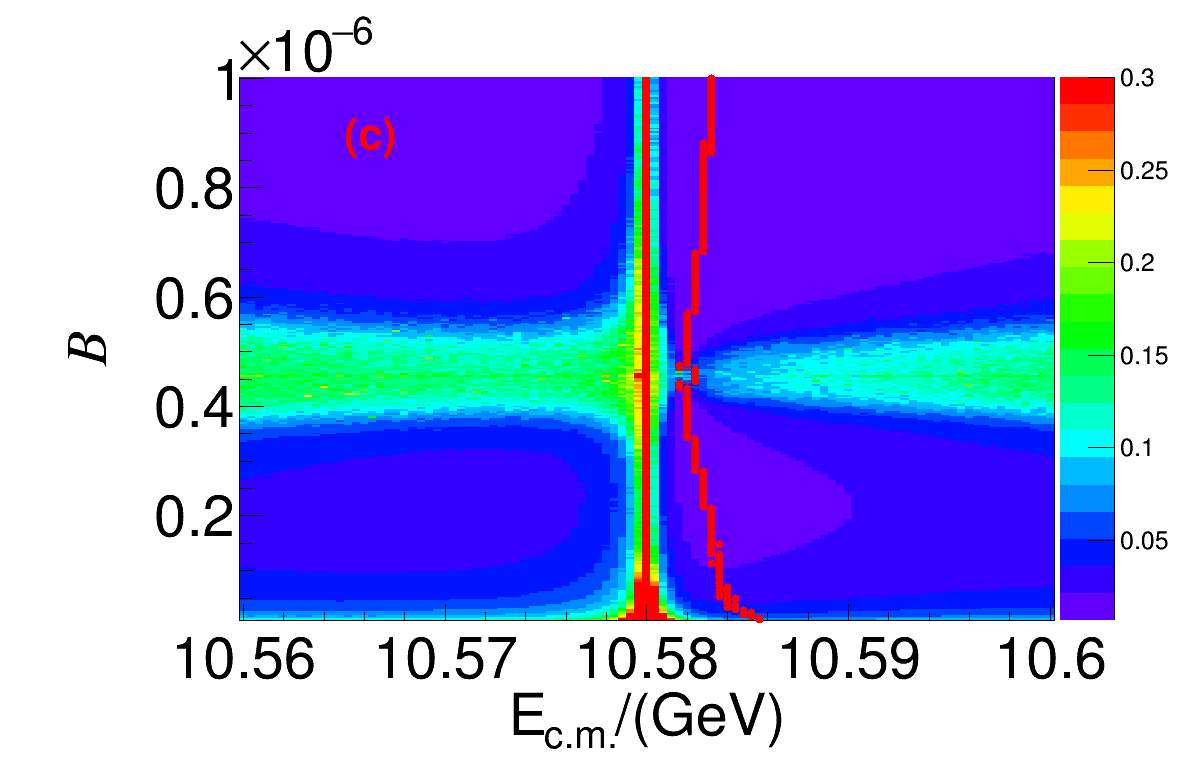} 
\includegraphics[width=4.0cm]{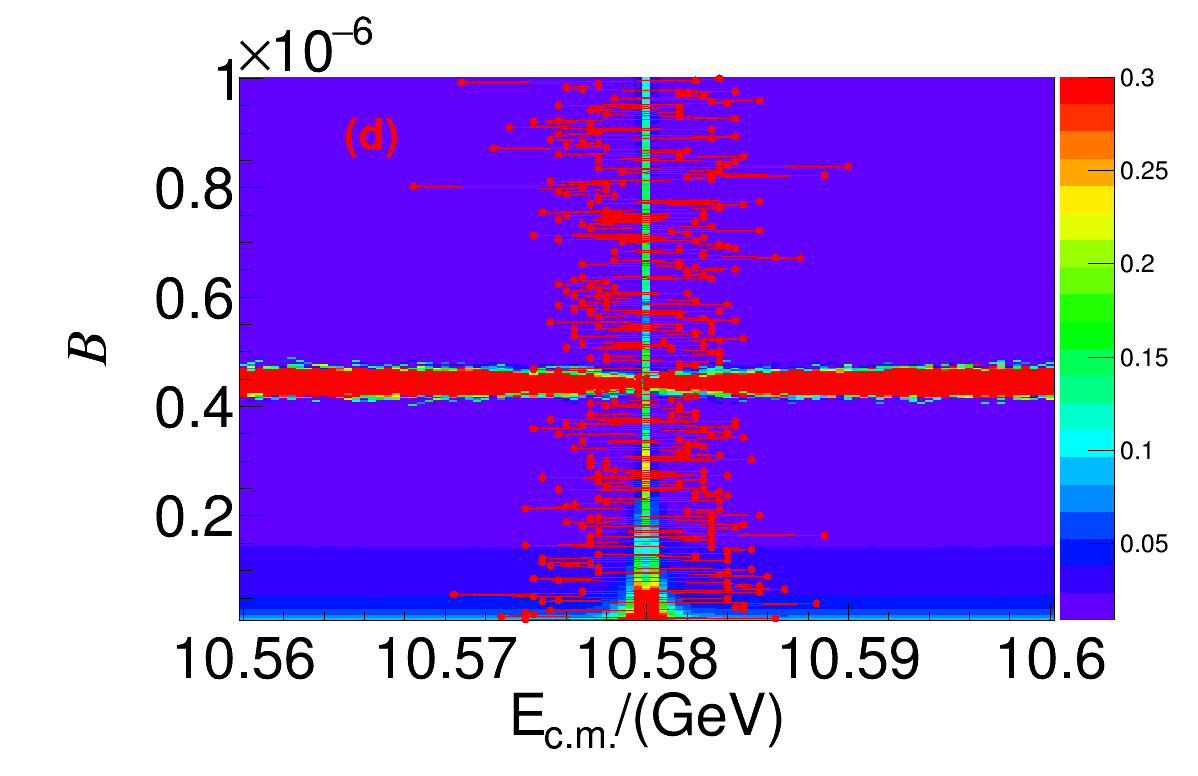} \\
\includegraphics[width=4.0cm]{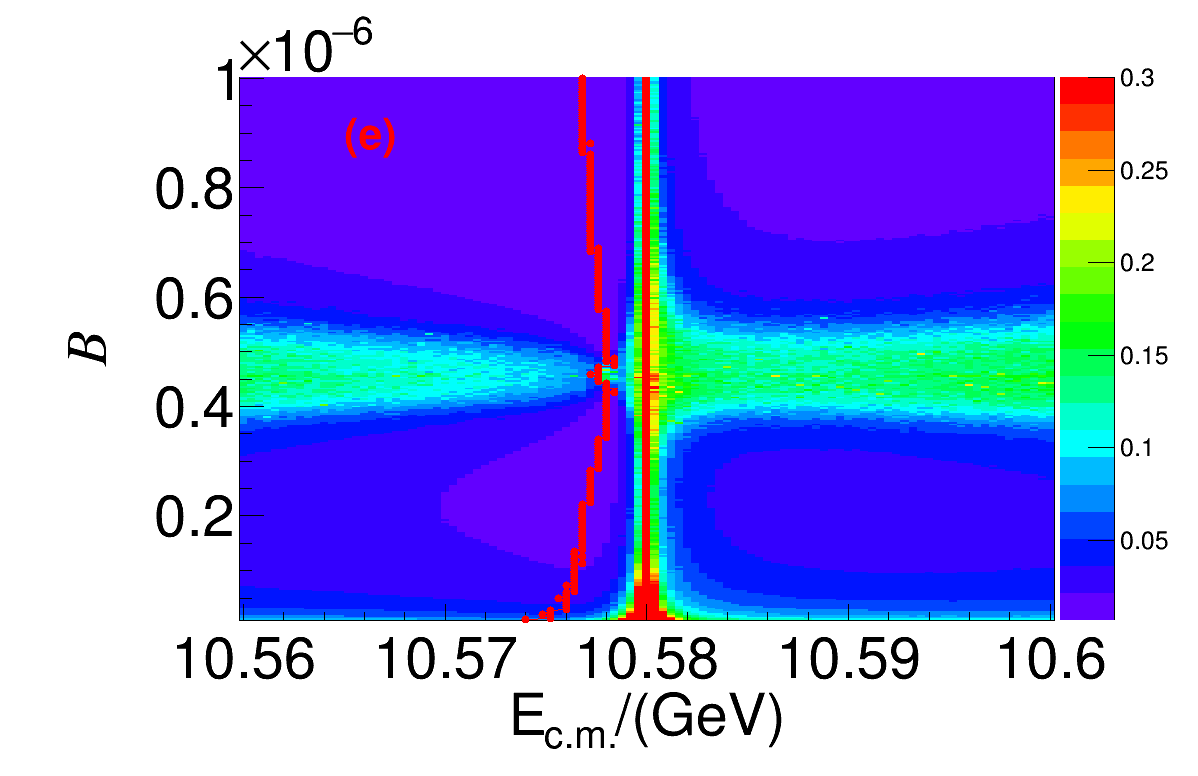} 
\includegraphics[width=4.0cm]{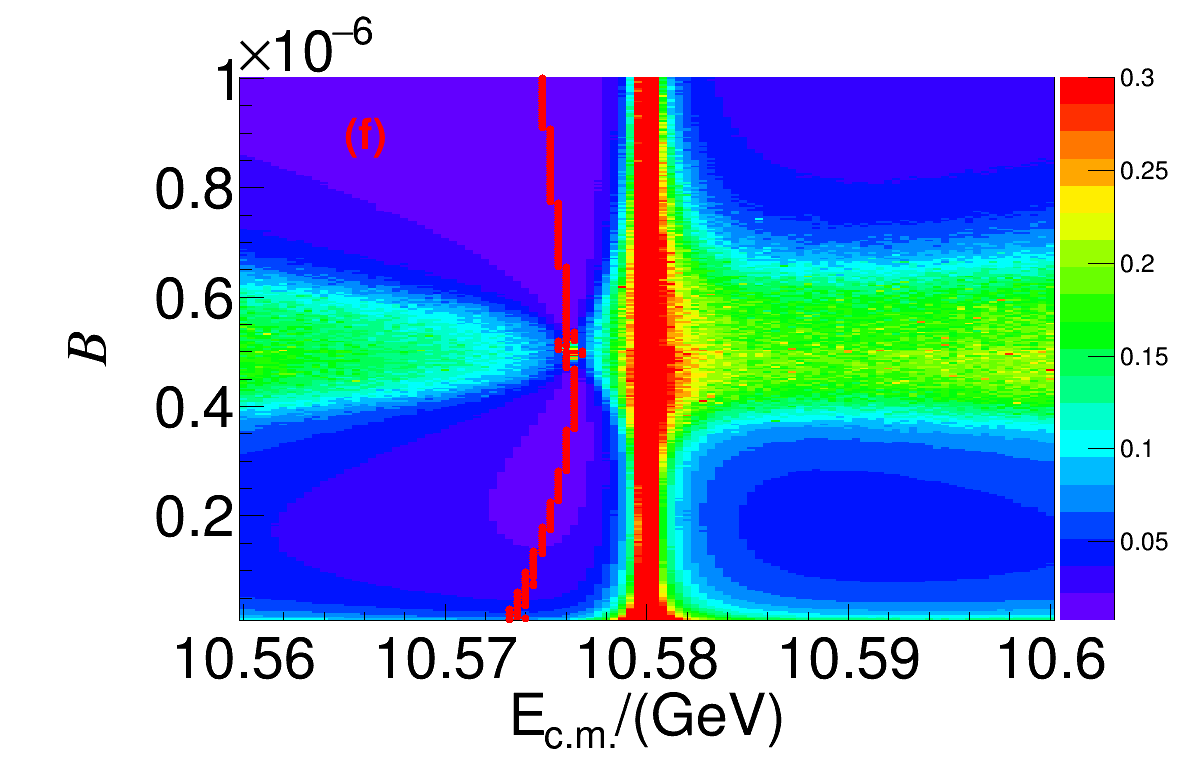} 
\includegraphics[width=4.0cm]{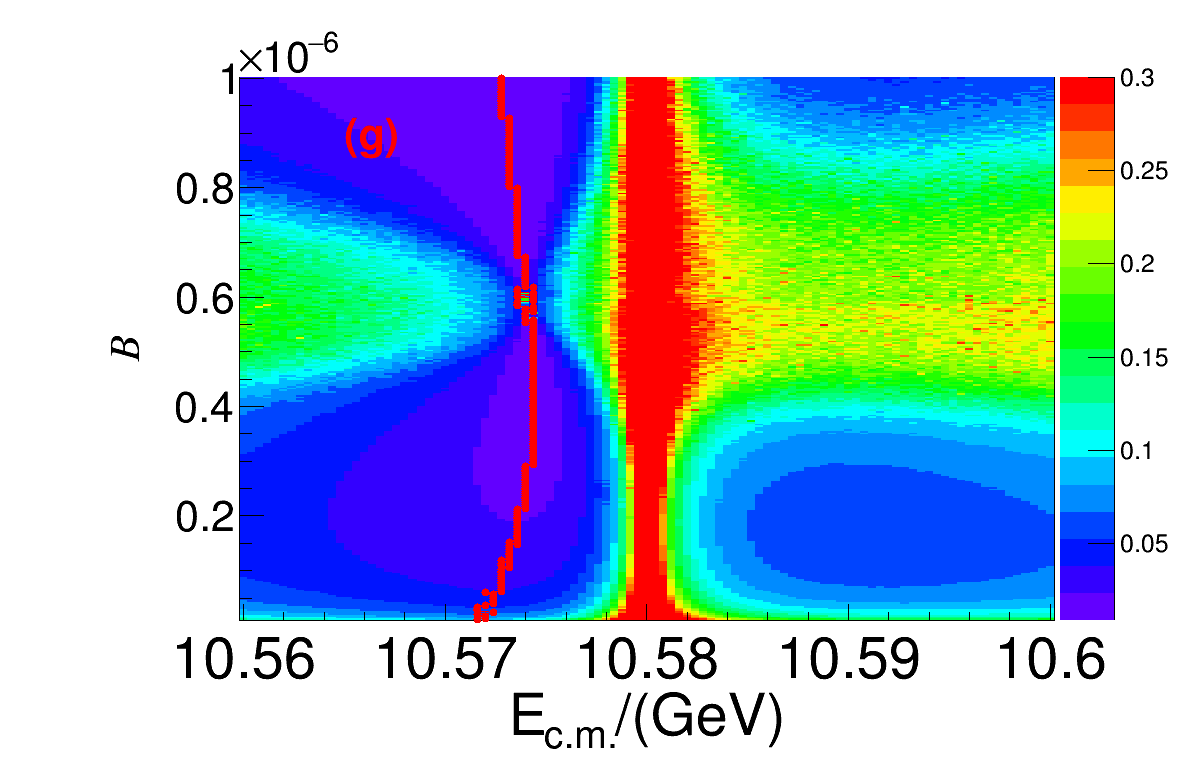}  
\caption{Distributions of $E(\BR)$ with $\ecm^i$ and the branching fraction for $\upsilonnn$ decays from
specific phase angles $\phi$: (a) $240^\circ$, (b) $250^\circ$, (c) $260^\circ$, (d) $270^\circ$, (e) $280^\circ$, (f)
$290^\circ$, and (g) $300^\circ$. The red lines show the best precision of $E(\BR)$ versus the branching ratio $\BR$.}
\label{4smin}
\end{center}
\end{figure}

\begin{figure}[htp]
\begin{center}
\includegraphics[width=8.0cm]{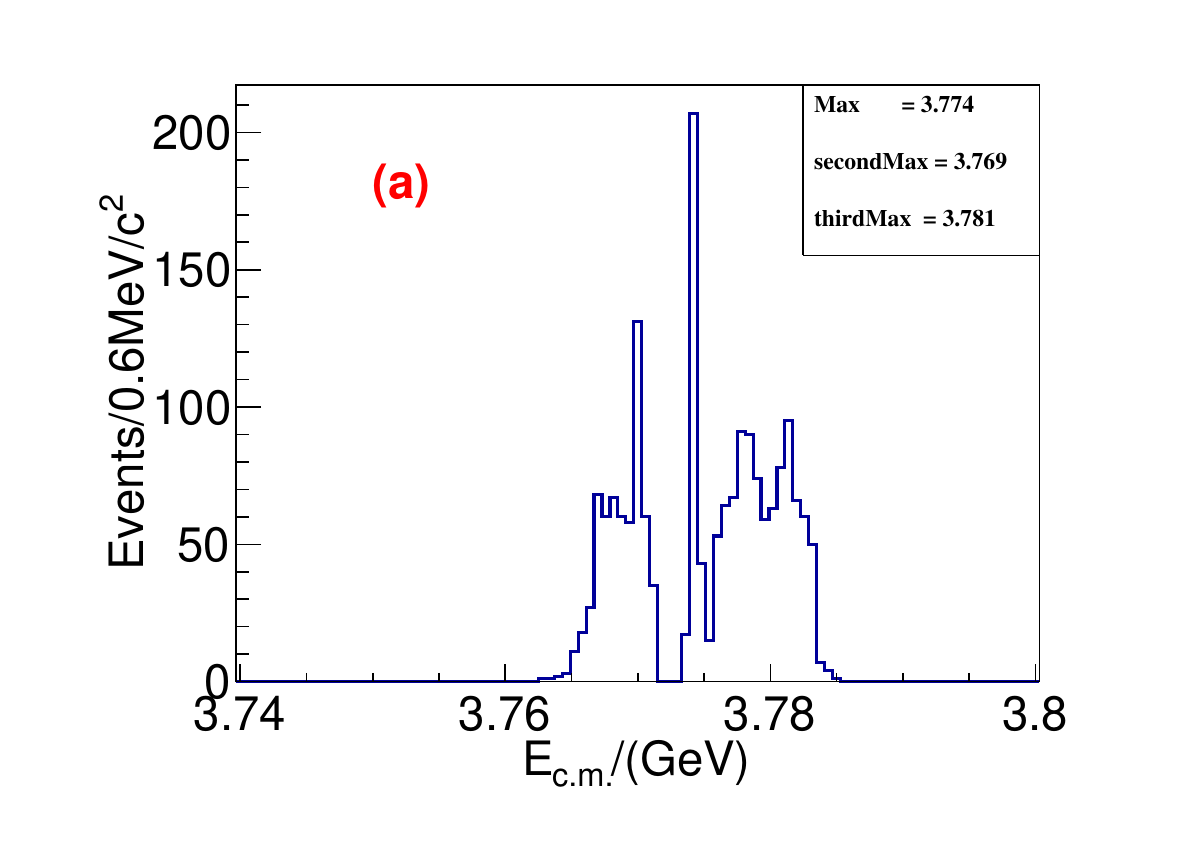} 
\includegraphics[width=8.0cm]{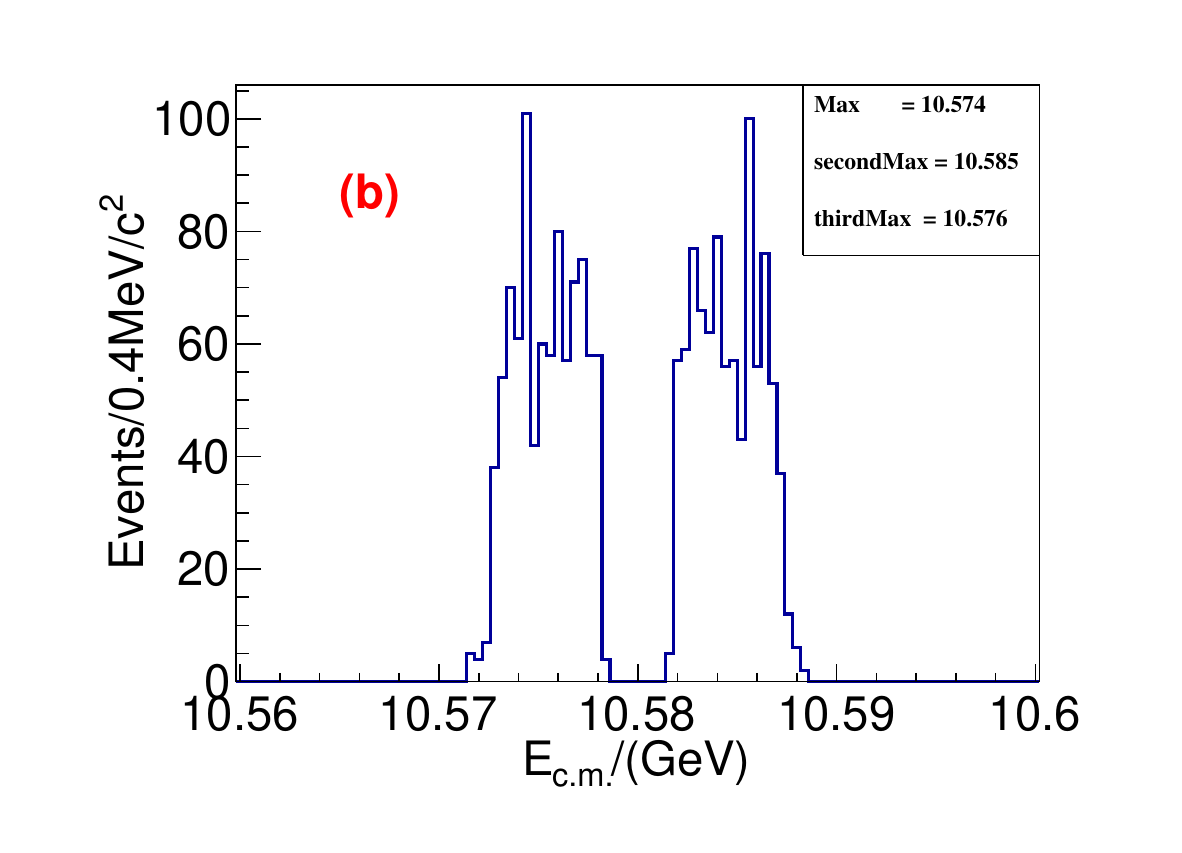}  
\caption{Distribution of optimal energy points projected onto the $\ecm^{i}$ axis from specific phase angle $\phi$ ranging from $240^\circ$ to $300^\circ$ for (a) $\psip$
decays and (b) $\ufs$ decays.}
\label{cou}
\end{center}
\end{figure}

Drawing from previous studies, a critical question emerges: Where should the optimal energy point be? In addressing
this inquiry, Ref.~\cite{3773phase} indicates that the relative phase $\phi$ is approximately $270^{\circ}$.
Accordingly, Figs.~\ref{poimin} and \ref{4smin} present the fitting results for $\phi$ ranging from $240^\circ$ to
$300^\circ$ in $10^{\circ}$ increments. The red line in each plot represents the optimal energy point corresponding to
the best precision of $E(\BR)$ as determined using the branching fraction. 
For the phase angles ranging from $240^{\circ}$ to $300^{\circ}$, the optimal energy points are projected onto the $\ecm^i$ axis, as shown in Fig.~\ref{cou}. 
This figure shows that the optimal energy points for $\psinn$ decays are $3.769~\gev$ and
$3.781~\gev$, whereas those for $\upsilonnn$ decays are $10.574~\gev$ and $10.585~\gev$. We should note that the
optimal energies will differ if the phase $\phi$ is in different ranges.

\section{Integrated luminosity and uncertainty}

\subsection{Integrated luminosity requirements for data taking strategy}
\label{sec5a}

\begin{figure}[htp]
\begin{center}
\includegraphics[width=7.5cm]{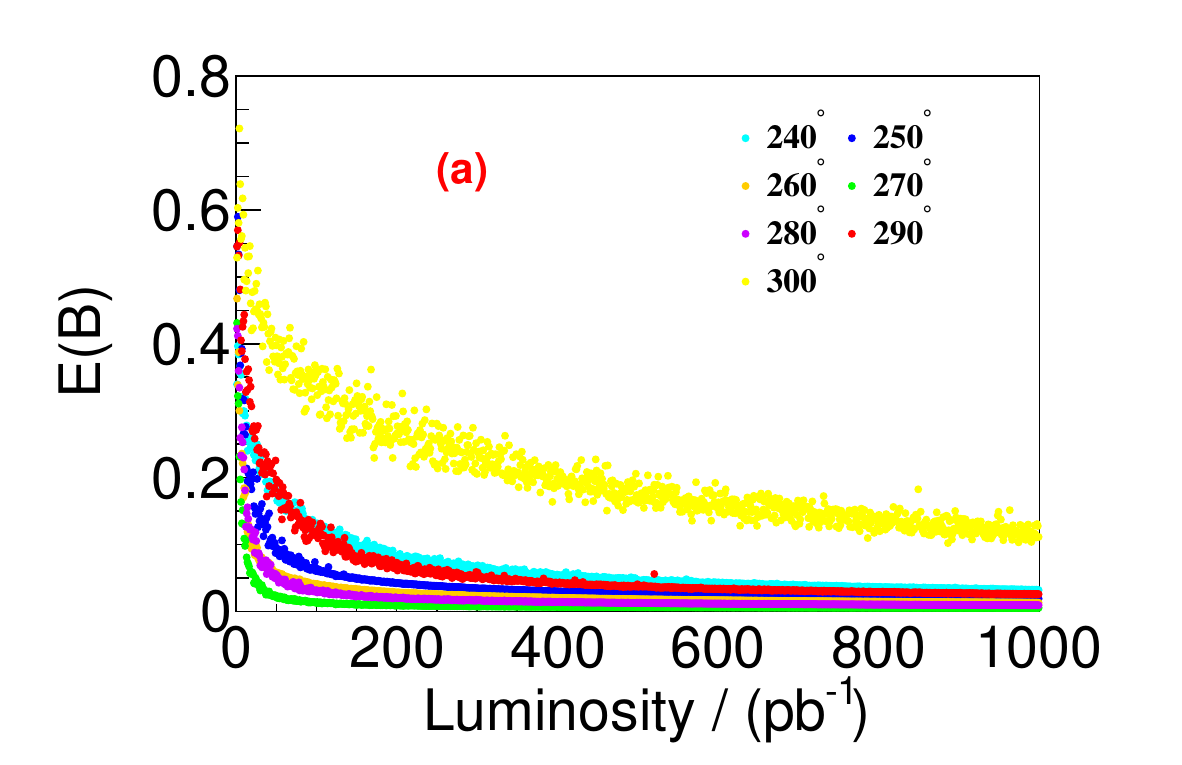}
\includegraphics[width=7.5cm]{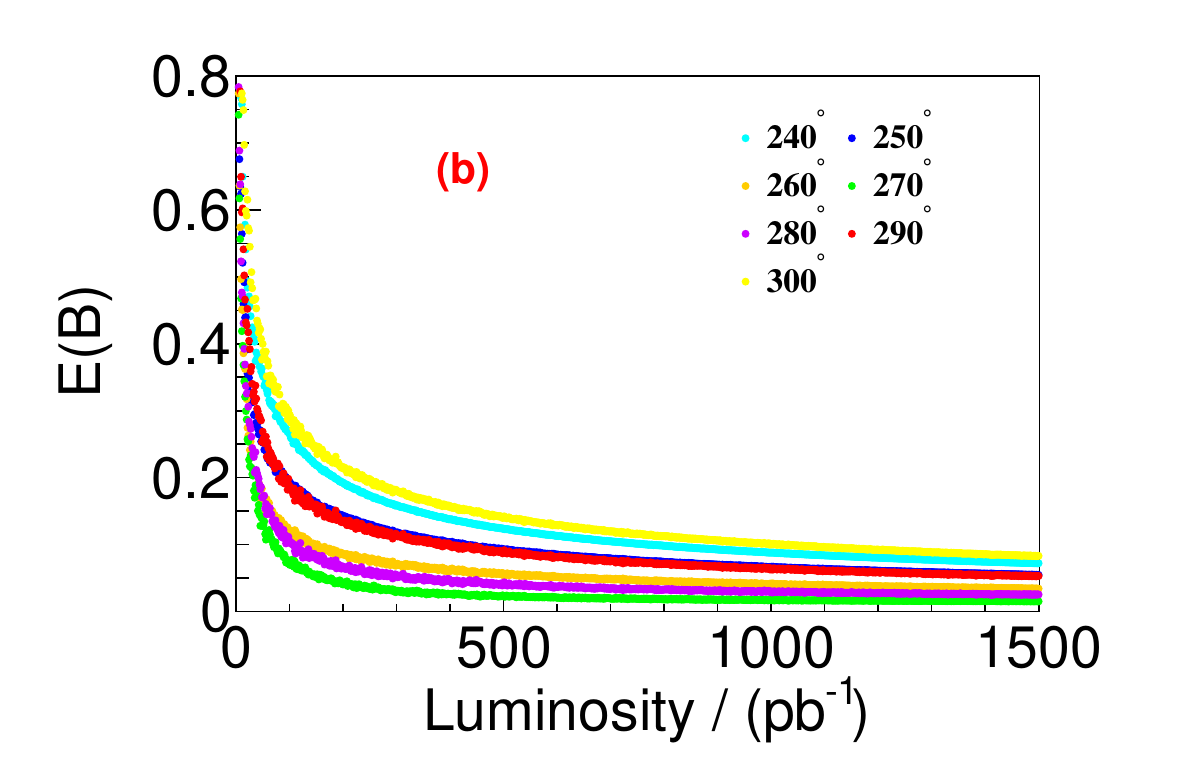}  \\
\includegraphics[width=7.5cm]{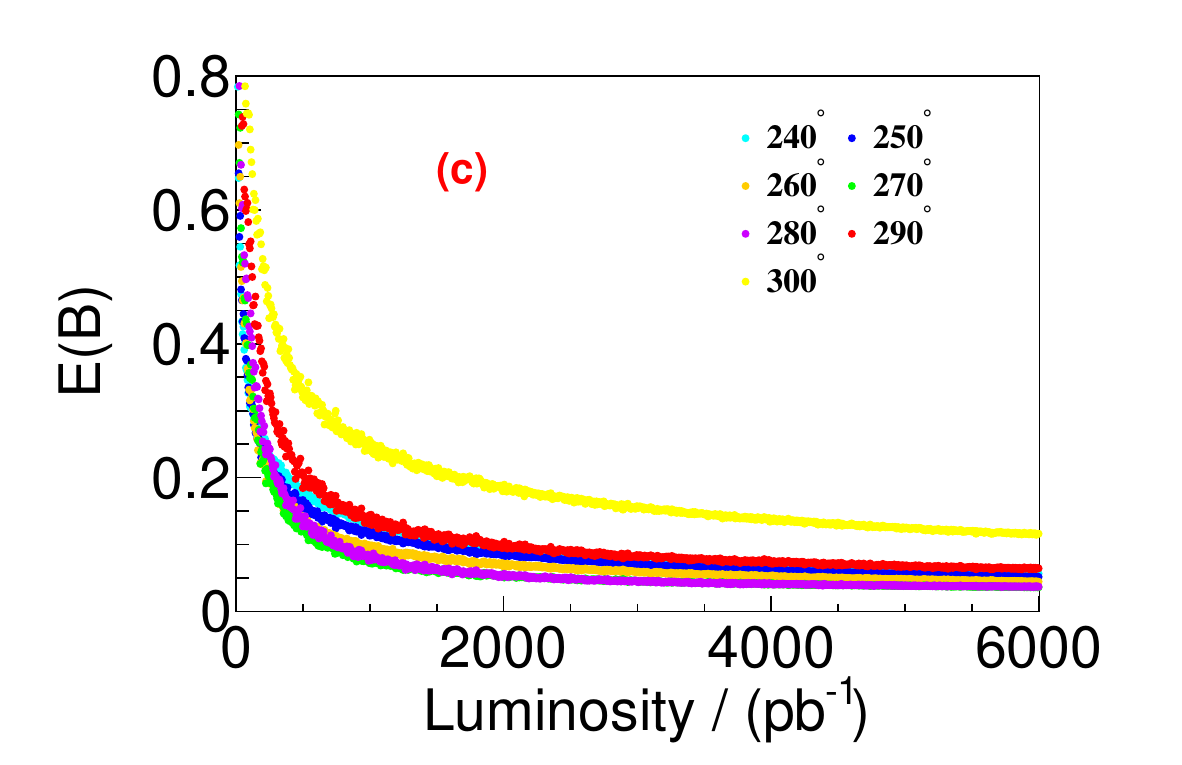}
\includegraphics[width=7.5cm]{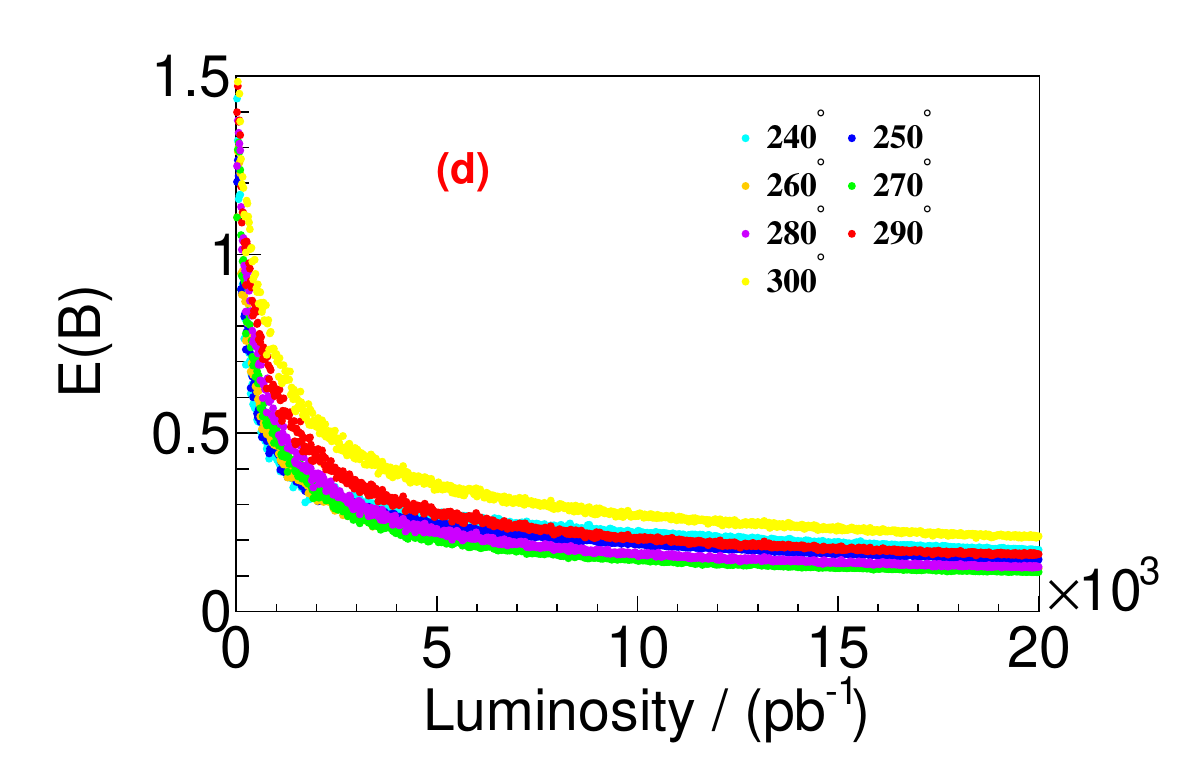}
\caption{Distributions of $E(\BR)$  with the integrated luminosity of data sample for $\psinn$ decays
with branching fractions of (a) $1 \times 10^{-3}$, (b) $1\times10^{-4}$, (c) $1\times 10^{-5}$, and (d)
$1\times10^{-6}$.}
\label{lum3773}
\end{center}
\end{figure}

\begin{figure}[htp]
\begin{center}
\includegraphics[width=7.5cm]{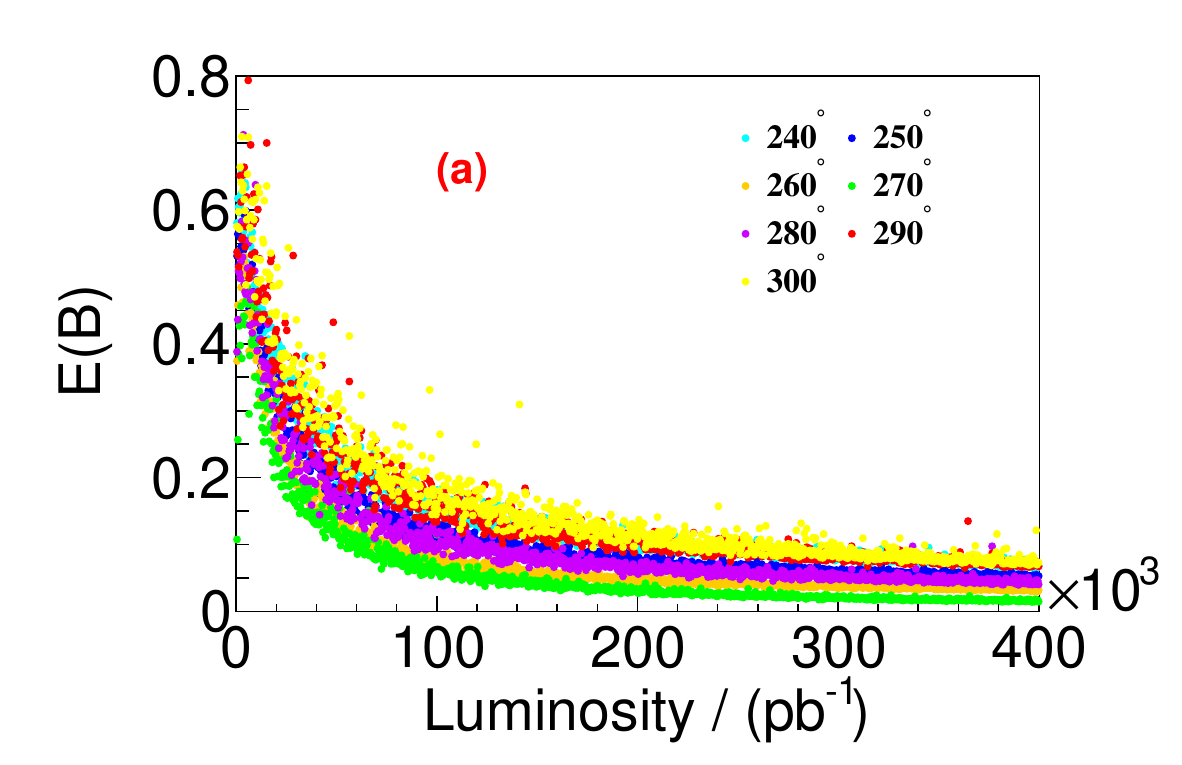}
\includegraphics[width=7.5cm]{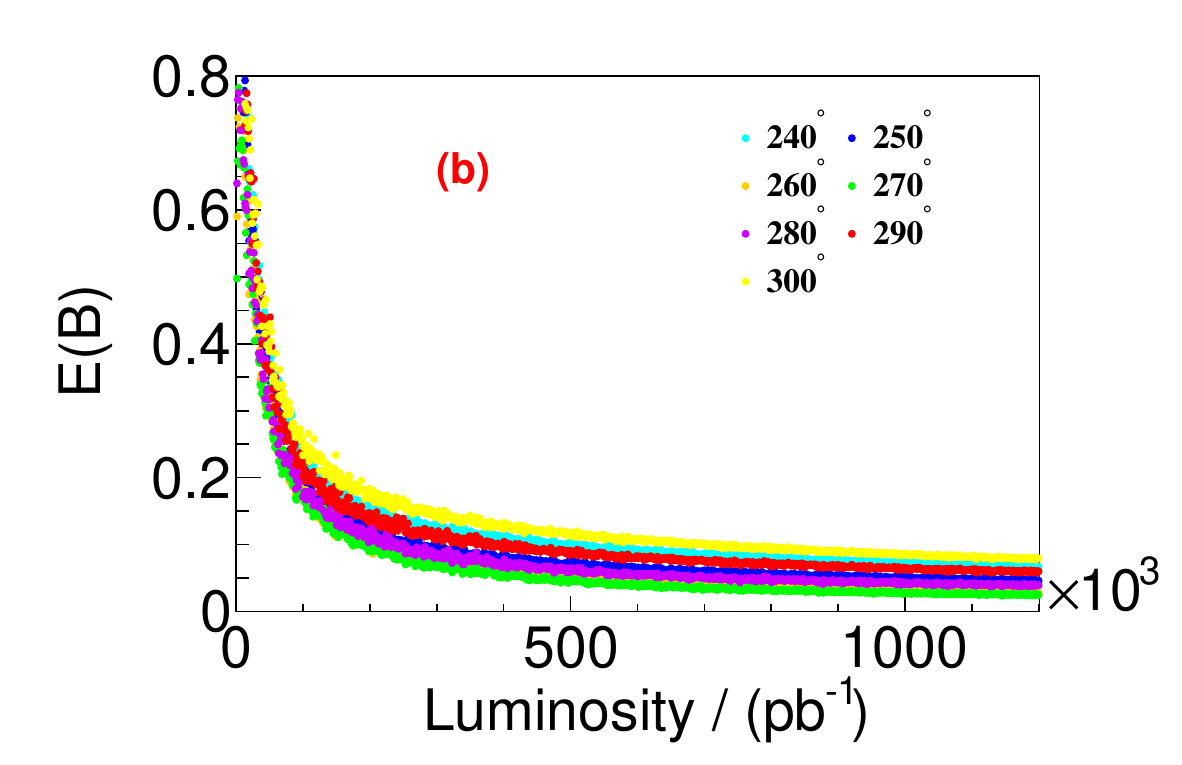}\\
\includegraphics[width=7.5cm]{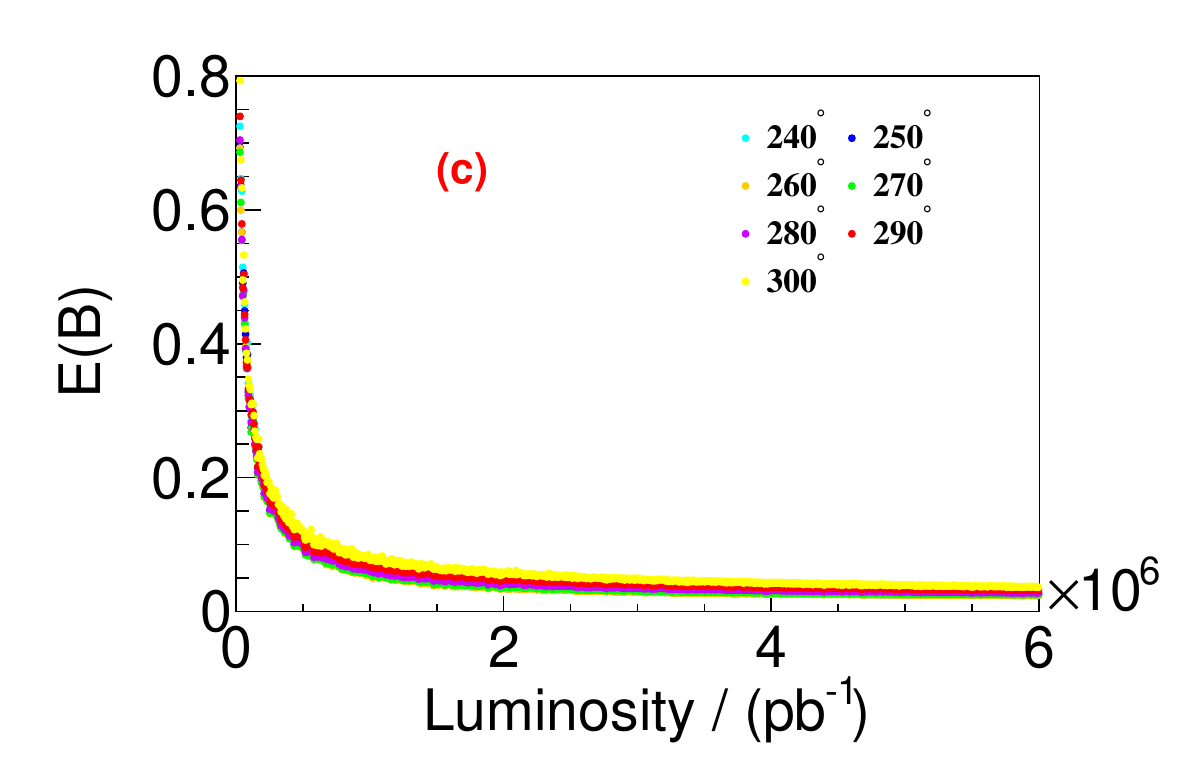}
\caption{Distributions of $E(\BR)$ with the integrated luminosity of data sample for $\upsilonnn$ decays
with branching fractions of (a) $1 \times 10^{-6}$, (b) $1 \times 10^{-7}$, and (c) $1\times 10^{-8}$.}
\label{lumufs}
\end{center}
\end{figure}

The last question is the relationship between the integrated luminosity and the precision of branching fraction
measurements. For the fitting procedure involving branching fractions that span several orders of magnitude, we
assume an equal allocation of integrated luminosity between the low-energy ($\llow$) and high-energy ($\lhigh$) data
acquisition points. The results of the integrated luminosity optimization for $\psip$ are shown in Fig.~\ref{lum3773},
where seven curves represent different relative phases ($\phi = 240^{\circ}, ~250^{\circ}, ~260^{\circ}, ~270^{\circ},
~280^{\circ}, ~290^{\circ}, ~300^{\circ}$). Similarly, Fig.~\ref{lumufs} illustrates a corresponding integrated
luminosity optimization scheme for $\ufs$. Table~\ref{lum} summarizes the integrated luminosity requirements for
an effective data acquisition strategy aimed at achieving a projected 10\% precision of $E(\BR)$ at both energy
points. The research results indicate that, when considering the maximum value of $\BR$ based on conservative estimates, a minimum integrated luminosity of $500~\inpb$ is required for $\psip$ decays,
whereas $200~\infb$ is necessary for $\ufs$ decays.

\begin{table}[htp]

\caption{Relationship between integrated luminosity and $\BR$ with $E(\BR)$ fixed to 10\%.}
\begin{tabular}{c c c c c}
\hline 
$\BR$            &  $\psinn$    &  $\BR$              & $\upsilonnn$ \\\hline
$1\times10^{-3}$ &  $500~\inpb$   &  $1\times10^{-6}$   & $200~\infb$ \\
$1\times10^{-4}$ &  $800~\inpb$   &  $1\times10^{-7}$   & $800~\infb$ \\
$1\times10^{-5}$ &  $3~\infb$     &  $1\times10^{-8}$   & $5~\inab$ \\
$1\times10^{-6}$ &  $14~\infb$    &                     & \\\hline
\end{tabular}
\label{lum}
\end{table}

\subsection{Optimization of integrated luminosity allocation}

\begin{figure}[htp]
\begin{center}
\includegraphics[width=7.5cm]{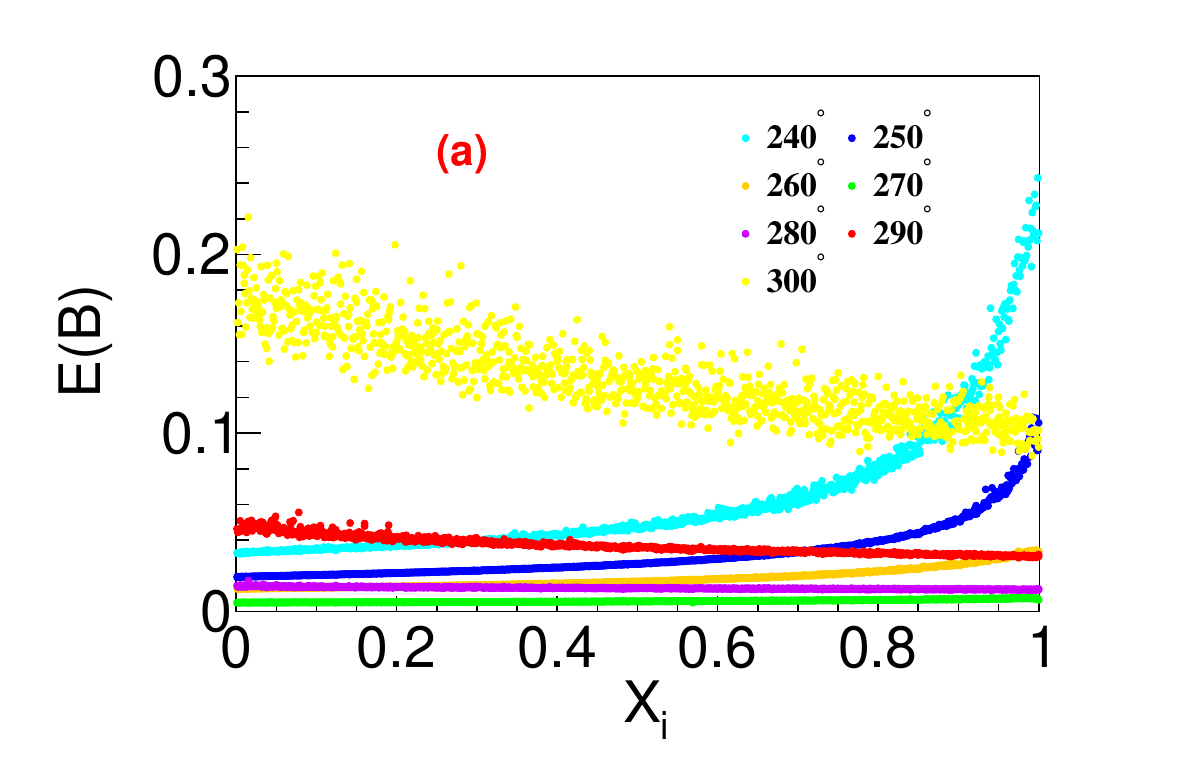}
\includegraphics[width=7.5cm]{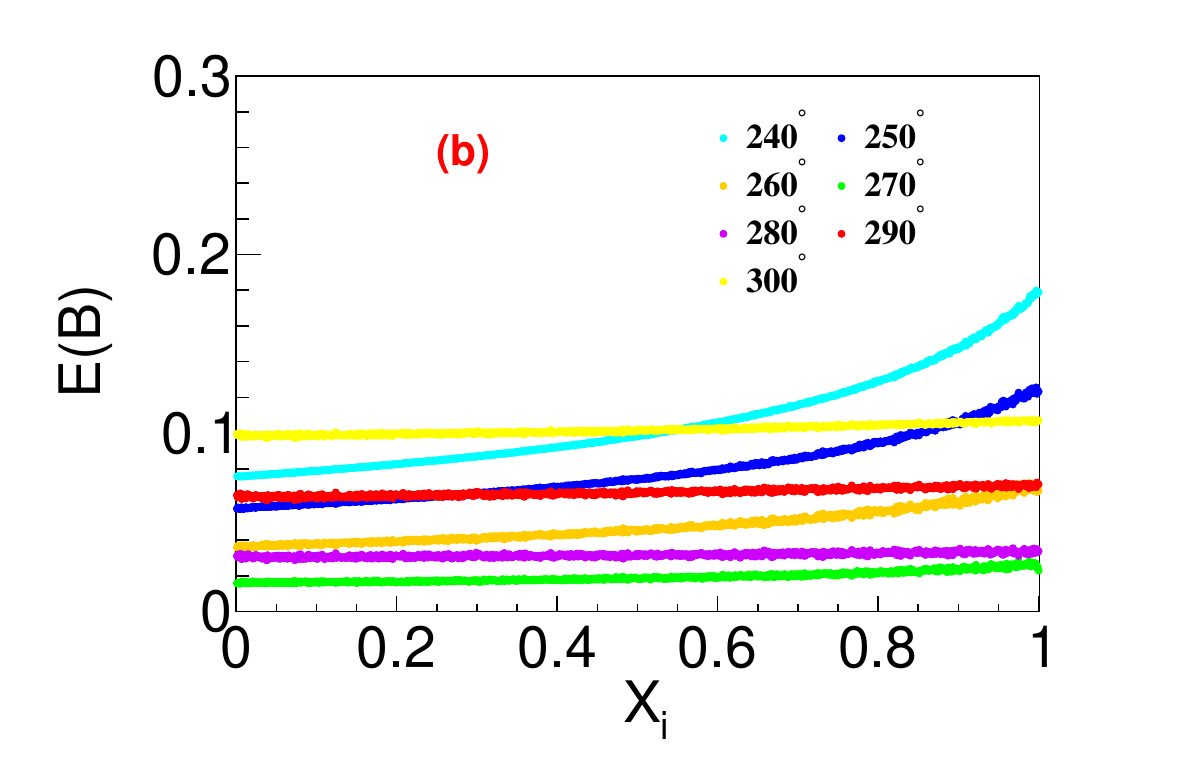}  \\
\includegraphics[width=7.5cm]{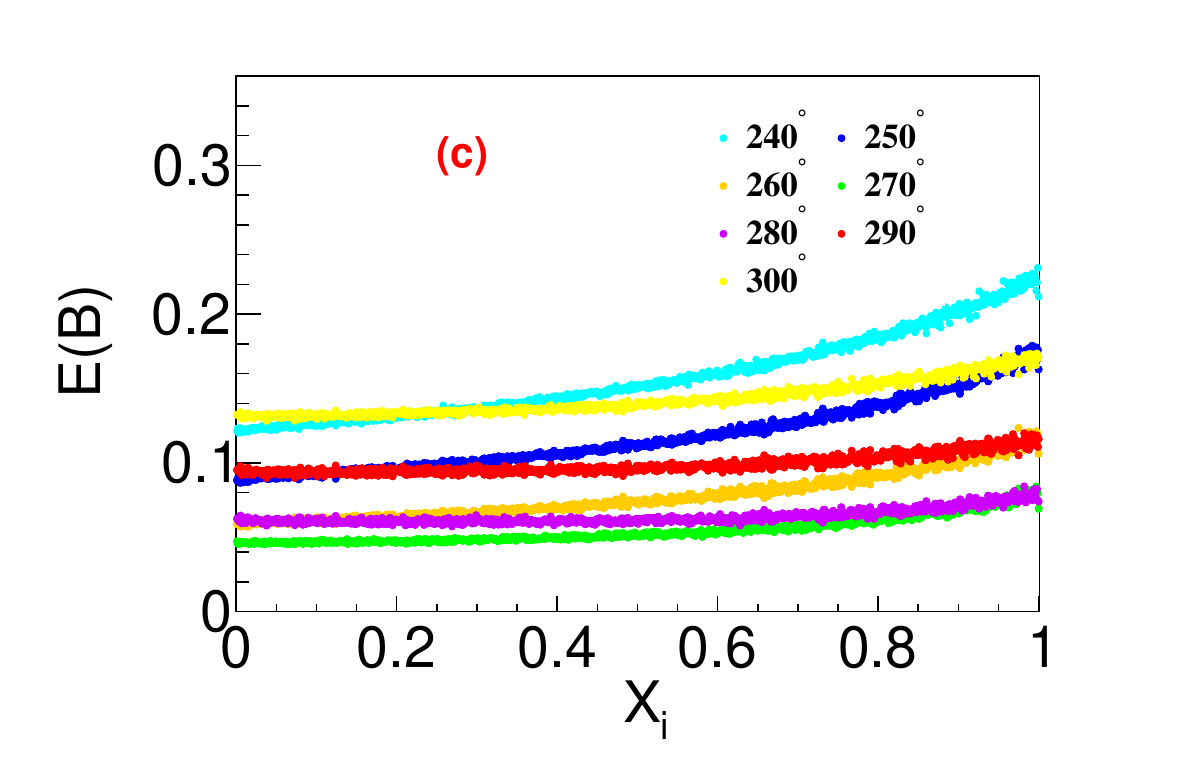}
\includegraphics[width=7.5cm]{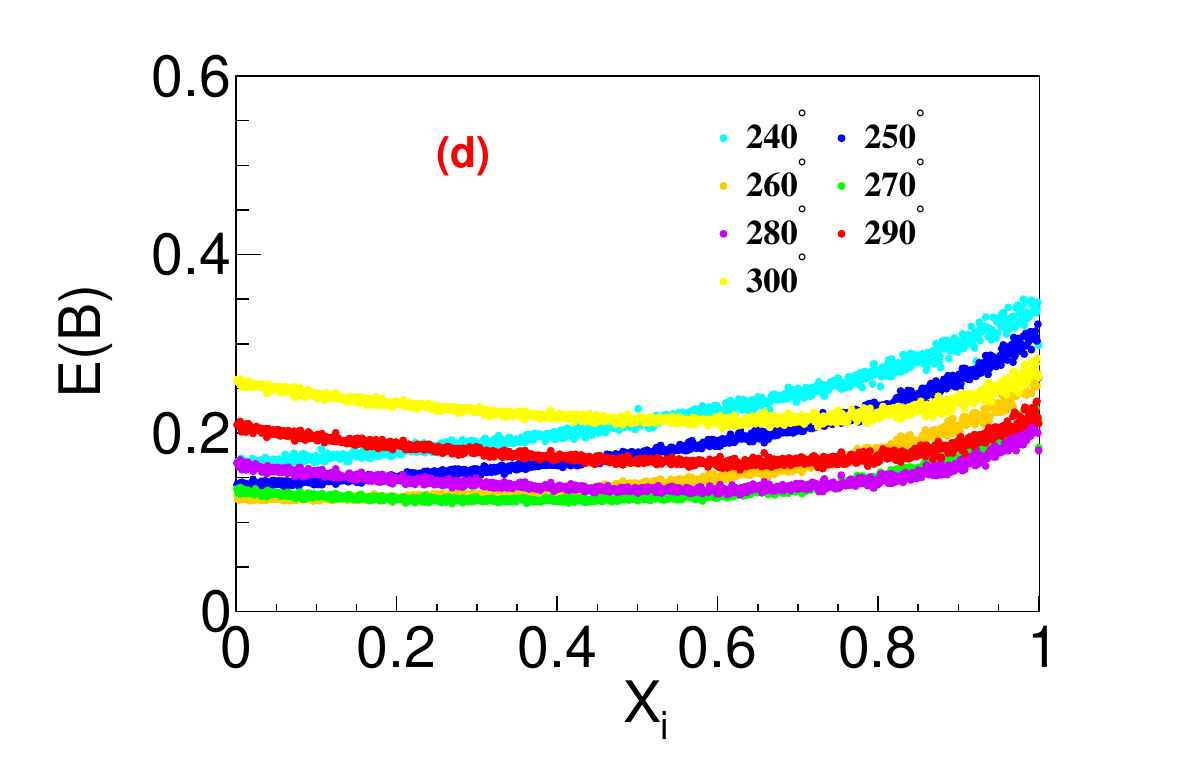}
\caption{Distributions of $E(\BR)$ with the integrated luminosity allocation ratio $X_i$ for $\psinn$
decays with branching fractions of (a) $1\times10^{-3}$, (b) $1\times10^{-4}$, (c) $1\times 10^{-5}$, and (d)
$1\times10^{-6}$. The total luminosities and $\BR$s are fixed according to Table~\ref{lum}.}
\label{lum3773_2}
\end{center}
\end{figure}

\begin{figure}[htp]
\begin{center}
\includegraphics[width=7.5cm]{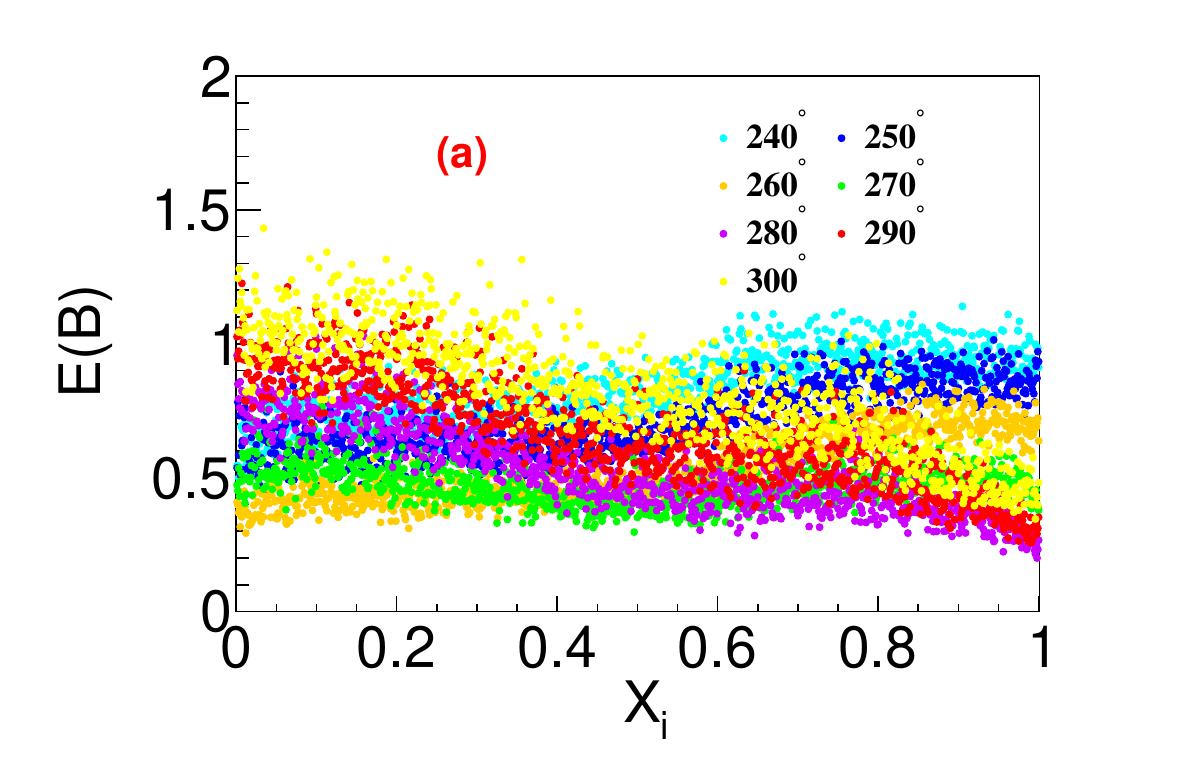}
\includegraphics[width=7.5cm]{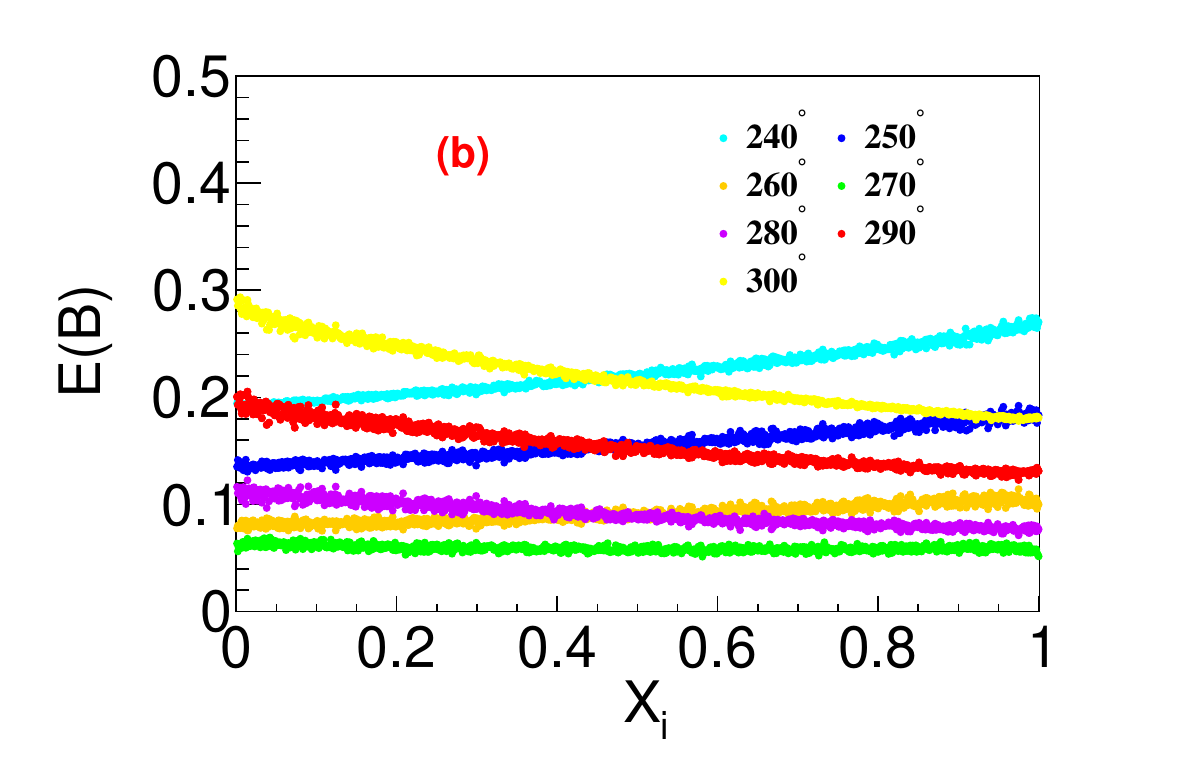}\\
\includegraphics[width=7.5cm]{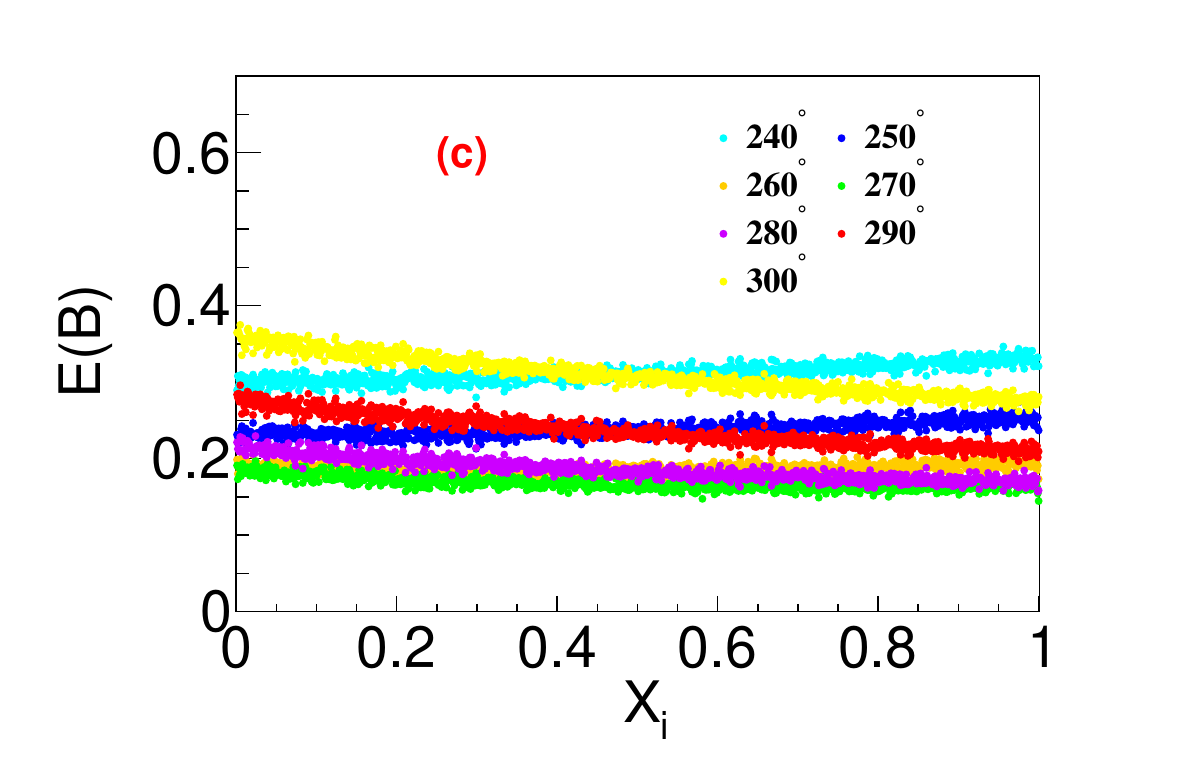}
\caption{Distributions of $E(\BR)$ with the integrated luminosity allocation ratio $X_i$ for
$\upsilonnn$ decays with branching fractions of (a) $1\times10^{-6}$, (b) $1\times10^{-7}$, and (c)
$1\times 10^{-8}$. The total integrated luminosities and $\BR$s are fixed according to Table~\ref{lum}.}
\label{lumufs_2}
\end{center}
\end{figure}

\begin{figure}[htp]
\begin{center}
\includegraphics[width=7.5cm]{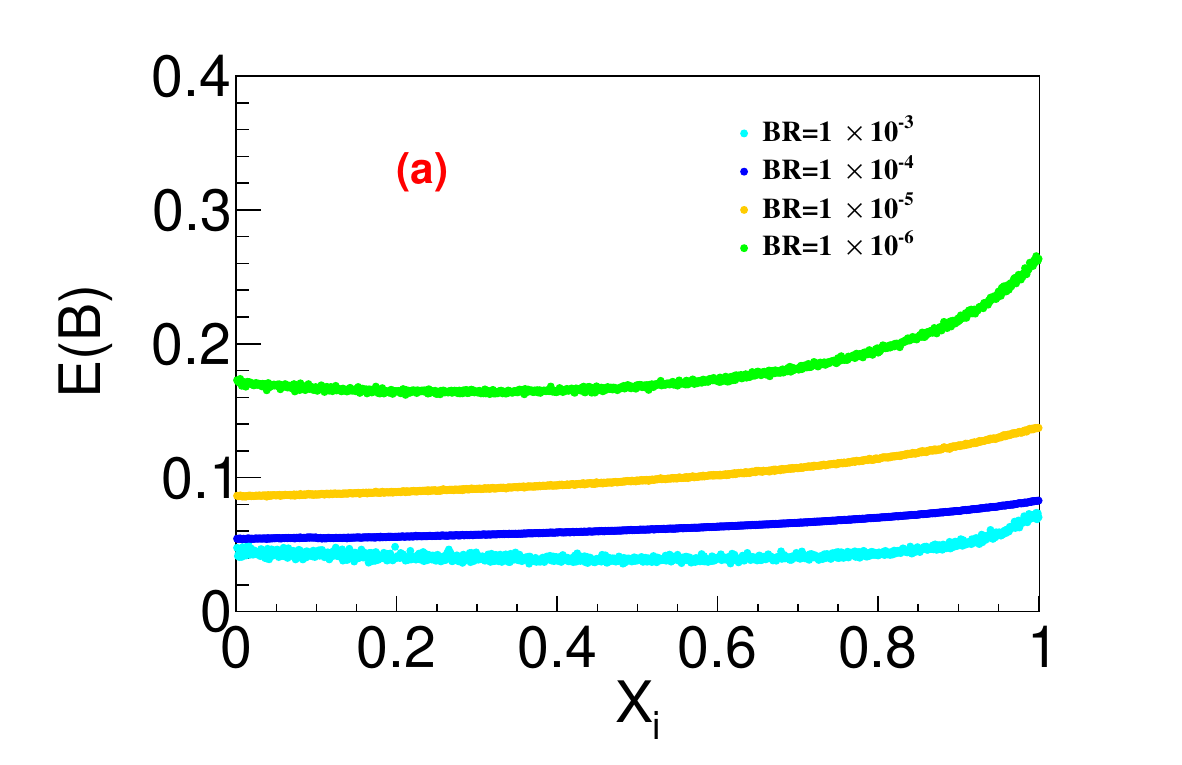}
\includegraphics[width=7.5cm]{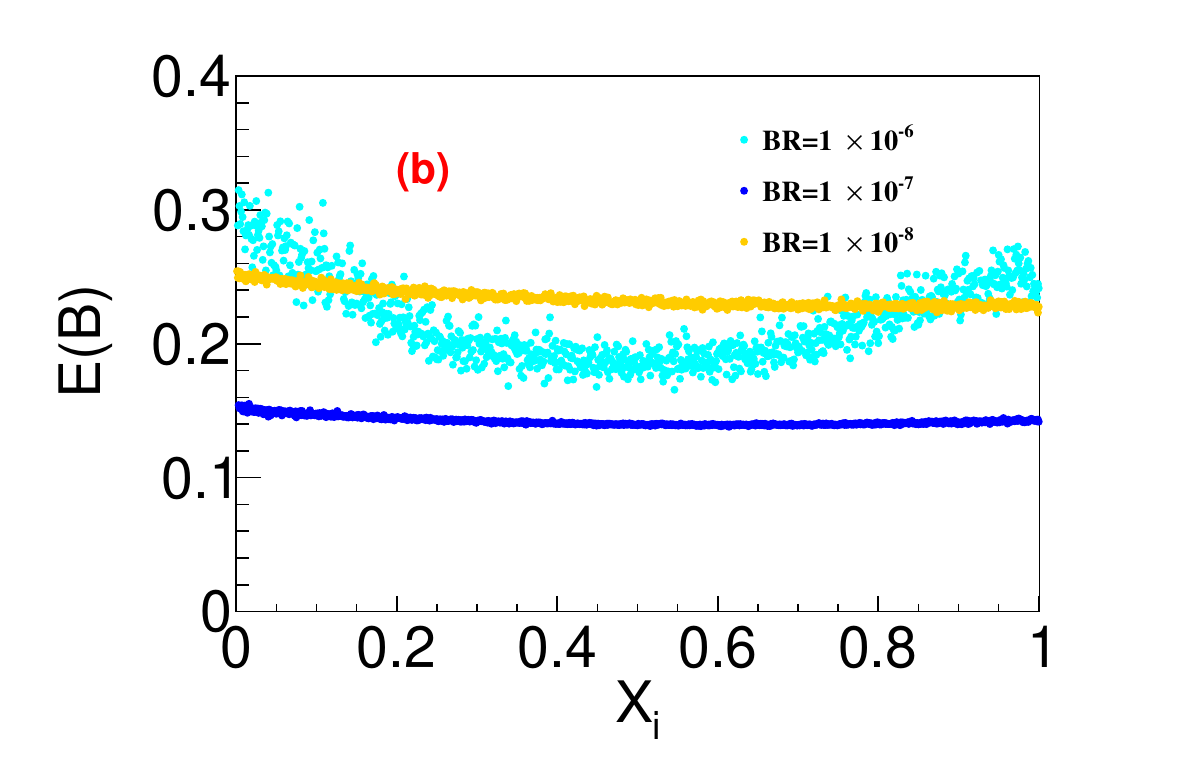}\\
\caption{Distributions of $E(\BR)$ with the integrated luminosity allocation ratio $X_i$, obtained by assigning equal weights to different phase angles at the same branching ratio, for (a) $\psinn$ decays and (b) $\upsilonnn$ decays.}
\label{fig1}
\end{center}
\end{figure}

Based on the findings presented in Sec.~\ref{sec5a}, we initially assumed an integrated luminosity ratio of
$\llow : \lhigh = 1:1$ without considering the effects of different integrated luminosities at different energy
points. To achieve optimal precision in the branching fraction measurements, we vary the parameter $\rm{X_{i}}\equiv
\llow/(\llow + \lhigh)$ to determine the corresponding $E(\BR)$. The optimized $\rm{X_i}$ values for both energy
points related to $\psip$ are shown in Fig.~\ref{lum3773_2}. Fig.~\ref{lumufs_2} illustrates a similar integrated
luminosity optimization scheme for $\ufs$. 
We assign equal weights to different phase angles corresponding to the same branching ratio, thereby achieving optimal results for that branching ratio. This is illustrated in Fig.~\ref{fig1}, where the minimum value of $E(\BR)$ occurs at $\rm{X_i=0}$ for $\psinn$ decays and $\rm{X_i=0.5}$ for
$\upsilonnn$ decays, which correspond to integrated luminosity ratios of $\llow : \lhigh = 0:1$ and $1:1$,
respectively.
 
\section{Discussion}

The minor deviations, either higher or lower than the continuum cross sections observed in Refs.~\cite{ref::ana0,
ref::ana3, ref::ana4,ref::ana5,ref::ana6}, may have indicated nonzero $\psip$ decays into light hadron final states,
and more data are required to confirm the observations. In the BESIII experiment, the peak luminosity is approximately
$1 \times 10^{33}~\cm^{-2}\s^{-1}$ at $\sqrt{s} =3.773~\gev$. A data acquisition period of 12 days is expected to be
sufficient to achieve a projected precision of 10\% in the measurements of the $\psinn$ branching fraction.

In June 2022, SuperKEKB, the asymmetric energy accelerator that provides $\EE$ collisions inside the Belle II
detector, achieved a new luminosity record of $4.7 \times 10^{34}~\cm^{-2}\s^{-1}$ and recorded $15~\infb$ data per
week. SuperKEKB aims to achieve a luminosity higher than $1 \times 10^{35}~\cm^{-2}\s^{-1}$ in the near future. This
accomplishment enables measurements with a branching fraction uncertainty of 10\% to be completed within two months.
If this larger data sample is collected slightly above the $\ufs$ peak, it will allow for the direct
measurement of the $B^{+}$ to $B^{0}$ production ratio at a new energy point, and thereby enhance the constraints on
the energy dependence of this ratio. Additionally, the study will explore potential molecular states near the
$B^{\ast} \bar{B}^{\ast}$ or $B \bar{B}^{\ast} $ threshold, search for inelastic channels such as $\pp\Upsilon(1S,2S)$
and $\eta h_b(1P)$, and investigate the violation of the isospin symmetry, among other topics.

\section{Summary}
We have employed MC simulation methodology and Fisher information to investigate various data taking
strategies for precisely measuring the branching fractions of $\psinn$ and $\upsilonnn$ decays. 
Assuming an expected phase angle $\phi$ close to $270^{\circ}$ and utilizing the cross section parameters specified in Table~\ref{para}, this analysis has
enabled us to identify an optimal measurement scheme. In response to the questions outlined in Sec.~\ref{sec:1}, we
present the following answers:
\begin{itemize}
\item The optimal energies for data taking are located at $3.769~\gev$ and $3.781~\gev$ for $\psinn$ decays and
$10.574~\gev$ and $10.585~\gev$ for $\upsilonnn$ decays;
\item Two optimal energy points are sufficient to achieve a small uncertainty in the measurements of the branching
fractions;
\item To achieve an expected precision of 10\% when considering the maximum value of $\BR$ based on conservative estimates, minimum integrated luminosities of $500~\inpb$ and $200~\infb$ with recommended allocation of $\llow : \lhigh = 0:1$ and $1:1$ are required for $\psinn$ and
$\upsilonnn$ decays, respectively.
\end{itemize}

We emphasize that the above conclusions are for an expected phase angle $\phi$ close to $270^\circ$. The outcomes
may vary significantly if the phase angle deviates significantly from this range. Nonetheless, the MC simulation and
Fisher information techniques discussed in this article can be applied similarly across all scenarios. This approach
enables us to determine optimal energies and integrated luminosity allocations in various contexts.

\section{Acknowledgments}

This work is supported by the National Key R\&D Program of China under Contract Nos. 2022YFA1601903 and 2020YFA0406300,
and National Natural Science Foundation of China under Contract Nos. 12175041, 12335004, and 12361141819.

\end{document}